\documentclass[twocolumn,showpacs,superscriptaddress,preprintnumbers,amsmath,amssymb,longbibliography]{revtex4-1}
\usepackage[bookmarks=true,colorlinks=true,citecolor=blue,linkcolor=blue,urlcolor=magenta]{hyperref}
\usepackage{dcolumn}
\usepackage{multirow}
\usepackage{dcolumn}
\usepackage{bm}
\usepackage{xfrac}
\usepackage{calc}
\usepackage{graphicx}
\usepackage[tight,footnotesize]{subfigure}
\usepackage{color}
\usepackage{longtable}
\usepackage{dashrule}
\usepackage{multirow}%
\usepackage{rotating}
\usepackage{float}
\usepackage{perpage}
\usepackage{morefloats}
\usepackage{latexsym}
\usepackage{gensymb} 
\usepackage{bbm}
\usepackage[dvipsnames]{xcolor}
\ifpdf
\graphicspath{
	{figures/}
	{figures/ActDensity/}
	{figures/QQpotential/}
	{figures/SelfInt/}	
	{figures/stringWidth/}		
	{figures/6updates_WidthFits/}
}
\else
\graphicspath{
	{figures/}
	{figures/ActDensity/}	
	{figures/QQpotential/}	
	{figures/SelfInt/}
	{figures/stringWidth/}	
	{figures/6updates_WidthFits/}
}
\fi

\DeclareFontFamily{OT1}{pzc}{}
\DeclareFontShape{OT1}{pzc}{m}{it}{<-> s * [0.900] pzcmi7t}{}
\DeclareMathAlphabet{\mathpzc}{OT1}{pzc}{m}{it}

\begin{document}

\preprint{AIP/123-QED}

\title[]{On QCD strings beyond non-interacting model}
\author{A. S. Bakry}
\affiliation{Institute of Modern Physics, Chinese Academy of Sciences, Gansu 730000, China}
\author{M. A. Deliyergiyev}\email[]{maksym.deliyergiyev@ujk.edu.pl}
\affiliation{Institute of Physics, The Jan Kochanowski University in Kielce, 25-406, Poland}
\affiliation{Institute of Modern Physics, Chinese Academy of Sciences, Gansu 730000, China}
\author{A. A. Galal}
\affiliation{Department of Physics,  Al Azhar University, Cairo 11651, Egypt}
\author{M. KhalilA Williams}
\affiliation{Department of Physics,  Al Azhar University, Cairo 11651, Egypt}
\affiliation{Department of Mathematics, Bergische Universit\"at Wuppertal, 42097 Germany}
\affiliation{Department of Physics, University of Ferrara, Ferrara 44121, Italy}
\affiliation{Research and computing center, The Cyprus Institute, Nicosia 2121, Cyprus}

\date{January 12, 2020}

\begin{abstract}

  We investigate the implications of Nambu-Goto (NG), Lüscher-Weisz (LW) and Polyakov-Kleinert (PK) string actions for the Casimir energy of the QCD flux-tube at one and two loop order at finite temperature. We perform our numerical study on the 4-dim pure SU(3) Yang-Mills lattice gauge theory at finite temperature and coupling $\beta=6.0$. The static quark-antiquark potential is calculated using link-integrated Polyakov loop correlators. At a high temperature-close to the critical point- We find that the rigidity and self-interactions effects of the QCD string to become detectable. The remarkable feature of this model is that it retrieves a correct dependency of the renormalized string tension on the temperature. Good fit to static potential data at source separations $R \ge 0.5$ fm is obtained when including additional two-boundary terms of (LW) action. On the other-hand, at a lower temperature-near the QCD plateau- We detect signatures of two boundary terms of the Lüscher-Weisz (LW) string action. The (LW) string with boundary action is yielding a static potential which is in a good agreement with the lattice Monte-Carlo data, however, for color source separation as short as $R=0.3$ fm.
  
\end{abstract}


\maketitle

\section{Introduction}
   The formulation of a string theory for hadrons has been an attractive proposal since the phenomenological success in explaining Venziano formula~\cite{Veneziano:1968yb} even before the formulation of quantum chromodynamics (QCD). Despite the difficulties encountered in the quantization scheme in a fundamental string theory, the proposal to describe the long-distance dynamics of strong interactions inside the hadrons by a low energy effective string~\cite{Luscherfr,Luscher:2002qv} has remained an alluring conjecture. 

  String formation~\cite{FUKUGITA1983374, Cea:1993pi, Cea:2015wjd, FLOWER1985128, Otto:1984qr, AM, 10.1143, PhysRevD.36.3297, SOMMER1987673, etde_6934022, DIGIACOMO1990441, Bali:1994de} is realized in many strongly correlated systems and is not an exclusive property of the QCD color tubes~\cite{PhysRevB.36.3583,PhysRevB.78.024510,2007arXiv0709.1042K,Nielsen197345,Lo:2005xt}. The normalization group equations imply that the system flows towards a roughening phase where the transverse string oscillations to the classical world sheet effects becomes measurable and can be verified in numerical simulations of lattice gauge theories (LGT).  
  
  In the leading Gaussian approximation of the NG action, the quantum fluctuations of the string bring forth a universal quantum correction to the linearly rising potential well known as the L\"uscher term in the mesonic configurations. In the baryon a geometrical L\"uscher-like terms~\cite{Jahn2004700,deForcrand:2005vv} ought to manifest.

    The width due to the quantum delocalizations of the string grows logarithmically~\cite{Luscher:1980iy} as the two color sources are pulled apart. The character of logarithmic broadening is expected for the baryonic junction~\cite{PhysRevD.79.025022} as well. 
    Precise lattice measurements of the $Q\bar{Q}$ potential in $SU(3)$ gauge model are in consistency with the L\"uscher subleading term for color source separation commencing from distance $R = 0.4$ fm~\cite{Luscher:2002qv}. The effective description is expected to hold over distance scale  $1/T_c$~\cite{Caselle:2012rp} where the effects of the intrinsic thickness~\cite{Vyas:2010wg} of the flux tube diminish. Many gauge models have unambiguously identified the L\"uscher correction to the potential with unprecedential accuracy~\cite{Juge:2002br, HariDass2008273, Caselle:2016mqu,caselle-2002,Pennanen:1997qm,Brandt:2016xsp}.

    In addition, the string model predictions for the logarithmic broadening~\cite{Luscher:1980iy} of the mean-square width of the string at very low temperatures has been observed in several lattice simulations corresponding to the different gauge groups~\cite{Caselle:1995fh,Bonati:2011nt,HASENBUSCH1994124,Caselle:2006dv,Bringoltz:2008nd,Athenodorou:2008cj,Juge:2002br,HariDass:2006pq,Giudice:2006hw,Luscher:2004ib,Pepe:2010na,Bicudo:2017uyy}
  
    In the high temperature regime of QCD, the overlap of the string's excited states spectrum would lead to a new quantum state with different characteristics. The free string approximation implies a decrease in the slop of the $Q\bar{Q}$ potential or in other-words a temperature-dependent effective string tension~\cite{PhysRevD.85.077501, Kac}. The leading-order correction for the mesonic potential turns into a logarithmic term of the Dedekind $\eta$ function which encompasses the L\"uscher term as a zero temperature $T=0$ limiting term ~\cite{Luscher:2002qv,Gao,Pisarski}. With respect to the string's width the logarithmic broadening turns into a linear pattern before the deconfinement is reached from below~\cite{allais,Gliozzi:2010zv,Caselle:2010zs,Gliozzi:2010jh}.

   Nevertheless, this non-interacting model of the bosonic string derived on the basis of the leading order formulation of NG action poorly describes the numerical data in the intermediate distances at high temperatures. For instance, substantial deviations~\cite{PhysRevD.82.094503,Bakry:2010sp,Bakry:2011zz,Bakry:2012eq} from the free string behavior have been found for the lattice data corresponding to temperatures very close to the deconfinement point. 

   The comparison with the lattice Monte-Carlo data supports the validity of the leading-order approximation at source separations larger than $R>0.9$ fm~\cite{PhysRevD.82.094503,Bakry:2010sp,Bakry:2012eq} for both the $Q\bar{Q}$ potential and the color-tube width profile. The descripancies casts over source separation distances at which the leading-order string model predictions are valid~\cite{Luscher:2002qv} at zero temperature. In the baryon ~\cite{Bakry:2014gea,Bakry:2016uwt,Bakry:2011kga}, taking into account the length of the Y-string between any two quarks, we found a similar behavior~\cite{refId0}.  

  The fact that the lattice data substantially deviate from the free string description at the intermediate distances and high temperatures has induced many numerical experiments to verify the validity of higher-order model-dependent corrections for the NG action~\cite{Caselle:2004jq,Caselle:2004er} before the string breaks~\cite{Bali:2005bg}.

  In the Nambu-Goto (NG) framework it seems that there is no reason to believe that all orders of the power expansion are relevant to the correct behavior of QCD strings~\cite{Giudice:2009di}. For example, a first-order term deviating from the universal behavior has been determined unambiguously in 3D percolation model ~\cite{Giudice:2009di}, no numerical evidence indicating universal features of the corrections beyond the L\"uscher term have been encountered among  $Z(2)$, $SU(2)$ and $SU(3)$ confining gauge models~\cite{Caselle:2004jq}. Numerical simulations of different gauge models in different dimensions may culminate in describing both the intermediate and long string behavior with different effective strings~\cite{Caselle:2004jq}. 


  However, the modeling of QCD flux-tubes beyond the free string approximation may suggest considering other possiblities such as string's resistance to bending. These strings with rigid structure ought to exhibit smooth fluctuations~\cite{POLYAKOV19, Kleinert:1986bk}. The idea that QCD strings may be rigid appeared long ago and been extensively scrutinized by Kleinert and German.

  The perturbative two loop potential at $T=0$, the exact potential in the large dimension limit~\cite{Braaten:1987gq,Kleinert:1989re}, the dynamical generation of the string tension~\cite{Kleinert:1988vq} have been studied, for example. The theory has other well-founded thermodynamical characteristics such as the partition function~\cite{Elizalde:1993af}, free energy and string tension at finite temperature~\cite{Viswanathan:1988ad,German:1991tc,Nesterenko:1997ku} and the deconfinement transition point~\cite{Kleinert:1987kv}.
    
    In this effective string theory only the smooth flux-sheets over long distance are favored and the sharply creased surfaces are excluded. This implies peculiar geometrical chracteristics that is being controlled via the extrinsic curvature or shape tensor of the surfaces.

    The string's rigidity can intuitively understood in relevance to the vortex line picture of the string which indicates a repulsive~\cite{PhysRevD.51.1842} nature among the flux tubes. This interpretation seems consisent with flux network within baryon~\cite{Bakry:2014gea,Bakry:2016uwt,Bakry:2011kga} which indicates that the sharply-creased flux sheets are energtically unfavorable. The strings appears aligning itself, either via temperature change or the color sources location, such that the angles between the three flux tubes are equally divided into $120^{o}$~\cite{Bakry:2014gea,Bakry:2016uwt,Bakry:2011kga}.

    
   Reviving interests appeared recently to address the rigidity of the QCD flux-tube in the numerical simulations of the confining potential. In fact, both $U(1)$ compact gauge group~\cite{Caselle:2014eka, Caselle:2016mqu} and $SU(N)$ gauge theories in $3D$~\cite{Brandt:2017yzw} has been reported.

    The contribution of the boundary action to the open string partition function come into play to recover the symmetry breaking by the cylindrical boundary condition. Two-variant formulas for the Lorentz-invariant boundary corrections to the static $Q\bar{Q}$ potential have been calculated in both the Wilson and the Polyakov-loop correlators cases~\cite{Caselle:2014eka,Aharony:2010cx, Aharony:2009gg, Aharony:2011gb}

    These corrections are hoped to reflect some features of the fine structure of the profile of QCD flux-tube~\cite{Bakry:2018kpn,Bakry:2017utr,Brandt:2017yzw, Brandt:2010bw, Battelli:2019lkz,Dubovsky:2013gi} at relatively short distances/low temperatures, larger distances/high temperature~\cite{Juge:2002br} or the excited spectrum of the flux-tubes~\cite{Juge:2002br, Bicudo:2018yhk,Juge1998543}. Indeed, detectable effects for the boundary corrections to the static quark potential have been shown viable on mitigating the deviations from predictions of the effective string and numerical outcomes~\cite{Billo:2011fd,Caselle:2014eka, Brandt:2017yzw}. We report similar findings in regard to the width profile of the QCD flux-tube near the critical point~\cite{}.
     
   One goal of the present paper is to examine the Lorentz-invariant boundary terms in Lüscher-Weisz (LW) effective string action for open string with Dirichlet boundary condition on a cylinder. The analytic estimate laid out for the static potential resulting from two boundary terms at the order of fourth and six derivative could be compatible with the energy fields set up by a static mesonic configurations.
    
   The pure $SU(3)$ Yang-Mills theory in four dimensions is the closest approximation to full QCD. Even though, we are lacking detailed understanding for the string behavior at high temperature and short distance scale. In this region the deviations from the free string behavior occurs on scales that is relevant to full QCD before the string breaks~\cite{Bali:2005bg}. The nature of the QCD strings at finite temperatures can be very relevant to many portrayals involving high energy phenomena~\cite{Caselle:2015tza,GIDDINGS198955} such as mesonic spectroscopy~\cite{Bali:2013fpo,Kalashnikova:2002zz,Grach:2008ij}, glueballs~\cite{Caselle:2013qpa,Johnson:2000qz} and string fireballs~\cite{Kalaydzhyan:2014tfa}, for example. This calls for a discussion of the validity of string effects beyond the free bosonic Nambu-Goto string which is the target of this report.

  The map of the paper is as follows: In section(II), we review the most relevant string model to QCD and discuss the lattice data corresponding to the Casimir energy versus different approximation schemes. In section(III), the numerical investigation is focused on the width profile of the energy density and the predictions of the Nambu-Goto (NG) and Polyakov-Kleinert (PK) strings. Concluding remarks and summary are provided in the last section.
  

\section{String actions and Casmir energy}
  The conjecture Yang-Mills (YM) vacuum admits the formation of a very thin string-like object~\cite{Nambu:1974zg, Nambu1979372, Luscherfr} has originated in the context of the linear rise property of the confining potential between color sources. The intuition is in consistency~\cite{Olesen:1985pv} with the dual superconductivity~\cite{Mandelstam76, Bali1996, DiGiacomo:1999a, DiGiacomo:1999b, Carmona:2001ja, Caselle:2016mqu, Pisarski} property of the QCD vacuum. The color fields are squeezed into a confining thin string dual to the Abrikosov line by the virtue of the dual Meissner effect.
  
  The formation of the string condensate spontaneously breakdown the transnational and rotational symmetries of the YM-vacuum and entails the manifestation of (D-2) massless transverse Goldstone modes in addition to their interactions~\cite{goddard, Low:2001bw}.

  To establish an effective string description, a string action can be constructed from the derivative expansion of collective string co-ordinates satisfying Poincare and parity invariance. One particular form of this action is the L\"uscher and Weisz ~\cite{Luscherfr,Luscher:2002qv}, in the physical gauge~\cite{Dubovsky:2012sh,Aharony:2011ga}, which encompasses built-in surface/boundary terms to account for the interaction of an open string with boundaries. The L\"uscher and Weisz~\cite{Luscher:2004ib} (LW) effective action up to four-derivative term read  

  \begin{equation}
  \begin{split}  
  S^{\rm{LW}}[X]&=\sigma_{0} A+\dfrac{\sigma_{0}}{2} \int d^2\zeta \Bigg[\left(\dfrac{\partial \bm{X}}{\partial \zeta_{\alpha}} \cdot \dfrac{\partial \bm{X}}{\partial \zeta_{\alpha}}\right)\\
    &+ \kappa_2 \left(\dfrac{\partial \bm{X}}{\partial \zeta_{\alpha}} \cdot \dfrac{\partial \bm{X}}{\partial \zeta_{\alpha}}\right)^2 +\kappa_3 \left(\dfrac{\partial \bm{X}}{\partial \zeta_{\alpha}} \cdot \dfrac{\partial \bm{X}}{\partial \zeta_{\beta}}\right)^2\Bigg]\\
  &+\gamma \int d^2\zeta \sqrt{g} \mathcal{R} +\alpha_{r} \int d^2 \zeta \sqrt{g} \mathcal{K}^2+S^{b}
\end{split}
\label{LWaction} 
\end{equation}
  with the physical gauge $X^{1}=\zeta_{0}, X^{4}=\zeta_{1} $ which restricts the string fluctuations to transverse directions  ${\cal C}$. The vector $X^{\mu}(\zeta_{0},\zeta_{1})$  maps the region ${\cal C}\subset \mathbb{R}^{2}$ into $\mathbb{R}^{4}$ and couplings $\kappa_1$, $\kappa_2$ are effective low-energy parameters. Invariance under party transform would keep only even number derivative terms. 


  The open-closed duality~\cite{Luscher:2004ib} imposes constraint on the kinematically-dependent couplings $\kappa_2, \kappa_3$
\begin{equation}
\kappa_2 + \kappa_3 = \dfrac{-1}{8\sigma_{0}}.
\label{couplings2}
\end{equation}
which can be shown~\cite{Billo:2012da,Aharony:2009gg} through a nonlinear realization of Lorentz transform is valid in any dimension $d$.
   The above condition Eq.~\eqref{couplings2} implies that all the terms with only first derivatives in the effective string action Eq.~\eqref{LWaction} coincide with the corresponding one of Nambu-Goto action in the derivative expansion. The Nambu-Goto action is the most simple form of string actions proportional to area of the world-sheet

\begin{equation}
\begin{split}  
  S^{\rm{NG}}[X]&=\sigma_{0} \,\int \, d^{2}\zeta  \,\sqrt{\left(\,{1+\left(\frac{\partial X}{\partial \zeta_{0}} \right)^{2}+\left(\frac{\partial X}{\partial \zeta_{1}}\right)^{2}} \right) }\\
  S^{\rm{NG}}_{\rm{\ell o}}[X]&=\sigma_{0} A+\dfrac{\sigma_{0}}{2} \int d\zeta^{2} \left(\dfrac{\partial \bm{X} }{\partial \zeta_{\alpha} } \cdot \dfrac{\partial \bm{X} }{\partial \zeta_{\alpha}} \right),\\ 
S^{\rm{NG}}_{\rm{n\ell o}}[X]&=\sigma_{0} \int d\zeta^{2} \left[ \left(\dfrac{\partial \bm{X}}{\partial \zeta_{\alpha}} \cdot \dfrac{\partial \bm{X}}{\partial \zeta_{\alpha}}\right)^2 + \left(\dfrac{\partial \bm{X}}{\partial \zeta_{\alpha}} \cdot \dfrac{\partial \bm{X}}{\partial \zeta_{\beta}}\right)^2\right],
\label{NGpert}
\end{split}
\end{equation}
where $g$ is the two-dimensional induced metric on the world sheet embedded in the background $\mathbb{R}^{4}$. On the quantum level the Weyl invariance of the NG action is broken in four dimensions; however, the anomaly is known to vanish at large distances \cite{Olesen:1985pv}.\\

  The boundary term $S^{b}$ describes the interplay between the effective string with the Polyakov loops~\cite{Luscher:2004ib} at the fixed ends of the string and is given by
\begin{equation}
\begin{split}
S^{b}&=\int_{\partial \Sigma} d\zeta_0 \Bigg[b_1 \frac{\partial \bm{X}}{\partial \zeta_1} \cdot \frac{\partial \bm{X}}{\partial \zeta_{1}} + b_2 \frac{\partial \partial \bm{X}}{\partial \zeta_{1} \partial \zeta_{0}} \cdot \frac{\partial \partial \bm{X}}{\partial \zeta_{1} \zeta_{0}}\\
  &+b_3 \left( \frac{ \partial \bm{X} } {\partial \zeta_1 } \cdot \frac{\partial \bm{X} }{ \partial \zeta_1} \right)^{2}+b_4   \frac{\partial^{2} \partial \bm{X}}{\partial \zeta_{0}^2 \partial \zeta_{1}} \cdot \frac{\partial^{2} \partial \bm{X}}{\partial \zeta_{0}^2 \partial \zeta_{1}}  \Bigg].
\end{split}
\label{boundaryaction}  
\end{equation}
    where $b_i$ are the couplings of the boundary terms. Consistency with the open-closed string duality~\cite{Luscher:2004ib} which implies a vanishing value of the first boundary coupling $b_{1}=0$, the leading-order correction due to second boundary terms with the coupling $b_2$ appears at higher order than the four derivative term in the bulk.

   An interesting generalization of the Nambu-Goto string ~\cite{Arvis:1983fp,Alvarez:1981kc,Olesen:1985pv} has been proposed by Polyakov~\cite{POLYAKOV19} and Kleinert~\cite{kleinert} to stabilize the NG action in the context of fluid membranes. The Polyakov-Kleinert string is a free bosonic string with additional Poincare-invariant term proportional to the extrinsic curvature of the surface as a next order operator after NG action~\cite{kleinert,POLYAKOV19}. That is, the surface representation of the Polyakov-Kleinert (PK) string depends on the geometrical configuration of the embedded sheet in the space-time. The bosonic free string action is equiped with the extrinsic curvature as a next-order operator after NG action~\cite{POLYAKOV19,Kleinert:1986bk}.

   The model preserves the fundamental properties of QCD of the ultraviolet (UV) freedom and infrared (IR) confinement properties ~\cite{POLYAKOV19,Kleinert:1986bk}, and is consistent with the formation of the glueballs~\cite{Kleinert:1988hz,Viswanathan:1987et}, and a real ($Q\bar{Q}$) potential~\cite{German:1989vk,Nesterenko1992,Ambjorn:2014rwa} with a possible tachyonic free spectrum~\cite{Kleinert:1996ry} above some critical coupling~\cite{Viswanathan:1987et}. 

   The action of the Polyakov-Kleinert (PK) string with the extrinsic curvature term reads
\begin{equation}
  S^{\rm{PK}}[X]= S_{\rm{\ell o}}^{\rm{NG}}[X]+S^{R}[X],
\label{PKaction}    
\end{equation}
  with $S^{R}$ defined as 
\begin{equation}
  S^{\rm{R}}[X]=\alpha_{r} \int d^2\zeta \sqrt{g}\, {\cal K}^2.
\label{Ext}  
\end{equation}  
  The extrinsic curvature ${\cal K}$ is defined as 
\begin{equation}
{\cal K}=\triangle(g) \partial_\alpha [\sqrt{g} g^{\alpha\beta}\partial_\beta],
\end{equation}
where $\triangle$ is Laplace operator and $M^{2}=\frac{\sigma_{0}}{2 \alpha_{r}}$ is the rigidity parameter. The term satisfies the Poincare and the parity symmetries and can also be considered~\cite{Caselle:2014eka} in the general class of (LW) actions~\eqref{LWaction}.

 The perturbative expansion ~\cite{German:1989vk} of the rigidity term Eq.~\eqref{Ext} reads 

\begin{equation}
S^{\rm{R}}[X]=S_{\rm{\ell o}}^{R}[X]+S_{\rm{n \ell o}}^{R}[X]+...,    
\end{equation}
has the leading term is given by
\begin{equation}
S_{\rm{\ell o}}^{R}= \alpha_{r} \int_{0}^{L_T} d\zeta_0 \int_{0}^{R} d\zeta_1  \left[\left(\dfrac{\partial^{2} \bm{X}} {\partial \zeta_{1}} \right)^2 + \left(\dfrac{\partial^{2} \bm{X}} {\partial \zeta_{0}^{2}}  \right)^2 \right]
\label{S_extlo}
\end{equation}


   The rigidity parameter is tuned so as to weigh favorably the smooth surface configuration over the creased worldsheets. In non-abelian gauge theories this ratio is expected to remain constant in the continuum limit~\cite{Caselle:2014eka}.

  In the following, prior to drawing a comparison between the numerical Yang-mills lattice data and the various models of the Casimir energy, we review the corresponding confining potential due to each string action in the remaining of this section.

  The Casimir energy is extracted from the string partition function as
\begin{equation}
V(R,T)=-\dfrac{1}{L_T} \log(Z(R,T)).
\label{Casmir}
\end{equation}
   The partition function of the NG model in the physical gauge is a functional integrals over all the world sheet configurations swept by the string
\begin{equation}
  Z(R,T)= \int_{{\cal C}} [D\, \bm{X} ] \,\exp(\,-S( \bm{X} )).
\label{PI}
\end{equation}  
  For a periodic boundary condition along the time direction such that  
\begin{align}
X (\zeta_{0}=0,\zeta_{1})= X (\zeta_{0}=L_T,\zeta_{1}),
\label{bc1}
\end{align}
  with an extent equals to the inverse of the temperature $L_{T}=\frac{1}{T}$ and Dirichlet boundary condition at the sources position given by  
\begin{align}
X(\zeta_{0},\zeta_{1}=0)= X(\zeta_{0},\zeta_{1}=R)=0, 
\label{bc2}
\end{align}
  
  the eigenfunctions are given by
\begin{equation}
\phi_{mn}=e^{2\pi i\left(\frac{m}{R}+\frac{n}{L_{T}} \right)},
\label{7}
\end{equation}
and eigenvalues of $-\triangle$ are given by
\begin{equation}
\Gamma_{nm}=\left(\frac{2\pi n}{L_{T}} \right)^{2}+\left(\frac{2\pi m}{R}\right)^{2}.
\label{8}
\end{equation}
   The determinant of the Laplacian after $\zeta$ function regularization~\cite{PhysRevD.27.2944} reads
\begin{equation}
\begin{split}
    \rm{Det}\left(-\triangle\right)&=(q^{\frac{1}{24}} \prod_{n=1}^{\infty}(1-q^{n}))^{2},
\end{split}
\label{15}
\end{equation}
where $q=e^{2\pi\tau}$ and $\tau=\frac{L_{T}}{2 R}$ is the modular parameter of the cylinder. The path integral Eq.~\eqref{PI} and Eq.~\eqref{Casmir} yields the static potential for the leading order contribution of the NG action $S^{NG}_{\ell o}$. The partition function and the static potential are respectively given by
\begin{equation}
    Z^{(NG)}_{\ell o}=e^{-\sigma R T-\mu(T)} [\rm{Det}\left(-\triangle\right)]^{-\frac{(d-2)}{2}},\\
\end{equation}

\begin{equation}
\label{Pot_NG_LO}
   V^{\rm{NG}}_{\rm{\ell o}}(R,T)= \sigma_{0} R+(d-2)T\, \log \eta \left(\tau \right)+\mu T,
\end{equation}  
\noindent where $\mu$ is a UV-cutoff and $\eta$ is the Dedekind $\eta$ function defined on the real axis as

\begin{equation}
  \eta(\tau)=q^{\frac{1}{24}} \prod_{n=1}^{\infty}(1-q^{n}).
\end{equation}
  The second term on the right hand side encompasses the L\"uscher term of the interquark potential. This term signifies a universal quantum effect which is a characteristic of the CFT in the infrared free-string limit and is independent of the interaction terms of the corresponding effective theory. One can extract the string tension dependency on temperature from the slop of the linear terms in $R$. Considering the modular transform of the Eq.~\eqref{Pot_NG_LO} $\tau \rightarrow 1/\tau$ and taking the limit of long string, the renormalized string tension to leading order is given by
\begin{equation}
\sigma(T)=\sigma_{0}-\dfrac{\pi(d-2)}{6} T^{2}+O(T^4).
\label{Tension_NG_LO}
\end{equation}

  Deitz and Filk ~\cite{PhysRevD.27.2944} extracted the next to leading order term~\cite{PhysRevLett.67.1681} of the Casimir potential from the explicit calculation of the two-loop approximation using the $\zeta$ regularization scheme. The static potential of NG string at second loop order Eq.~\eqref{NGpert} is given by 
\begin{equation}
\label{Pot_NG_NLO}
\begin{split}
  V^{\rm{NG}}_{\rm{n\ell o}}(R,T)&= \sigma_{0} R+(d-2)T\, \log \eta \left(\tau \right)- T \log \bigg(1-\frac{T}{R^{3}}\\
  &\dfrac{(d-2)\pi^{2}}{1152 \sigma_{0}} \left[2 E_4(\tau) +(d-4)E_{2}^{2}(\tau)\right] \bigg)+\mu T,
\end{split}
\end{equation} 
\noindent with $E_{2}$ and $E_{4}$ are the second and forth-order Eisenstein series defined as 
\begin{eqnarray}
  E_{2}(\tau)&=&1-\dfrac{1}{24} \sum^{\infty}_{n=1} \dfrac{n\,q^{n}}{1-q^{n}},\\\nonumber
  E_{4}(\tau)&=&1+\dfrac{1}{240} \sum^{\infty}_{n=1} \dfrac{n^{3}\,q^{n}}{1-q^{n}},\\
\end{eqnarray}
respectively.
With the modular transform $\tau \rightarrow 1/\tau$ of Eq.~\eqref{Pot_NG_NLO} and considering the limit of long string, the string tension, which defines the slop of the leading linear term of the potential, as a function of the temperature reads
\begin{equation}
\sigma(T)=\sigma_{0}-\dfrac{\pi(d-2)}{6} T^{2}-\dfrac{\pi^2(d-2)^2}{72 \sigma_{0}}T^{4}+O(T^6).
\label{Tension_NG_NLO}
\end{equation}
   where $\sigma_{0}$ denotes the string tension of the string at zero temperature. The coefficient of the next higher-order terms $T^{6}$ can be induced~\cite{Arvis:1983fp,Giudice:2009di} from the expansion of NG action and leads to the exact NG string tension given by
\begin{align}
\sigma(T)=\sigma_{0}\sqrt{1-\frac{\pi (d-2) T^{2}}{3 \sigma_{0}}}.
\label{Tension_NG_Exact}
\end{align}

   The boundary term $S^b$ in L\"uscher-Weisz action due to the symmetry breaking by the cylindrical boundary conditions by the Polyakov lines. In Refs.~\cite{Aharony:2010cx,Billo:2012da} the first and second nonvanishing Lorentz-Invariant boundary terms contribution to the potential have been calculated. The modification to the potential received when considering Dirichlet boundary condition are given by

\begin{equation}
\begin{split}  
 V^{B}&=V^{b_{2}}+V^{b_{4}},\\
 V^{b_{2}}&=b_2 (d-2) \dfrac{\pi^{3} L_T}{60 R^{4}}E_{4}\left(q\right),\\
\end{split}
\label{Pot_Boundb2}
\end{equation}
   The contribution to the partition function coming from the action
\begin{equation}
  Z=\int DX e^{-S_{\ell o}^{NG}-S_{b_2}-S_{b_4}} 
\end{equation}
Expanding around the free action yields
\begin{equation}
  Z=Z^{0} \left( \left(1-\left\langle S_{b_2} \right\rangle-\left\langle S_{b_4} \right\rangle \right)+\frac{1}{2}  \left\langle \left(S_{b_2}+S_{b_4}+....\right)^2 \right\rangle \right)+...
\end{equation}  

  The Lorentzian-Invariance imply $b_1=0,b_3=0$, the next two non-vanishing Lorentzian-Invariant terms come at order four and six derivative terms at coupling $b_2$

\begin{equation}
\left\langle S_{b_2} \right\rangle =b_2 \int_{\partial \Sigma} d\zeta_{0} \left\langle \partial_{0}\partial_{1} X \cdot \partial_{0} \partial_{1} X \ \right\rangle,
\end{equation}

and coupling $b_4$ 

\begin{equation}
\left\langle S_{b_4} \right\rangle= b_4 \int_{\partial \Sigma} d\zeta_{0} \left\langle \partial_{0}^{2}\partial_{1}X \cdot \partial_{0}^{2} \partial_{1} X \right\rangle, 
\end{equation}
respectively.

The spectral Green function corresponding to solution of Laplace equation with Dirichelet boundary conditions on cylinder 
\begin{equation}
\begin{split}
&G(\zeta_1,\zeta_0,\zeta_1',\zeta_0')=\\
&\frac{2}{\pi^{2} R L_{T}} \sum_{m,n}\sum_{m',n'} \sin\left(\frac{n \pi \zeta_1}{R}\right)\sin\left( \frac{n' \pi \zeta_1'}{R} \right) \frac{e^{ \frac{2 \pi i m}{L_{T}} (\zeta_0-\zeta_0')}}{\frac{n^2}{R^2}+\frac{4 m^2}{L_{T}^{2}}}.
\end{split}
\label{SpectralGreen}
\end{equation}

  The correlator $\left\langle S_{b_4} \right\rangle$, which is the line integral over each of the Polyakov loops

\begin{equation}
\left\langle S_{b_4} \right\rangle=\int_{\partial \Sigma}  \, \partial_{0} \partial_{0} \partial_{1} \partial_{0}^{'}\partial_{0}^{'} \partial_{1}^{'} G(\zeta_1,\zeta_0,\zeta_1',\zeta_0'), 
\end{equation}
  becomes after substituting the spectral Green function Eq.~\eqref{SpectralGreen},
\begin{equation}
  \begin{split}
  \left\langle S_{b_4} \right\rangle&= \frac{32 \pi^{2} }{R^2 L_{T}^{4}}\Bigg[\sum_{m n}  \frac{m^2 n}{\frac{n^2}{R^{2}}+\frac{4 m^2}{L_{T}^2}}\sum_{m' n'} \frac{m'^2 n'}{\frac{n'^2}{R^{2}}+\frac{4 m'^2}{L_{T}^2}}\\
  & \left\{ \lim_{\zeta_1 \to 0}+\lim_{\zeta_1 \to R} \right\} \int_{\partial \Sigma} d\zeta_{0} \cos\left(\frac{n \pi \zeta_1}{R} \right) \cos \left(\frac{n' \pi \zeta_1}{R} \right) \Bigg].
\end{split}
  \label{S4}
\end{equation}
 Making use of the $\zeta$ function regularization of the series sum
\begin{equation}
\sum_{m,n}\frac{n^{l} m^{2k}}{\frac{n^2}{R^2}+\frac{4 m^2}{L_{T}^2}}=(-1)^k \pi R^2 \left(\frac{L_{T}}{2 R}\right)^{2k+1}\zeta(1-l-2k) E_{2k+l}(q).
\end{equation}

  The correlator Eq.~\eqref{S4} yields the following correction to the static potential from the boundary term at coupling $b_4$, 
\begin{equation}
 V^{b_{4}}= \dfrac{-b_4 (d-2)\pi^{5}L_{T}}{126  R^{6}} \text{E}_{6}\left(\tau\right).\\
\label{Pot_Boundb4}
\end{equation}

  The modular transforms of Eq.\eqref{Pot_Boundb2} and Eq.\eqref{Pot_Boundb4} do not yield a linear term proportional to $R$ which changes the slop of the potential. 

  The static potential for smooth open strings was evaluated in ~\cite{German:1989vk,German:1991tc,Braaten:1987gq,Nesterenko:1991qp}. Employing $\zeta$ function regularization~\cite{Elizalde:1993af} the finite-temperature contribution is calculated in Ref.~\cite{Viswanathan:1988ad} without subtraction~\cite{Nesterenko:1997ku} to the first loop. Here we show in more detail the calculation of the determinant of the Laplacian to unambiguously show the formulation of the static potential and link between different forms used in the literature.

 The partition functions due to the leading order contribution from NG action and rigidity terms 

\begin{equation}
  Z=Z^{(NG)}_{\ell o} Z^{(R)}_{\ell o},
\label{1}
\end{equation}
are given by 
\begin{equation}
\begin{split}
Z &=\int DX \exp[-S^{\rm{NG}}_{\rm{\ell 0}}S_{\rm{\ell o}}^{\rm{R}}],\\
  &= \int D \bm{X} \exp -\sigma[ 1+\frac{1}{2} \bm{X}(1-\frac{\triangle}{2 M^2})(-\triangle)\bm{X}].
\end{split}  
\label{2}
\end{equation}
With the use of the transformation 
\begin{equation}
\bm{X}'=\frac{\triangle}{2 M^2}\bm{X},
\label{3}
\end{equation}
  the partition function can be decoupled into the leading-order NG partition function Eq.~\eqref{15} multiplied by the corresponding rigidity contribution which appears as the Jacobian of the transformation

\begin{equation}
Z^{(NG)}_{\ell o}=e^{-\sigma R T-\mu(T)} [\rm{Det}\left(-\triangle\right)]^{-\frac{(d-2)}{2}},
\label{4}
\end{equation}
and
\begin{equation}
  Z^{(R)}=[\rm{Det}\left(1-\frac{\triangle}{M^{2}}  \right)]^{-\frac{(d-2)}{2}}.
\label{5}
\end{equation}
  The eigenvalues of the trace of the triangle operators can be calculated from the eigenvalue equations corresponding to periodic and Dirichlet boundary condition defined as
\begin{equation}
(-\triangle)\psi_{nm}=\lambda_{nm}\psi_{nm}.
\label{6}
\end{equation}
  The eigen functions and eigenvalues are given by
\begin{equation}
\psi_{mn}=e^{2\pi i\left(\frac{m}{R}+\frac{n}{L_{T}} \right)},
\label{7}
\end{equation}
and
\begin{equation}
\lambda_{nm}=(2\pi n T)^{2}+\left(\frac{2 \pi m}{R}\right)^{2},
\label{8}
\end{equation}
respectively. The traces then read
\begin{equation}
\begin{split}  
&\ln\left(\rm{Det}\left(1-\frac{\triangle}{M^{2}}  \right)\right)\\
= & -\lim_{s \rightarrow 0}\frac{d}{ds}\sum_{n,m} \left(\frac{4\pi^{2}}{M^{2}} \left[\frac{m^{2}}{R^2}+\frac{n^2}{L_{T}^{2}} + \frac{M^{2}}{4 \pi^{2}} \right]\right)^s,
\label{9}
\end{split}
\end{equation}

where $s$ is an auxilary parameter. In the following we illustrate the $\zeta$ function regularization scheme used for the evaluation of the two summations appearing in the above integral.

  The sum over $m$ is the Epstein-Hurwitz $\zeta$ function~\cite{PhysRevD.37.974}. Using Sommerfeld-Watson transform~\cite{polchinski1986} the summation over $n$ in the last expression can be presented as a contor integral in the complex plane $t$ as

\begin{equation}
\begin{split}  
\ln(Z^{R})=&\frac{(d-2)}{2} \Bigg(\lim_{s \rightarrow 0}\frac{d}{ds}\sum_{n,m} \Bigg[ \left(\frac{4\pi^{2}}{M^{2}R^2}\right)^{-s}\Bigg( \oint_{ c}dt \sum_{m}\\
&\left(\frac{e^{i\pi t}}{2i \sin(\pi t)}+\frac{1}{2}\right) \left(m^2+t^2+\frac{M^{2}R^2}{4\pi^{2}} \right)^{-s} \Bigg]\Bigg).\\
\label{10}
\end{split}
\end{equation}
 
   Solving the above integral with the contor $(+\infty+i\epsilon)$ to $(-\infty+i\epsilon)$ yields the two terms

\begin{equation}
  \begin{split}
 \frac{2}{2-d} \log(Z^{R})&=4\sum_{m=0}^{\infty}\log(1-e^{2 \pi (m^2+\frac{M^{2} R^{2}}{\pi^{2}})})\\
 & -\lim_{s \rightarrow 0}\frac{d}{ds}\left( \frac{4\pi^{2}}{M^{2}\,R^{2}} \right)^{-s} \frac{\sin(\pi s)}{\cos(\pi s)}(2\tau)^{1-2s}\\
 & \frac{\Gamma^{2}(1-s)}{\Gamma(2-2s)} \sum_{m} \frac{1}{(m^2+M^{2}/4\pi^2)^{s-1/2}}.
 \label{11}
\end{split}
\end{equation}

  The second sum over $m$ is also the Epstein-Hurwitz $\zeta$ function. However, we proceed in the regularization using different integral representation~\cite{Nesterenko:1997ku} to bring the final closed form in terms of the standard mathematical functions. Each term in this sum can be transformed into the integral presentation corresponding to Euler-Gamma function~\cite{Lambiase:1995st}.

For general $s$ the Epstein-Hurwitz $\zeta$ function reads
\begin{equation}
  \zeta(s,M/4\pi)=\frac{1}{\Gamma(s)}\int_{0}^{\infty} t^{s-1}\sum_{n=1}^{\infty} e^{-t(n^2+M^{2}\,R^2)} dt,
  \label{12}
\end{equation}
   then with the modular transform of $t \longrightarrow 1/t$ of the Jacobi $\vartheta$, $\theta(t)=\sum_{n=1}^{\infty} e^{-n^2 t}$ appearing in the above sum, the zeta function turns into the integral presentation

\begin{equation}
  \begin{split}
\zeta(s,M/4\pi)&=\frac{-(M^2 R^2)^{-2s}}{2}+\frac{\sqrt{\pi} \Gamma(s-\frac{1}{2})}{2\Gamma(s)}(M^2 R^2)^{-s+\frac{1}{2}}\\
& +\frac{\sqrt{\pi}}{\Gamma(s)}\sum_{n=1}^{\infty} \int_{0}^{\infty} t^{s-\frac{3}{2}} \exp\left(-t \frac{M^{2}R^{2}}{\pi^{2}}-\frac{\pi^{2}n^{2}}{t}\right)dt.
  \end{split}
  \label{13}
\end{equation}

  The integral in the last term can be presented in terms of sum over the modified Bessel functions~\cite{pearce_1949} of the second kind  $\sum_{n}n^{s-1/2}K_{s-1/2}(2 n M\,R)$. Employing Eq.~\eqref{12} in Eq.~\eqref{13} in the partition function $Z^{R}$ Eq.~\eqref{13} the resultant
expression would read

\begin{equation}
\begin{split}  
Z^{R} =\exp \Bigg[ \frac{(d-2)M}{2\pi}\sum_{n=1}n^{-1}K_{1}( 2 n M\,R) \Bigg]\\
\times \prod^{\infty}_{n=0}\left( 1-e^{2\pi\tau(n^{2}+M^{2})^{\frac{1}{2}}}  \right).
\end{split}
\label{14}
\end{equation}

 The potential corresponding to the total partion function $Z^{NG}_{\ell o} Z^{R}_{\ell o}$ is thus

 \begin{equation}
  \begin{split}
    V^{R}_{\ell o}(R,T)&=\sigma_{0} R-(d-2)T \log\left(\eta(\tau)\right)-\lambda(T)\\
    &+\frac{(d-2)M}{2\pi}\sum_{n=1}^{\infty}n^{-1}K_{1}( 2 n M\,R)]\\
    &+(d-2) T \sum^{\infty}_{n=0} \Bigg( \log \left( 1-e^{2\pi \tau \sqrt{n^{2} + M^{2}}}\right)\ \\
    &. 
  \end{split}
  \label{16}
\end{equation}

  The massless limit corresponding to $M^2=\sigma_0/2\alpha_{r}\rightarrow 0$ in the above equation yields the doublet degeneracy of the free bosonic string modes or the well-known doubling of L\"uscher term in the zero temperature limit,

\begin{equation}
V^{R}_{\ell o}(R,T) = \sigma_{0} R- 2(d-2)T \log\left(\eta(\tau)\right)-\nu(T).
\label{17}
\end{equation}

  Let us define a string model for the quark-antiquark potential with the leading extrinsic curvature term in conjunction with two subsequent orders of the NG perturbative expansion corresponding to the free and the leading self-interacting components

\begin{equation}
  Z=Z^{NG}_{\ell o}\,Z^{NG}_{n\ell o}\, Z^{R}_{\ell o}.
  \label{18}
\end{equation}

 The potential of the rigid string with the next to leading order NG contribution is

\begin{equation}
\begin{split}
&V_{n\ell o}^{R_{(\ell o)}}(R,T) =  V_{\ell o}^{R_{(\ell o)} }(R,T)\\
&-T  \ln  \Bigg( 1-\dfrac{(d-2)\pi^{2}T}{1152 \sigma_{o}R^{3}}  \left[2 E_4(\tau)+(d-4)E_{2}^{2}(\tau)\right] \Bigg).
\label{StiffNLO}
\end{split}
\end{equation}

   The string tension form this model can be calculated from the power expansion of $T$ in the large $R$ limit and it turns out to be

\begin{equation}
  \begin{split}
    &\lambda =\frac{2 \alpha_r  \sigma_{\rm{ren}} T^2}{4 \pi ^2},\\
    &S(\alpha_r,  \sigma_r, T)=\frac{1}{\lambda}\sum _{n=1}^{\infty} (\sqrt{\lambda +n^2}-\frac{\lambda }{2 n}-n),\\
    &\sigma(T)=\sigma_{r}-(d-2)\sigma_{r} \Bigg(\sqrt{\frac{\alpha_r }{\sigma_r}} T+ \frac{\pi }{6 \sigma_{r}} T^2\\
    &+ \frac{\alpha_r}{2 \pi }  S(\alpha_r,  \sigma_r, T)\Bigg). 
  \end{split}
\label{StringTensionRigid}
\end{equation}

  In the limit of high and low temperatures the series sum in Eq.~\eqref{StringTensionRigid} $S$ takes the asymptotic forms

\begin{equation} 
  S=-\frac{  \zeta (3)}{8}\lambda+\frac{ \zeta(5)}{16}\lambda ^2-\frac{5 \zeta(7)}{128}\lambda ^3+.....;
\label{HT}  
\end{equation}
and
\begin{equation}
  S=-\frac{1}{\pi } \frac{1}{\sqrt{\lambda }} \sum _{n=1}^{\infty} \frac{1}{n} K_{1}\left(2 \pi n \sqrt{\lambda }  \right).
\label{LT}    
\end{equation}

  At short distances, renormalization corrections to the zero temperature string tension have to be taken into account $\sigma_{0}$~\cite{German:1989vk,German:1991tc}. The lattice spacing, however, naturally introduces cutoff-scale which affects the value of the returned fit parameters over large source separation distances.

  The free NG string is known to reproduce ~\cite{Luscher:2002qv} the subleading aspects of the QCD string over long distances $R \ge 0.5$ but low temperatures. The increase in the temperature may result in the onset of the string self-interactions at all distance scales. Moreover,  other potential aspects of the confining strings may come into play. These include the geometrical constraints on the string fluctuations which reflects on its rigidity/stiffness structure; or in otherwords, its resistance to transverse bending and  the role of the Lorentzian-invariant ~\cite{Aharony:2010cx, Billo:2012da} boundary terms at short distances in particular.
  Considering such models~\cite{Caselle:2014eka}, from the pure formal point of view, would introduce corrections to the static potential on both the linearly rising part in addition to the inverse higher powers in the source separation $R$. The rigidity effects may not be dominant but can not be neglected when discussing static potential of spectrum of excited states~\cite{Juge:2002br}.
  
   In the following, we numerically measure the two point Polyakov-loop correlators and explore to what extent each of the above string actions can be a sufficiently good description for the potential between the two static color sources.    

\section{Lattice $Q\bar{Q}$ potential and string phenomenology}
 \begin{table}[!hpt]
	\begin{center}
		\begin{ruledtabular}
			\begin{tabular}{ccccccccccc}
				\multirow{2}{*}{} & &\multirow{2}{*}{$V_{Q\bar{Q}}(R,T)$ $fm^{-1}$} &\multirow{2}{*}{$e(R)$}\\
				&$n=R/a$ &&&\\
				\hline                                             
				\multirow{12}{*}{\begin{turn}{90}$~T/T_{c}=0.9~~~$ \end{turn}}		
				& 1 &-5.9003 &0.000246\\
				\multicolumn{1}{c}{} &2 &-4.7622&0.000313  \\				
				\multicolumn{1}{c}{} &3 &-3.7976&0.000499 \\				
				\multicolumn{1}{c}{} &4 &-3.1342&0.000657 \\				
	            \multicolumn{1}{c}{} &5 &-2.5985&0.000713 \\				
				\multicolumn{1}{c}{} &6 &-2.1313&0.000725  \\				
				\multicolumn{1}{c}{} &7 &-1.7064&0.000758  \\				
				\multicolumn{1}{c}{} &8 &-1.3134&0.000766  \\				
				\multicolumn{1}{c}{} &9 &-0.9475&0.000772  \\				
				\multicolumn{1}{c}{} &10 &-0.6092&0.000827  \\
                \multicolumn{1}{c}{} &11 &-0.2932&0.000857 \\
                \multicolumn{1}{c}{} &12 &~0.0000&0.000899  \\

			\end{tabular}		
		\end{ruledtabular}
	\end{center}
	\caption{The quark-antiquark potential at each color source separation $R$ in lattice units, $\beta=6.0$, spatial volume $N_s=36^3$ and $N_t=8$ time slices and temperature scale $T/T_{c}=0.9$.}
\label{T1_Pot_Cor_T09}		
 \end{table}
\subsection{Simulation setup} 

  At fixed temperature $T$, the Polyakov loop correlators address the free energy of a system of two static color charges coupled to a heatbath~\cite{polyakov:78}. Within the transfer matrix formalism~\cite{Luscher:2002qv} the two point Polyakov-loop correlators are the partition function of the string.

  The Monte-Carlo evaluation of the temperature dependent quark--antiquark potential at each $R$ is calculated through the expectation value of the Polyakov loop correlators  
\begin{align}
\mathcal{P}_{\rm{2Q}} =& \int d[U] \,P(0)\,P^{\dagger}(R)\, \mathrm{exp}(-S_{w}),  \notag\\
=& \quad\mathrm{exp}(-V(R,T)/T).
\label{PolyakovCorrelators}
\end{align}
with the Polyakov loop defined as 
\begin{equation}
  P(\vec{r}_{i}) = \frac{1}{3}\mbox{Tr} \left[ \prod^{N_{t}}_{n_{t=1}}U_{\mu=4}(\vec{r}_{i},n_{t}) \right],
\end{equation}

  Making use of the space-time symmetries of the torus, the above correlator is evaluated at each point of the lattice and then averaged.  We perform simulations on large enough lattice sizes to gain high statistics in a gauge-independent manner ~\cite{Bali} in addition to reduce correlations across the boundaries. The two lattices employed in this investigation are of a typical spatial size of $3.6^3$ $\rm{fm^3}$ with a lattice spacing $a=0.1$ fm.

  We choose to perform our analysis with lattices with temporal extents of $N_t = 8$, and $N_t = 10$ slices at a coupling of value $\beta = 6.00$. The two lattices correspond to temperatures $T/T_c =0.9$ just before the deconfinement point, and $T/T_c = 0.8$ near the end of QCD plateau~\cite{Doi2005559}. 

\begin{figure*}[!hptb]
\centering
\subfigure[]{\includegraphics[scale=0.35]{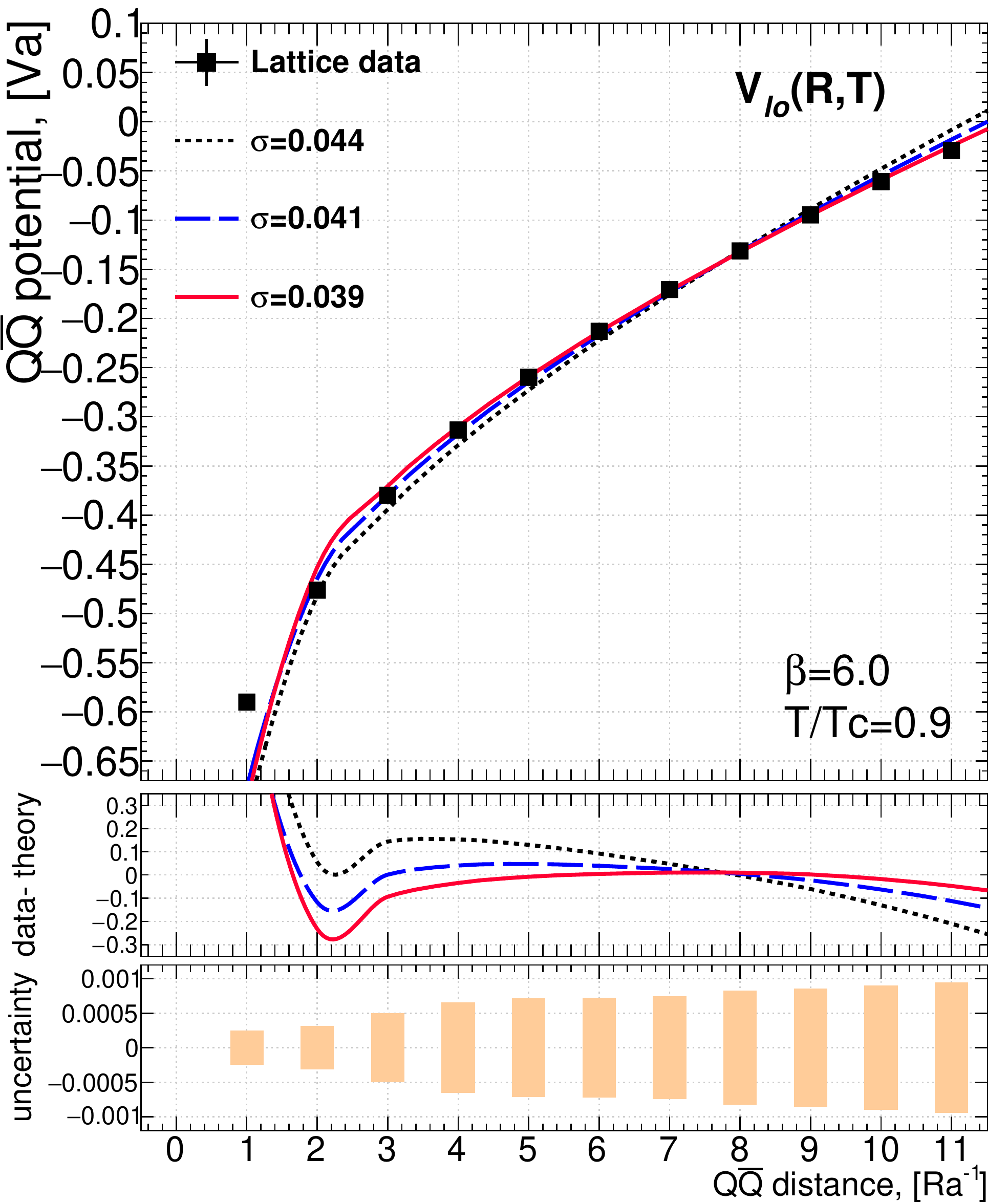}}
\subfigure[]{\includegraphics[scale=0.35]{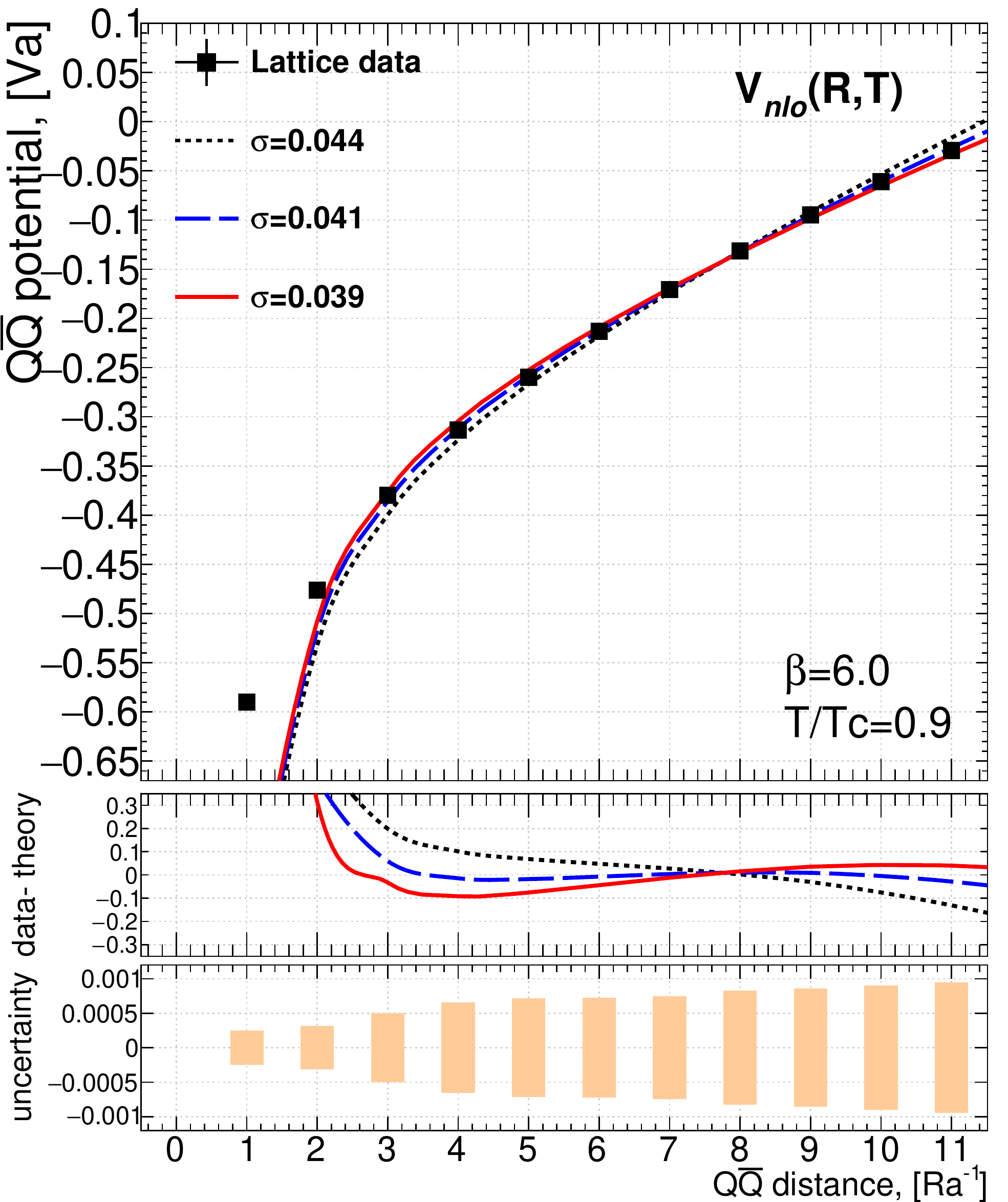}}
\caption{The quark-antiquark $Q\bar{Q}$ potential measured at temperature $T/T_c=0.9$, the left and right plots correspond to the fits to LO and NLO Nambu-Goto string Eq.\eqref{Pot_NG_LO} and Eq.~\eqref{Pot_NG_NLO} for the depicted values of string tension $\sigma a^{2}$, respectively. The returned fit parameters are inlisted in Table.~\ref{T3_Pot_NG_LO_NLO}}
\label{Fits_Pot_NG_LO_NLO}
\end{figure*}
\begin{table}[!hpt]
	\begin{center}
		\begin{ruledtabular}
			\begin{tabular}{cc|ccccccccc}
			  \multirow{2}{*}{$V_{NG}$} &Fit
				&\multicolumn{2}{c}{Fit Parameters, $T/T_{c}=0.9$} &\multicolumn{2}{c}{}\\
				&Interval
				&\multicolumn{1}{c}{$\sigma$} &\multicolumn{1}{c}{$\mu$} 
			        &\multicolumn{1}{c}{$\chi^{2}$}\\
				\hline
                                \multirow{10}{*}{\begin{turn}{90}\hspace{1cm}$V_{{\ell o}}$ \end{turn}}
			 &$[R_{m},11]$  &    &          &     \\
\multicolumn{1}{c}{} &[2,11]  & 0.041865(6)         & -0.42355(4)   & 111231.   \\
\multicolumn{1}{c}{} &[3,11]  & 0.039894(9)         & -0.40838(6)   & 17494.   \\
\multicolumn{1}{c}{} &[4,11]  &0.03901(1) &-0.4010(1)  & 4903.69     \\
\multicolumn{1}{c}{} &[5,11]  &0.0385(2)  &-0.396(2)   & 2142.17    \\
\multicolumn{1}{c}{} &[6,11]  &0.0381(2)  &-0.393(2)   & 1083.57   \\
\multicolumn{1}{c}{} &[7,11]  &0.03766(3) &-0.3889(2)  & 487.15   \\
\multicolumn{1}{c}{} &[8,11]  &0.0371(3)  &-0.383(3)   & 153.693   \\
\multicolumn{1}{c}{} &[9,11]  &0.0366(6)  &-0.378(6)   & 19.567      \\\hline
\multirow{10}{*}{\begin{turn}{90}\hspace{1cm}$V_{{n\ell o}}$ \end{turn}}
 &[2,11]  &0.036250(8)  &-0.35054(5)&627275.     \\
\multicolumn{1}{c}{} &[3,11]  &0.040406(9)  &-0.38757(7)&11136.     \\        
\multicolumn{1}{c}{} &[4,11]  &0.04111(1)   &-0.3938(1) &1786.73                       \\
\multicolumn{1}{c}{} &[5,11]  &0.0410(1)    &-0.393(1)  &1694.91     \\
\multicolumn{1}{c}{} &[6,11]  &0.0407(2)    &-0.390(2)  &1093.66     \\
\multicolumn{1}{c}{} &[7,11]  &0.04033(3)   &-0.3862(2) &525.181     \\
\multicolumn{1}{c}{} &[8,11]  &0.0398(3)    &-0.380(4)  &168.203     \\
\multicolumn{1}{c}{} &[9,11]  &0.0392(6)  &-0.374(6)    &21.832     \\
			\end{tabular}		
		\end{ruledtabular}
	\end{center}
	\caption{The $\chi^{2}$ values and the corresponding fit parameters returned from fits to the leading and the next-to-leading order (NLO) static potential of NG string Eqs.~\eqref{Pot_NG_LO} and ~\eqref{Pot_NG_NLO}, respectively.}
\label{T2_Pot_NG_LO_NLO}	
\end{table}

  The gauge configurations were generated using the standard Wilson gauge-action employing a pseudo-heatbath algorithm~\cite{Fabricius,Kennedy} updating to the corresponding three $SU(2)$ subgroup elements~\cite{1982PhLB..119..387C}. Each update step/sweep consists of one heatbath and 5 micro-canonical reflections. The gauge configurations are thermalized following 2000 sweeps. The measurements are taken on 500 bins. Each bin consists of 4 measurements separated by 70 sweeps of updates. 

  The correlator Eq.\eqref{COR} is evaluated after averaging the time links~\cite{Parisi} in Eq.\eqref{COR}
\begin{align}
\label{LI}
\bar{U_t}=\frac{\int dU U e^{-Tr(Q\,U^{\dagger}+U\,Q^{\dagger}) }}{\int dU e^{-Tr(Q\,U^{\dagger}+U\,Q^{\dagger}) }}.
\end{align}

   The temporal links are integrated out analytically by evaluating the equivalent contor integral of Eq.\eqref{LI} as detailed in Ref.~\cite{1985PhLB15177D}.  

  The lattice data of the ($Q\bar{Q}$) potential are extracted from the two point Polyakov correlator 
\begin{equation}
V_{Q\bar{Q}}(R)=-\dfrac{1}{T} \log \langle P(x)P(x+R)\rangle
\label{COR}
\end{equation}
   

\subsection{Temperature scale near critical point $T/T_c=0.9$}
   Thermal effects become dominant in the SU(3) Yang-Mills model~\cite{PhysRevD.85.077501,Doi2005559} under scrutiny if the temperature is scaled down close enough to the critical point $T/T_{c}=0.9$. The lattice data corresponding to the measured $Q\bar{Q}$ potential Eq.~\eqref{COR} at this temperature are inlisted in Table.~\ref{T1_Pot_Cor_T09}.

  In the following, we consider examining three possible ansatz of the string potential and draw comparisons between their possibly interesting combinations. The target is to understand the relevance of each model at the selected source separation intervals for this temperature scale.     

\subsubsection{Nambu-Goto string at leading and next-to-leading orders}
 
  The $Q\bar{Q}$ potential data are fitted to theoretical formulas of the NG string potential at the leading and next-to-leading orders Eqs.~\eqref{Pot_NG_LO} and Eq.~\eqref{Pot_NG_NLO}, respectively. We set the string tension $\sigma_{0}\, a^{2}$ and the renormalization constant $\mu(T)$ as a free fitting parameters. The same value of the string tension taken as a fit parameter as in Ref.~\cite{PhysRevD.85.077501} and ~\cite{Kac} is reproduced with the corresponding function of the static potential~\cite{Gao,Luscher:2002qv} and fit domain.
  
  Table~\ref{T2_Pot_NG_LO_NLO} enlists the returned value of the string tension $\sigma_{0}\, a^{2}$ and $\chi_{\rm{dof}}^2$ for various source separations commencing from $R=0.2, 0.5, 0.6$ and $0.7$ fm and extending to $R=1.2$ fm. The point at $R=0.1$ fm is excluded from the fits interval due to the overlap of the heatbath plaquettes which is well-known limitation of the link integration method.

  The fit of the numerical data to the leading order approximation Eq.~\eqref{Pot_NG_LO} produces similar reduction in the residuals by excluding short distance points. The fits return a minimal of $\chi^{2}$ at $\sigma_{0} a^{2}=0.039$ on  $R \in [0.5,1.2]$ fm and $\sigma_{0} a^{2}=0.037$ on $R \in [0.9,1.2]$ fm. However, the values of $\chi^{2}$ are outstandingly higher than the corresponding returned values considering the next-to-leading approximation Eq.~\eqref{Pot_NG_NLO}.

  Higher-order terms in the free energy provide fine corrections for the value of the returned free-parameter $\sigma_{0}\, a^{2}$. This parameter is interpreted as the zero temperature string tension, that is, the value that should be returned at zero or low temperature as in Ref.~\cite{Koma:2017hcm}. 
  
   To appreciate the role played by the string tension we disclose the fit behavior of the lattice data at this temperature scale We systematically inspect the returned values of $\chi^{2}$ for an interval of selected values of the string tension $\sigma_{0} a^{2} \in [0.035, 0.045]$. The residuals and normalization constant $\mu(T)$ for the corresponding $\sigma_{0} a^{2}$ are inlisted in Table.~\ref{T2_Pot_NG_LO_NLO} with plots in Fig.~\ref{Fits_Pot_NG_LO_NLO}. 

   Figure~\ref{ChiNG}-(a) show the stability of the fits and a well-defined global Minimal in the $(\sigma,\mu)$ parameter space. The gradual descend of the string tension parameter from 0.045 to 0.041 reduces dramatically the values of $\chi^{2}$ (as inlisted in Table~\ref{T3_Pot_NG_LO_NLO}) till a minimum is reached at $\sigma_{0} a^{2}=0.041$ for a fit interval from $R=[0.5,1.2]$ fm. The plot of the static potential owing to the (LO) and the (NLO) of the NG string at three selected values of the string tension $\sigma a^{2}$ is shown in Fig.~\ref{Fits_Pot_NG_LO_NLO}.

\begin{figure*}[!hptb]	
\centering
\subfigure[~~~~~~~~~~~~~~~~~~~~~~~~~~~~~~~~~~~~(b)]{\includegraphics[scale=0.7]{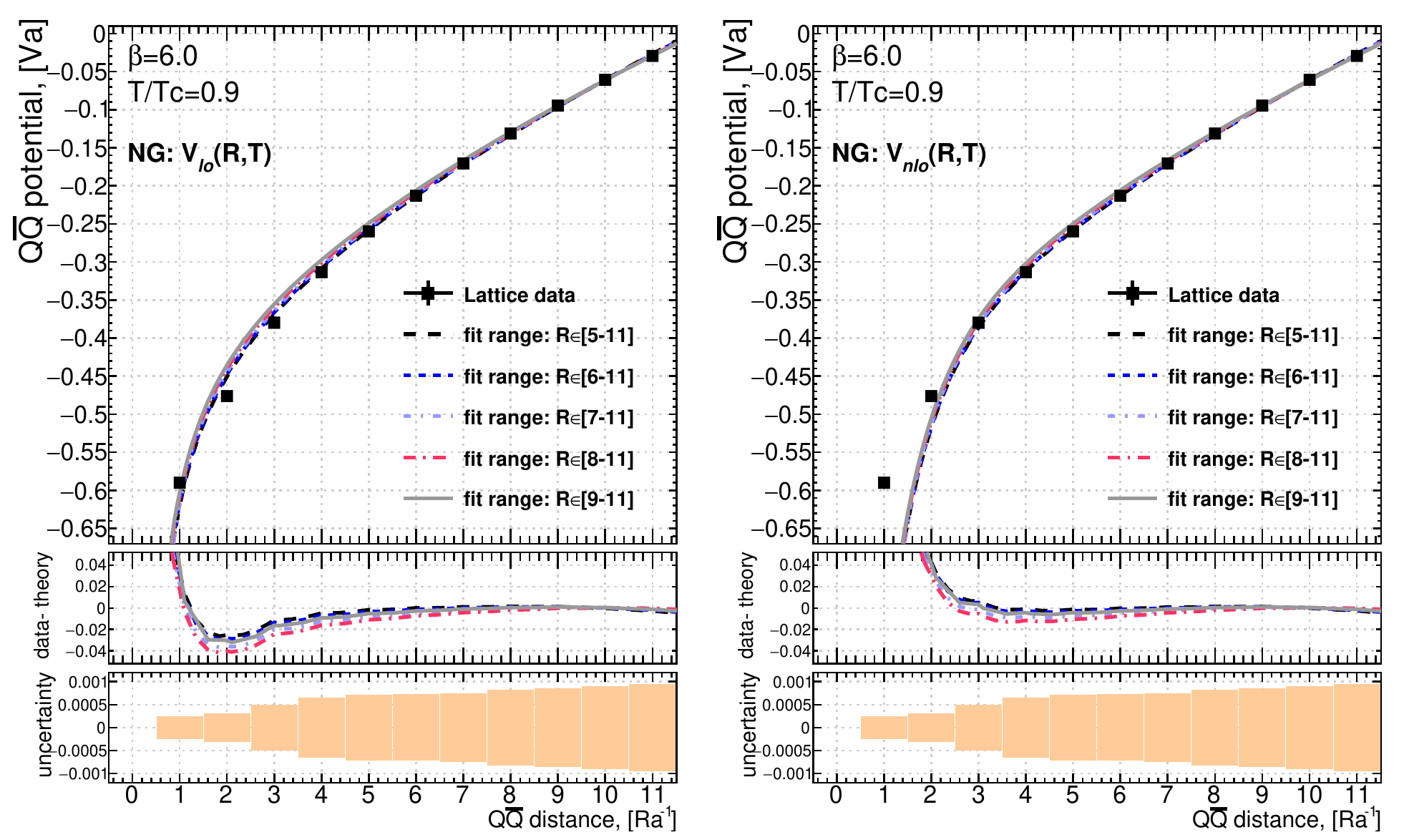}}
\caption{The fits to the quark-antiquark $Q\bar{Q}$ potential data measured at temperature $T/T_c=0.9$ Table.~\ref{T1_Pot_Cor_T09}, the lines correspond to the potential in accord to the leading Eq.~\eqref{Pot_NG_LO} and the next to leading NG Eq.~\eqref{Pot_NG_NLO} for the depicted fit ranges, the returned values of the string tension is enlisted in Table.\ref{T2_Pot_NG_LO_NLO}.}
\label{Fits_Pot_NG_LO_NLO2}
\end{figure*}

\begin{figure}[!hptb]
\centering
\subfigure[]{\includegraphics[scale=0.80]{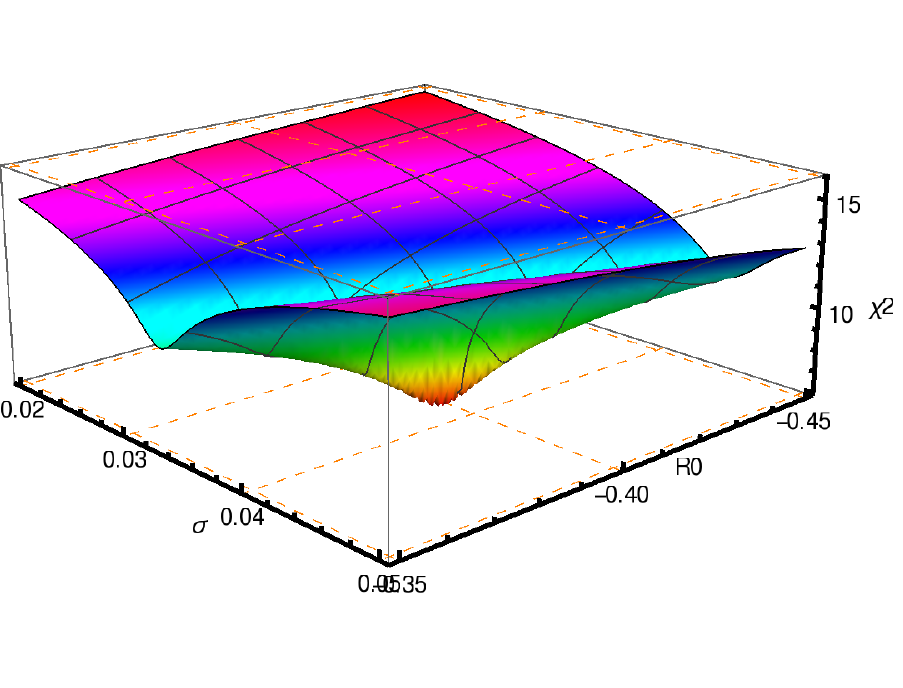}}
\subfigure[]{\includegraphics[scale=0.35]{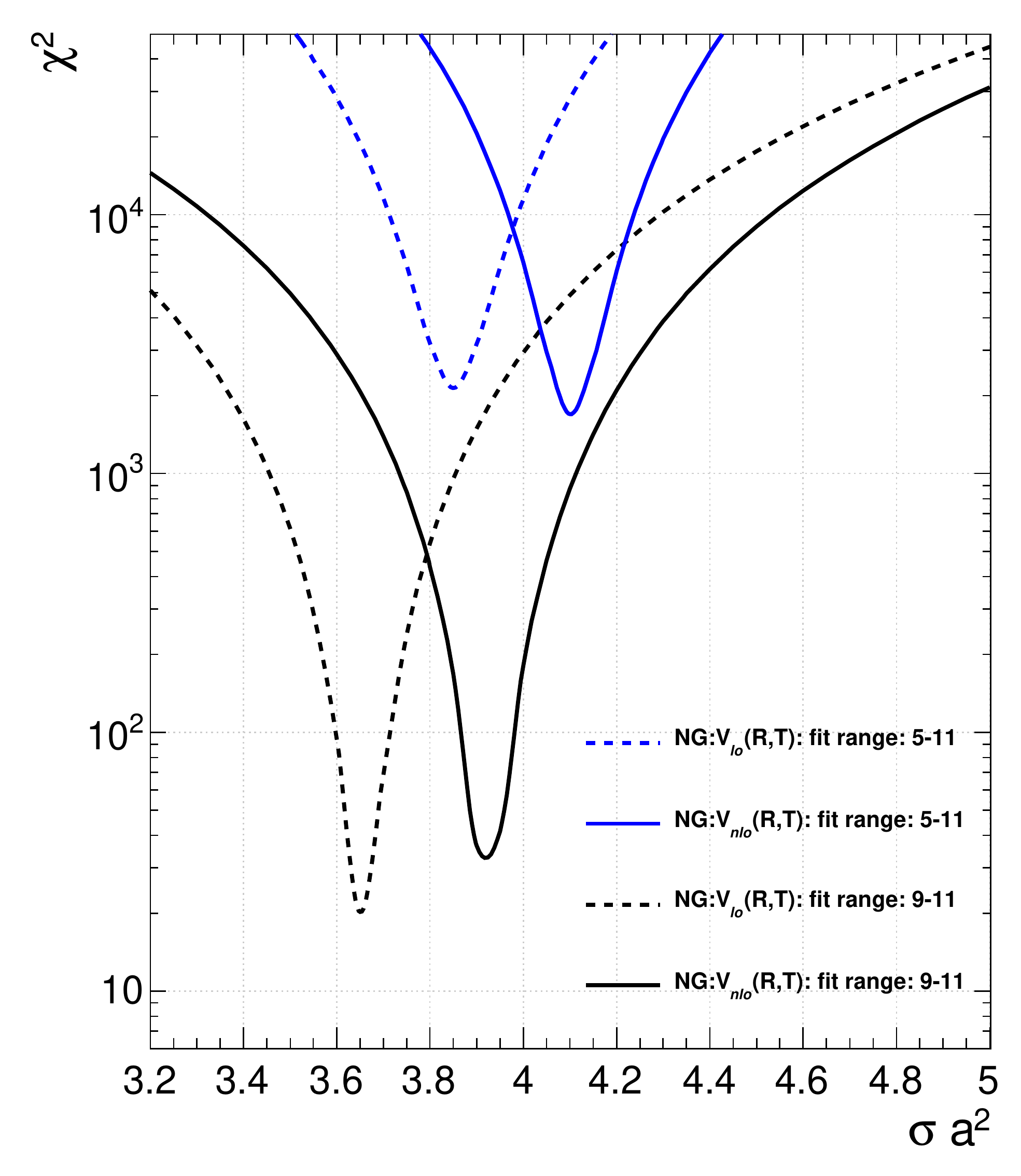}}
\caption{(a) The returned $\chi^{2}_{dof}$ versus the string tension $\sigma_{0} a^2$ and cutoff $\mu$, from the fits of $Q\bar{Q}$ potential to leading order approximation of Nambu-Goto string Eq.~\eqref{Pot_NG_LO} at $T/T_c=0.9$. (b)The returned $\chi^{2}_{dof}$ versus the string tension $\sigma_{0} a^2$; however, the fits are for the next-to-leading order approximation Eq.~\eqref{Pot_NG_NLO}. }
\label{ChiNG}
\end{figure}

   The two-dimensional version of Fig~\ref{ChiNG}-(b) depicts how excluding points at short distance, e.g, considering a fit interval $R \in [0.9,1.2]$, results in a smaller value of $\chi^{2}$ with a shifted minimal at $\sigma_{0} a^{2}=0.039$ as depicted in Fig.~\ref{Fits_Pot_NG_LO_NLO2}.

   Larger residuals $|Theory-data|$ at the (NLO) $V_{n\ell o}$ appear to be stringent at shorter distances $R<0.5$ fm at string tension value $\sigma_{0}a^{2}=0.044$ compared to the LO potential ansatz Eq.~\eqref{Pot_NG_NLO} of the NG string $V_{\ell o}$. This suggest that the string's self-interactions are more relevant to the string configurations swept over long distance scales. 

 
   The  effective description based only on Nambu-Goto model does not accurately match the $Q\bar{Q}$ potential data. In spite of  the fits beyond the free Gaussian approximation, the inclusion of the NLO terms does not provide an acceptable optimization for the potential data. Poor fits persist at both short distance and intermediate distances. The fitted string potential, at its minimal sum of the residuals $\chi^2$, produces deviation from the value of the measured value of the zero temperature string tension~\cite{Koma:2017hcm}.

   The fit to the Casimir energy of the self-interacting string returns a value of the zero temperature string tension $\sigma_{0} a^{2}=0.039$ which deviates at least by $11\%$ of that measured at zero temperature $\sigma_0 a^{2}=0.044$. The pure NG string with its free and next to leading self-interacting pictures does not provide the correct renormalization of the string tension by virtue of the thermal effects.
   
\begin{table}[!hpt]
	\begin{center}
		\begin{ruledtabular}
			\begin{tabular}{cccccccc}
				\multirow{2}{*}{$T/T_{c}$} & 
				&\multicolumn{2}{c}{$V_{{\ell o}}$} &\multicolumn{2}{c}{$V_{{n\ell o}}$}\\ 
				&&\multicolumn{1}{c}{$\sigma_{0} a^{2}$} &\multicolumn{1}{c}{$\chi^{2}$}& \multicolumn{1}{c}{$\sigma_{0}$} &\multicolumn{1}{c}{$\chi^{2}$} \\ \hline				
\multirow{10}{*}{\begin{turn}{90} Interval $R\in[9,11]$ \end{turn}}				
                     & &0.035& 612.657 &0.038&	437.806\\
\multicolumn{2}{c}{} &0.0355 &	292.088 &0.0385&168.529\\
\multicolumn{2}{c}{} &0.036  &	94.614&0.039&	36.5551\\
\multicolumn{2}{c}{} &0.0365 &	20.2356&0.0395&	41.4493\\
\multicolumn{2}{c}{} &0.037  &	68.9527&0.04&	182.793\\
\multicolumn{2}{c}{} &0.038  &	535.673&0.0405&	460.185\\
\multicolumn{2}{c}{} &0.0385 &	953.676&0.041&	873.24\\
\multicolumn{2}{c}{} &0.044  &	13676.&0.044&	6180.71\\
\multicolumn{2}{c}{} &0.045  &	17589.4&0.045&	9019.94\\
\end{tabular}		
\end{ruledtabular}
\end{center}
\caption{The values of $\chi^{2}$ returned from the fits for each corresponding value of the string tension, the table compares both values for fits to the leading order (LO) Eq.\eqref{Pot_NG_LO} and next-to-leading order (NLO) Eq.\eqref{Pot_NG_NLO}.}
\label{T3_Pot_NG_LO_NLO}		
\end{table}

  

\begin{figure*}[!hptb]
\centering
\subfigure[]{\includegraphics[scale=0.35]{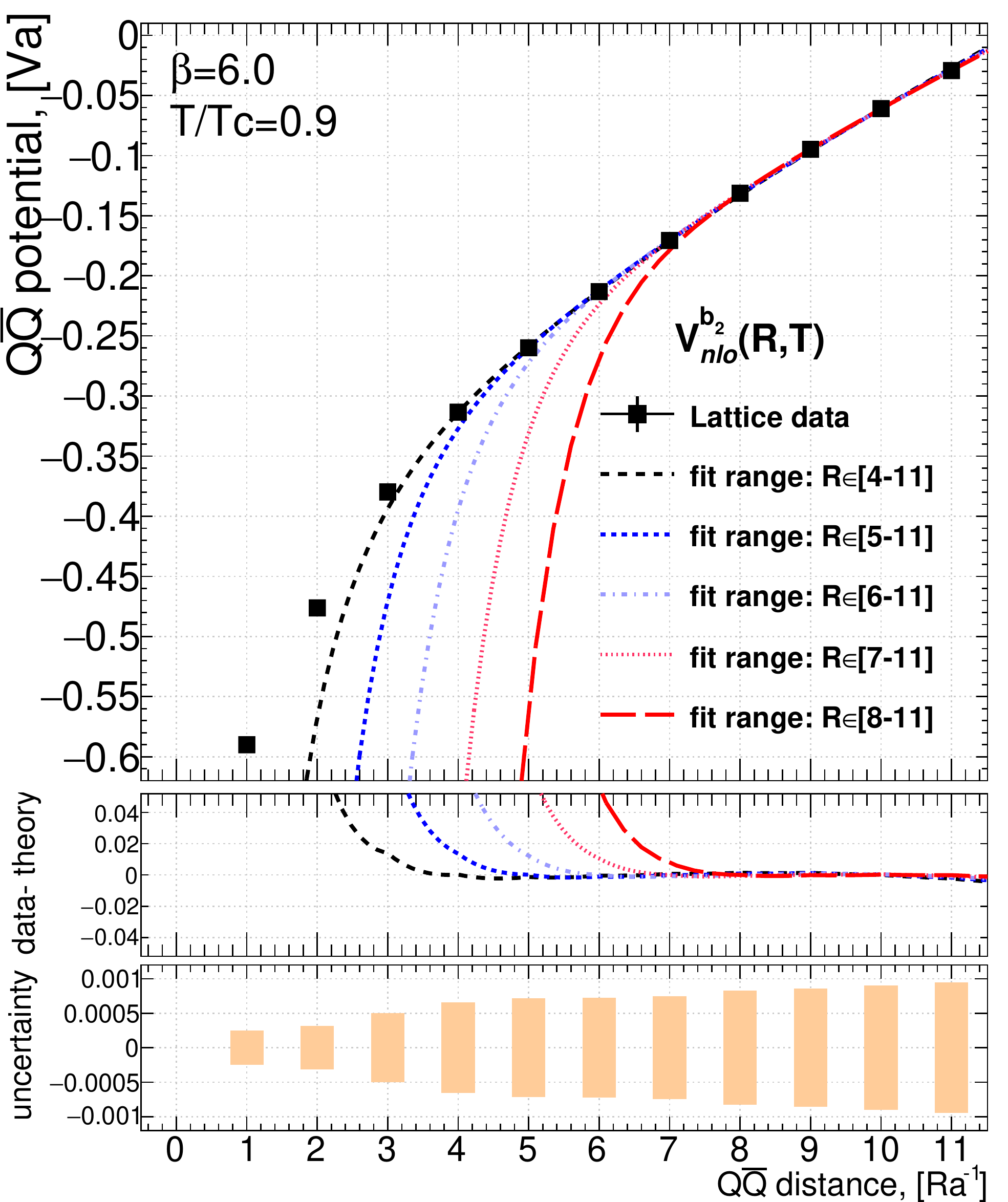}}
\subfigure[]{\includegraphics[scale=0.35]{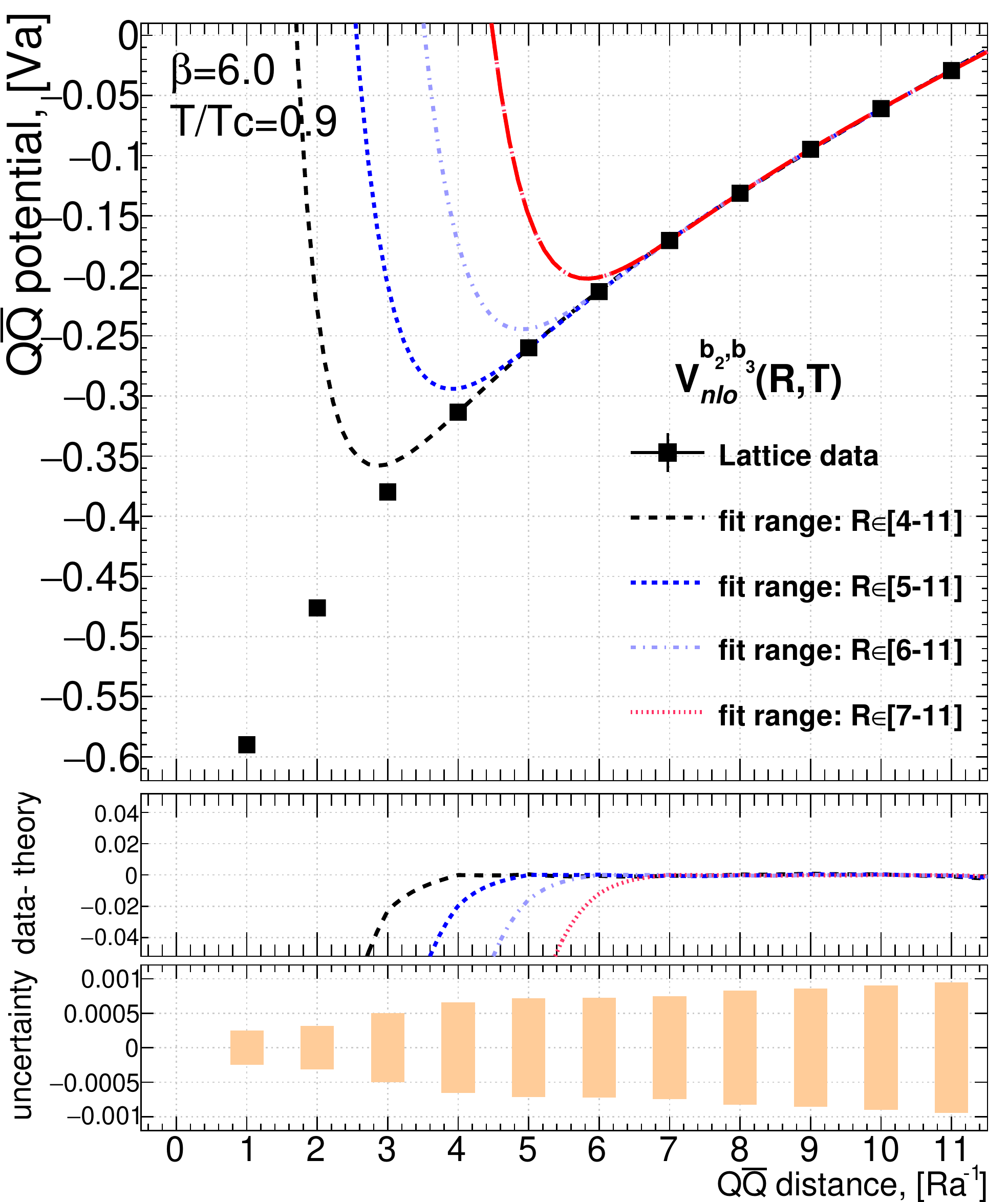}}
\caption{The quark-antiquark $Q \bar{Q}$ potential at temperature $T/T_c=0.9$, the lines correspond to the next to leading order Nambu-Goto string with two different boundary terms $V^{b2}_{n\ell o}$ and $V^{b4}_{n\ell o}$ Eqs.~\eqref{Pot_NG_LO_NLO_Boundb2} and Eq.~\eqref{Pot_NG_LO_NLO_Boundb2b3}, respectively, at $T/T_c=0.9$.}
\label{Fits_Pot_T09_NG_NLO_Boundb2_b3}
\end{figure*}

\begin{figure}[!hptb]
\centering
\includegraphics[scale=0.8]{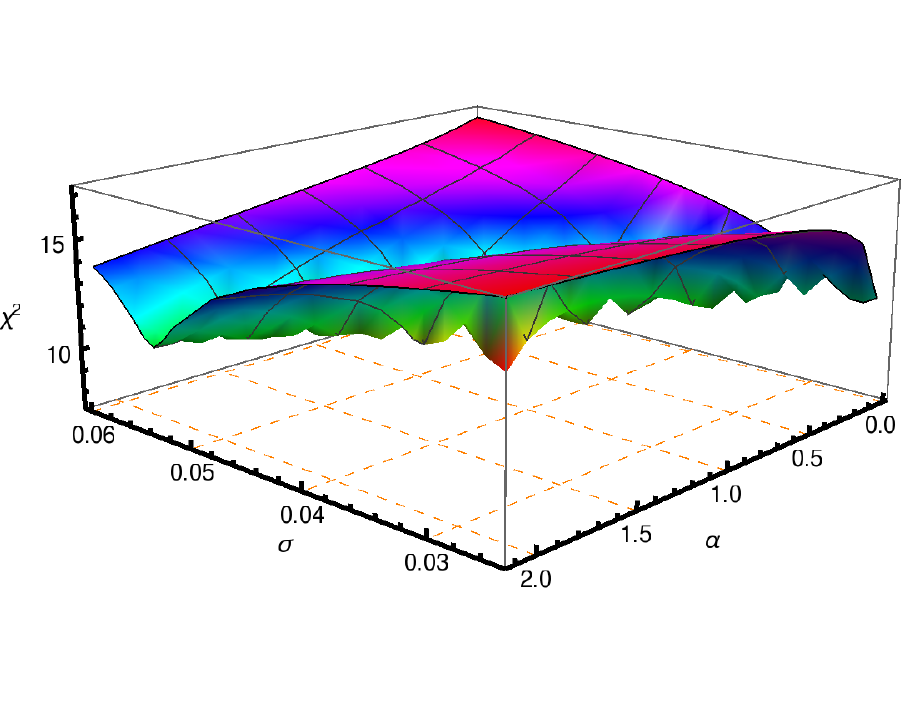}
\caption{Plot of $\chi^{2}_{dof}$ versus the string tension $\sigma_{0} a^2$ and the rigidity $\alpha$ from the fits of $Q\bar{Q}$ potential data to rigid string ansatz $V^{R}_{n \ell o} $Eq.~\eqref{Pot_NG_LO_NLO_Rigid} at $T/T_c=0.9$.}
\label{Chi2AlphaSigma}
\end{figure}

\begin{figure*}[!hptb]
\centering
\subfigure[]{\includegraphics[scale=0.37]{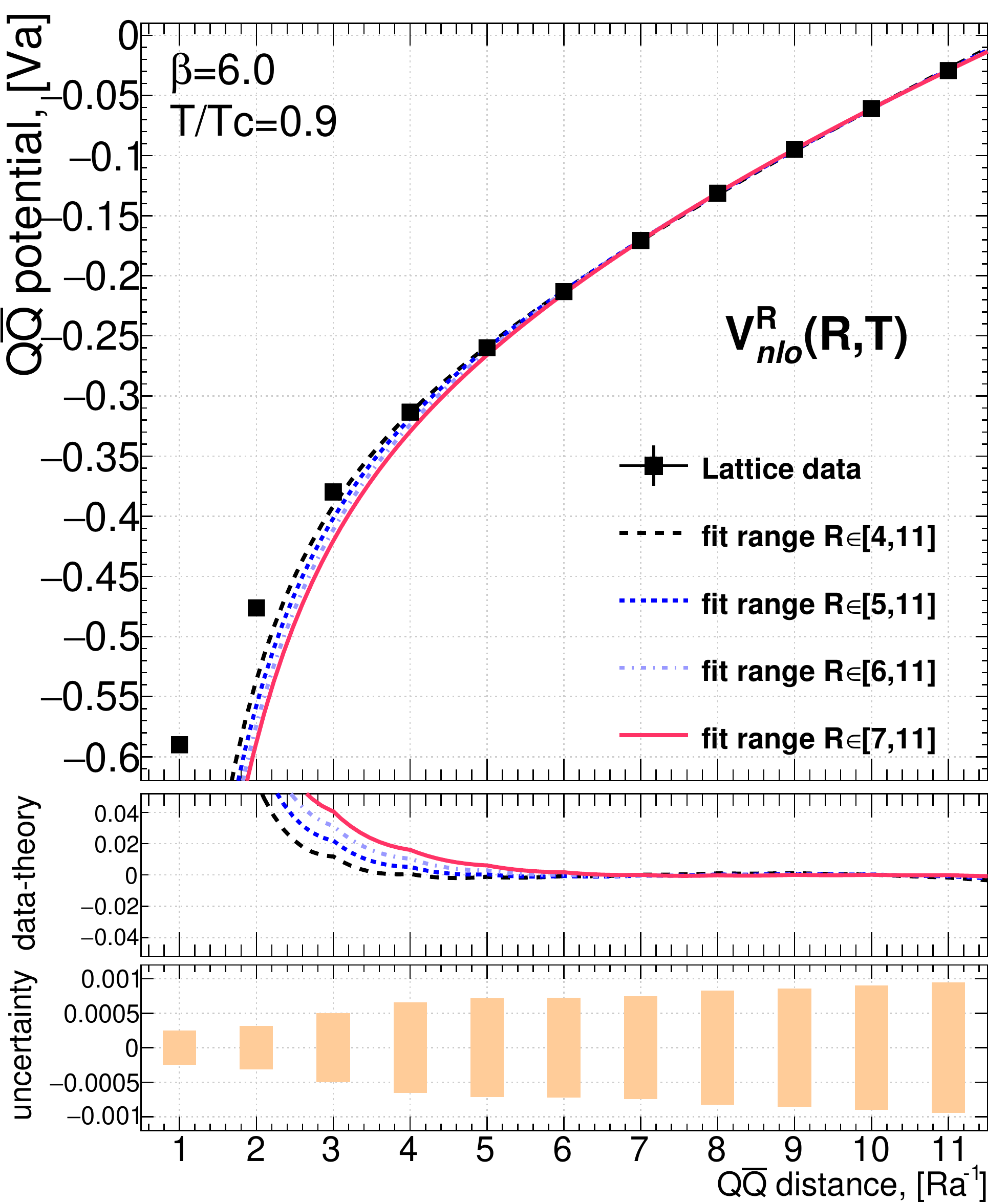}}
\subfigure[]{\includegraphics[scale=0.37]{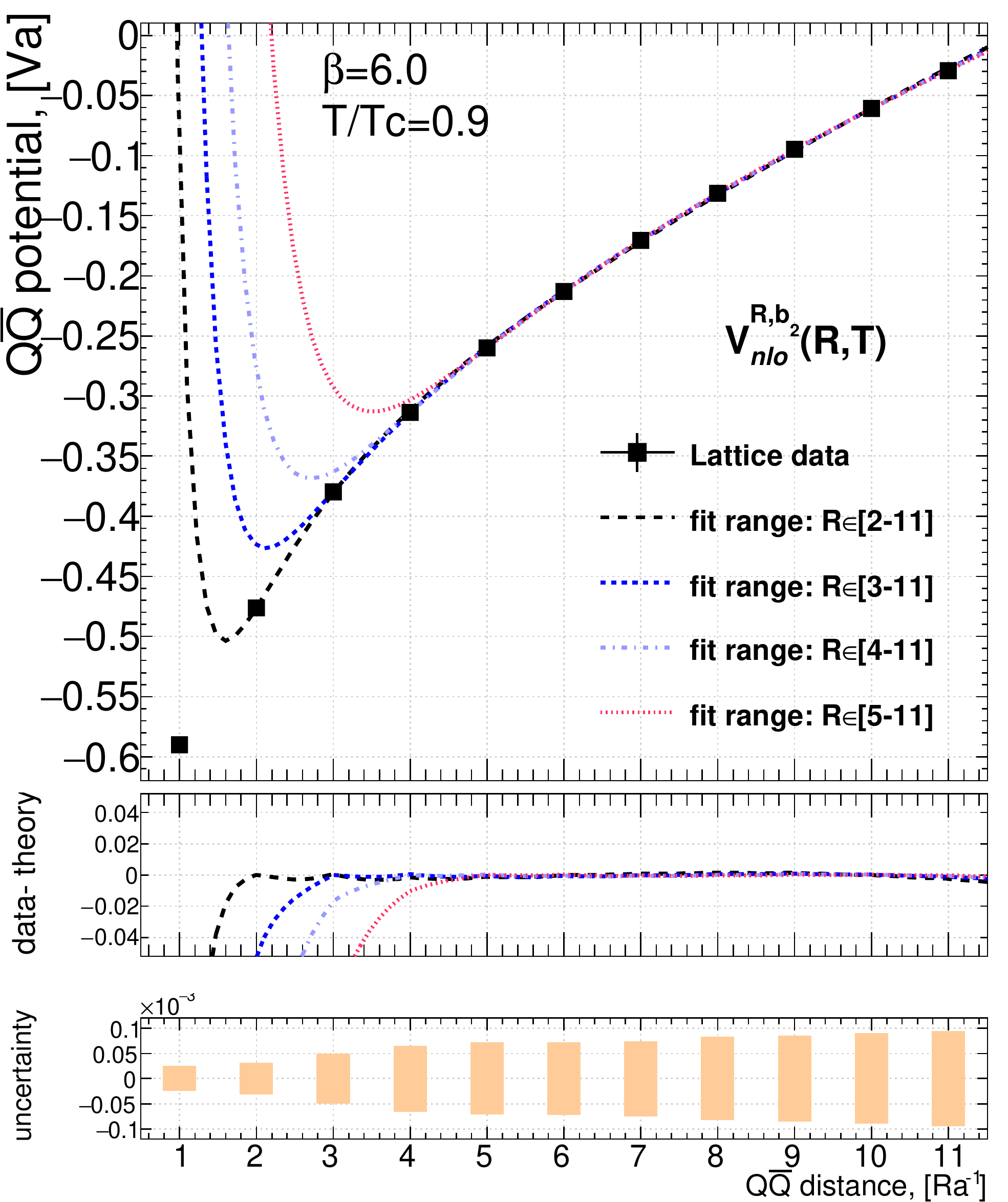}}
\caption{(a)The quark-antiquark $Q \bar{Q}$ potential  at $T/T_c=0.9$, the lines correspond to the fits (Table~\ref{T7_Pot_T09_NG_NLO_Rigid}) to the static potential of rigid string model $V_{n\ell o}^{R}$  of Eq.~\eqref{Pot_NG_LO_NLO_Rigid}. (b) Similar to (a), the lines correspond to the fits (Table~\ref{T7_Pot_T09_NG_NLO_Rigid_Boundb2}) to the rigid string but with the boundary term $b_2$ given by the model $V_{n\ell o}^{R,b2}$ of Eq.~\eqref{Pot_NG_LO_NLO_Rigid_Boundb2}.}
\label{Fig6_Fits_Pot_T09_NG_NLO_Rigid_Boundb2}
\end{figure*}


\begin{table}[!hpt]
	\begin{center}
		\begin{ruledtabular}
			\begin{tabular}{cc|ccccccccc}
			  \multirow{2}{*}{$V_{{n\ell o}}^{b_2}$} &Fit
				&\multicolumn{2}{c}{Fit Parameters, $T/T_{c}$=0.9} &\multicolumn{2}{c}{$$}\\
				&Interval
				&\multicolumn{1}{c}{$\sigma a^{2}$} &\multicolumn{1}{c}{$\mu$} 
				&\multicolumn{1}{c}{$b2$} &\multicolumn{1}{c}{$\chi^{2}$}\\
				\hline
                                \multirow{10}{*}{\begin{turn}{90} \hspace{1cm} $V_{{n\ell o}} + V^{b_{2}}$ \end{turn}}
                                &$[R_{m},R_{M}]$  &    &          &     \\
\multicolumn{1}{c}{} &[2,11]  &  0.0411327(9) &  -0.39693(1) & -0.1746(2)&2597.5   \\
\multicolumn{1}{c}{} &[3,11]  &  0.04124(1)   &  -0.39836(1) & -0.202(2)&2428.1     \\
\multicolumn{1}{c}{} &[4,11]  &  0.04094(2)   &  -0.3888(4)  &  0.20(1)&1572.9 \\
\multicolumn{1}{c}{} &[5,11]  &  0.0405(2)    &  -0.35(1)    &  2.3(8)&852.6    \\
\multicolumn{1}{c}{} &[6,11]  &  0.0401(3)    &  -0.19(7)    &  11(4)  &363.1   \\
\multicolumn{1}{c}{} &[7,11]  &  0.03954(5)   &   0.47(4)    &  52(3)&99.4      \\
\multicolumn{1}{c}{} &[8,11]  &  0.0389(2)    &   2.9(8)     &  205(50)&9.4    \\\hline
\multicolumn{1}{c}{} &$[R_{m},R_{M}]$  &    &             &     \\
\multicolumn{1}{c}{} &[2,5]  &0.04005(5) & -0.3926(2)& -0.1722(3)& 373.7  \\
\multicolumn{1}{c}{} &[3,6]  &0.04232(5) & -0.4062(4)& -0.278(4) & 0.049   \\
\multicolumn{1}{c}{} &[3,7]  &0.0423(4)  & -0.4057(3)& -0.274(3) & 2.2 \\
\multicolumn{1}{c}{} &[3,9]  &0.04189(2) & -0.4032(2)& -0.252(3) &205.2   \\
\multicolumn{1}{c}{} &[4,9]  &0.0417(3)  & -0.3994(4)& -0.10(2)	 &120.5   \\
\multicolumn{1}{c}{} &[5,10] &0.04102(3) & -0.368(2) &  1.46(9)  &268.5 \\
			\end{tabular}		
		\end{ruledtabular}
	\end{center}
	\caption{The $\chi^{2}$ values and the corresponding fit parameters $b2$ and $\mu$ returned from fits to the next to leading order (NLO) static potential with boundary terms $V_{n\ell o}^{b2}$ given by Eq.~\eqref{Pot_NG_LO_NLO_Boundb2}.}
\label{T4_Pot_T09_NG_NLO_boundb2}		
\end{table}
\begin{table*}[!hpt]
	\begin{center}
		\begin{ruledtabular}
			\begin{tabular}{cc|cccccccccc}
			  \multirow{2}{*}{\begin{turn}{90}\hspace{1cm}$ V_{{n\ell o}}^{b_2,b_4} $ \end{turn}  } &\tiny{Fit Interval}
				&\multicolumn{2}{c}{\tiny{Fit Parameters, $T/T_{c}=0.9$}} &\multicolumn{2}{c}{}\\
				&$R\in I$
				&\multicolumn{1}{c}{$\sigma_0 a^{2}$} &\multicolumn{1}{c}{$\mu $(LU)} 
			  &\multicolumn{1}{c}{$b_2$(LU)} &\multicolumn{1}{c}{$b_4$(LU)}&\multicolumn{1}{c}{$\chi^{2}$}\\
                          \\
				\hline
\multirow{10}{*}{\begin{turn}{90}\hspace{1cm}$V_{{n\ell o}}+V^{b_{2}}+V^{b_{4}}$ \end{turn}}				
&\tiny{$[R_{m},R_{M}]$} &           &           &     \\
\multicolumn{1}{c}{} &[2,11]  &0.04123(1)&-0.3991(2)&-0.236(5)&0.023(2)&2452.4\\
\multicolumn{1}{c}{} &[3,11]  &0.04091(2)&-0.332(2) &2.61(9)&-1.84(6)&1511.9\\
\multicolumn{1}{c}{} &[4,11]  &0.04053(2)&0.97(5) &58.1 (2.1)&-44.7(1.6)&805.8\\
\multicolumn{1}{c}{} &[5,11]  &0.0400(3) &27.7 (12.4)&1185.4 (520.8)&-948.9 (417.7)&336.6\\
\multicolumn{1}{c}{} &[6,11]  &0.0394(5) &553.3 (334.4)&23246.0 (14037.0)&-18790.0 (11352.0)&89.2\\
\multicolumn{1}{c}{} &\tiny{$[R_{m},R_{M}]$}  &     &           &         &        \\
\multicolumn{1}{c}{} &[2,7] &0.0423(4) &-0.4090(4) &-0.408(7)& 0.086(3) &4.8\\
\multicolumn{1}{c}{} &[3,8] &0.04198(4)&-0.393(3)  & 0.2(1)  &-0.31(8)  &19.4\\
\multicolumn{1}{c}{} &[4,9] &0.04142(4)&0.12(6)    &22.3 (2.5)&-17.2 (1.9)&41.3\\
			\end{tabular}		
		\end{ruledtabular}
	\end{center}
\caption{The $\chi^{2}$ values and the corresponding fit parameters $b_2$, $b_4$ and $\mu$ returned from fits to the next-to-leading order (NLO) static potential with boundary terms $V_{n\ell o}^{b2,b3}$ given by Eq.~\eqref{Pot_NG_LO_NLO_Boundb2b3}.}
\label{T5_Pot_T09_NG_NLO_boundb2b4}		
\end{table*}
\begin{table}[!hpt]
	\begin{center}
		\begin{ruledtabular}
			\begin{tabular}{cc|ccccccccc}
			  \multirow{2}{*}{$V_{{n\ell o}}^{R}$} &\tiny{Fit Interval}  
				&\multicolumn{2}{c}{\tiny{Fit Parameters, $T/T_{c}=0.9$}} &\multicolumn{2}{c}{}\\
				&$R\in I$
				&\multicolumn{1}{c}{$\sigma_0 a^{2}$} &\multicolumn{1}{c}{$\mu$ (LU)} 
			        &\multicolumn{1}{c}{$\alpha$} &\multicolumn{1}{c}{$\chi^{2} $}\\
                          	\multicolumn{1}{c}{} &$R_{m}$(LU)  &    &          &     \\
				\hline
                                \multirow{10}{*}{\begin{turn}{90}\hspace{1cm}$V_{{n\ell o}}+V^{R}$ \end{turn}}
				&\tiny{$[R_{m},11]$}  &    &          &     \\
\multicolumn{1}{c}{} &[2,11]  &0.0362498(8) & -0.35054(5)&0(3.6)  &627275.   \\
\multicolumn{1}{c}{} &[3,11]  &0.040(4)     & -0.39(1)   &0(3.7)  &11136           \\
\multicolumn{1}{c}{} &[4,11]  &0.04245(3)   & -0.3756(7) &0.059(2)&1148.11       \\
\multicolumn{1}{c}{} &[5,11]  &0.0435(2)    & -0.34(2)   &0.16(5) &319.657     \\
\multicolumn{1}{c}{} &[6,11]  &0.0440(2)    & -0.30(4)   &0.3(2)  &79.2561   \\
\multicolumn{1}{c}{} &[7,11]  &0.0442(3)    & -0.2(1)    &0.6(7)  &11.592    \\
\multicolumn{1}{c}{} &[8,11]  &0.044(2)     & -0.2(4)    &1(6)    &1.80228\\\hline
\multicolumn{1}{c}{} &\tiny{$[R_{m},R_{M}]$}  &      &       &     \\
\multicolumn{1}{c}{} &[2,5]  &0.02677(6) &-0.3095(3) & 0.000291    &460013. \\
\multicolumn{1}{c}{} &[3,6]  &0.038(8)   &-0.37(2)   & 0.002391    &5774.38     \\
\multicolumn{1}{c}{} &[3,7]  &0.039(5)   &-0.38(2)   & 0.002434	   &8395.19    \\
\multicolumn{1}{c}{} &[4,10] &0.040(4)   &-0.38(1)   & 0.002473	   &10933.4     \\
\multicolumn{1}{c}{} &[4,10] &0.04214(5) &-0.3879(8) & 0.033753    &487.035    \\
\multicolumn{1}{c}{} &[5,10] &0.04344(4) &-0.360(2)  & 0.108(5)    &119.563\\
			\end{tabular}		
		\end{ruledtabular}
	\end{center}
        \caption{The $\chi^{2}$ values and the corresponding fit parameters rigidity $\alpha$ and cutoff $\mu$ returned from fits to the rigid-self-interacting string potential $V_{n\ell o}^{R}$ given by Eq.~\eqref{Pot_NG_LO_NLO_Rigid}.}
\label{T7_Pot_T09_NG_NLO_Rigid}		
\end{table}
\begin{table}[!hpt]
	\begin{center}
		\begin{ruledtabular}
			\begin{tabular}{cc|cccccccccc}
			  \multirow{2}{*}{\begin{turn}{90}\hspace{1cm}$ V_{{n\ell o}}^{R,b_2} $ \end{turn}} &\tiny{Fit Interval}
				&\multicolumn{2}{c}{\tiny{Fit Parameters, $T/T_{c}=0.9$}} &\multicolumn{2}{c}{}\\
				&$R \in I$
				&\multicolumn{1}{c}{$ \sigma_0 a^{2}$} &\multicolumn{1}{c}{$\mu$ (LU)} 
			        &\multicolumn{1}{c}{$b_2$ (LU)} &\multicolumn{1}{c}{$\alpha_r$}&\multicolumn{1}{c}{$\chi^{2}$}\\
                          \\
				\hline
\multirow{10}{*}{\begin{turn}{90}\hspace{1cm}$V_{{n\ell o}}+V^{R}+V^{b_2}$ \end{turn}}				
 &\tiny{$[R_{m},R_{M}]$}  &     &          &     \\
\multicolumn{1}{c}{} &[2,11]  &4.169(4)&-0.3929(5)&0.025(1)     &-0.199(2) &2487.2  \\
\multicolumn{1}{c}{} &[3,11]  &4.343(2)&-0.359(1) &0.129(3)     &-0.533(6) &501.3  \\
\multicolumn{1}{c}{} &[4,11]  &4.395(2)&-0.337(2) &0.244(9)     &-1.1(3)   &161.6  \\
\multicolumn{1}{c}{} &[5,11]  &4.42(2) &-0.32(4)  &0.5(3)       &-3(2)     &38.8  \\
\multicolumn{1}{c}{} &[6,11]  &4.42(8) &-0.353(8) &1.0(2)       &-9(1)     &4.7  \\
\multicolumn{1}{c}{} &[7,11]  &4.39(4) &-0.48(4)  &3(2.9)       &-25(7)    &0.071  \\\hline
\multicolumn{1}{c}{} &\tiny{$[R_{m},R_{M}]$}  &    &          &    & \\
\multicolumn{1}{c}{} &[2,6] &0.041(1)   & -0.39(1) & 0.01(5) &-0.18(7) & 752.6\\
\multicolumn{1}{c}{} &[3,7] &0.0428(2)  & -0.400(3)& 0.03(1) &-0.31(3) & 0.74\\
\multicolumn{1}{c}{} &[3,9] &0.04327(4) & -0.385(1)& 0.066(3)&-0.40(1) &40.9 \\
\multicolumn{1}{c}{} &[4,10]&0.04394(3) & -0.355(2)& 0.173(9)&-0.97(4) &58.9 \\
\multicolumn{1}{c}{} &[5,10]&0.04430(3) & -0.343(4)& 0.31(3) &-2.6(2)  &14.7 \\
			\end{tabular}		
		\end{ruledtabular}
	\end{center}
 \caption{The $\chi^{2}$ values and the corresponding fit parameters; rigidity $\alpha_r$, boundary parameter $b_2$ and cutoff $\mu$, returned from fits to the rigid-self-interacting string potential $V_{n\ell o}^{R,b_2}$ given by Eq.~\eqref{Pot_NG_LO_NLO_Rigid_Boundb2}.}
\label{T7_Pot_T09_NG_NLO_Rigid_Boundb2}		
\end{table}

\begin{table}[!hpt]
	\begin{center}
		\begin{ruledtabular}
			\begin{tabular}{cc|cccccccccc}
			  \multirow{2}{*}{\begin{turn}{90}\hspace{1cm}$ V_{{n\ell o}}^{R,b_4} $ \end{turn}  } &\tiny{Fit Interval}
		          &\multicolumn{2}{c}{\tiny{Fit Parameters, $T/T_{c}=0.9$}} & \multicolumn{2}{c}{}\\
		          &$R\in I$
		          &\multicolumn{1}{c}{$\sigma_0 a^{2}$} &\multicolumn{1}{c}{$\mu$(LU)} 
			  &\multicolumn{1}{c}{$\alpha_r$} &\multicolumn{1}{c}{$b_4$(LU)}&\multicolumn{1}{c}{$\chi^{2}$}\\
                          \\
				\hline
\multirow{10}{*}{\begin{turn}{90}\hspace{1cm}$V_{{n\ell o}}+V^{R}+V^{b_4}$ \end{turn}}				
&\tiny{$[R_{m},R_{M}]$}  &     &          &     \\
\multicolumn{1}{c}{} &[2,11]  &0.0409(5) &-0.390(2)&0.01(3) &-0.067(7)&4460.22\\
\multicolumn{1}{c}{} &[3,11]  &0.04327(2)&-0.354(1)&0.109(2)&-0.323(4)&558.3\\
\multicolumn{1}{c}{} &[4,11]  &0.04393(2)&-0.312(2)&0.23(8) &-0.89(3) &154.9\\
\multicolumn{1}{c}{} &[5,11]  &0.04422(2)&-0.248(7)&0.45(3) &-2.6(1)  &36.02\\
\multicolumn{1}{c}{} &[6,11]  &0.04415(9)&-0.11(3) &1.1(2)  &-7.9(9)  &4.44\\
\multicolumn{1}{c}{} &\tiny{$[R_{m},R_{M}]$}  &    &          &   &      & \\
\multicolumn{1}{c}{} &[2,6]  & 0.0409(9)  & -0.390(4) & 0.01(6)  & -0.06(1)  &3835.6\\
\multicolumn{1}{c}{} &[3,8]  & 0.0427(1)  & -0.391(2) & 0.032(5) & -0.205(9) &13.5\\
\multicolumn{1}{c}{} &[4,10] & 0.0439(4)  & -0.335(3) & 0.162(8) & -0.71(3)  &56.3\\
\multicolumn{1}{c}{} &[5,10] & 0.0443(3)  & -0.284(9) & 0.30(3)  & -2.0(2)   &14.0\\
			\end{tabular}		
		\end{ruledtabular}
	\end{center}
        \caption{The $\chi^{2}$ values and the corresponding fit parameters $b_4$(LU) and $\mu$(LU) returned from fits to the next to leading order (NLO) static potential with boundary terms $V_{n\ell o}^{b4}$ given by Eq.\eqref{Pot_NG_LO_NLO_Rigid_Boundb3}.} 
 \label{T5_Pot_T09_NG_NLO_boundb3}		
\end{table}

\begin{table*}[!hpt]
	\begin{center}
		\begin{ruledtabular}
			\begin{tabular}{cc|cccccccccc|cccccccccc|}
			  \multirow{2}{*}{\begin{turn}{90}\hspace{1cm}$ V_{{n\ell o}}^{R,b_2,b_4} $ \end{turn}  } &\tiny{Fit Interval}
				&\multicolumn{2}{c}{\tiny{Fit Parameters, $T/T_{c}=0.9$}} &\multicolumn{2}{c}{}\\
				&$R\in I$
				&\multicolumn{1}{c}{$\sigma_0 a^{2}$} &\multicolumn{1}{c}{$\mu$(LU)} 
  &\multicolumn{1}{c}{$\alpha_r$} &\multicolumn{1}{c}{$b_2$(LU)}&\multicolumn{1}{c}{$b_4$(LU)}&\multicolumn{1}{c}{$\chi^{2}$}\\
                  \\
				\hline
\multirow{10}{*}{\begin{turn}{90}\hspace{1cm}\tiny{$V_{{n\ell o}}+V^{R}+V^{b_2}+V^{b_4}$} \end{turn}}				
&\tiny{$[R_{m},R_{M}]$}  &     &      &   &  & & &  \\
\multicolumn{1}{c}{} &[2,11]&0.043519, 0.0000183&-0.36374, 0.000937&-0.140876, 0.0032&-0.903213, 0.013&0.222968, 0.00464167&437.185\\
\multicolumn{1}{c}{} &[3,11]&0.0439943, 0.0000204&-0.404129, 0.00240&0.258954, 0.0100&-4.24603, 0.188&2.25785, 0.120102&145.932\\
\multicolumn{1}{c}{} &[4,11]&0.0442363, 0.0000234&-1.36259, 0.0873&0.502746, 0.038&-47.2974, 3.9&34.353, 3.07522&34.16\\
\multicolumn{1}{c}{} &[5,11]&0.0441876, 0.0000808&-13.6639, 2.41&1.03577, 0.206&-568.27, 102.6&689.797, 82.3587&3.96\\
\multicolumn{1}{c}{} &\tiny{$[R_{m},R_{M}]$}   &        &   &   & &&& \\
\multicolumn{1}{c}{} &[2,8]  &0.0432653, 0.0000787&-0.395082, 0.00194&0.0545925, 0.00554004&-0.569233, 0.0288&0.123925, 0.00867162&5.41\\
\multicolumn{1}{c}{} &[3,8]  &0.0437036, 0.000155&-0.401344, 0.00342&-0.0855039, 0.0161917&-1.42324, 0.399&0.628589, 0.23702&0.9\\
\multicolumn{1}{c}{} &[3,9]  &0.043877, 0.000071&-0.401914, 0.0030&0.120593, 0.0111493&-2.10003, 0.285&1.02458, 0.173838&7.4\\
\multicolumn{1}{c}{} &[4,9]  &0.0442943, 0.000106759&-0.78807, 0.146409&0.198267, 0.0383622&-19.6914, 6.71211&14.3284, 5.07934&0.7\\
\multicolumn{1}{c}{} &[4,10]  &0.0443398, 0.0000340&-1.08879, 0.109&0.330004, 0.0378015&-34.0812, 4.99&25.1703, 3.7973&12.5\\
			\end{tabular}		
		\end{ruledtabular}
	\end{center}
 \caption{The $\chi^{2}$ values and the corresponding fit parameters; rigidity $\alpha$, boundary parameters $(b_2,b_4)$ and string tension and cutoff $(\sigma,\mu)$, returned from fits to the rigid-self-interacting string potential with two boundary terms $V_{n\ell o}^{R,b_2,b_4}$ given by Eq.~\eqref{Pot_NG_LO_NLO_Rigid_Boundb2b3}.}
\label{T8_Pot_T09_NG_NLO_Rigid_Boundb2b3}		
\end{table*}

\subsubsection{Boundary terms in L\"uscher-Weisz action}

  Effects such as the interaction of the string with the Ployakov lines at the boundaries may be relevant to the discrepancies in the effective string description at the short and intermediate string length. The contribution to the Casmir energy due to the two next-to-leading nonvanshing term $S^b$ in L\"uscher-Weisz action Eq.~\eqref{LWaction} does not not affect the slop of the linearly rising part of the potential. However, the effects are received as inverse powers in $R$ and are given by $V_{n\ell o}^{b_2,b_3}$ of Eq~\eqref{Pot_NG_LO_NLO_Boundb2b3}.  
\begin{figure}[!hptb]
\includegraphics[scale=0.35]{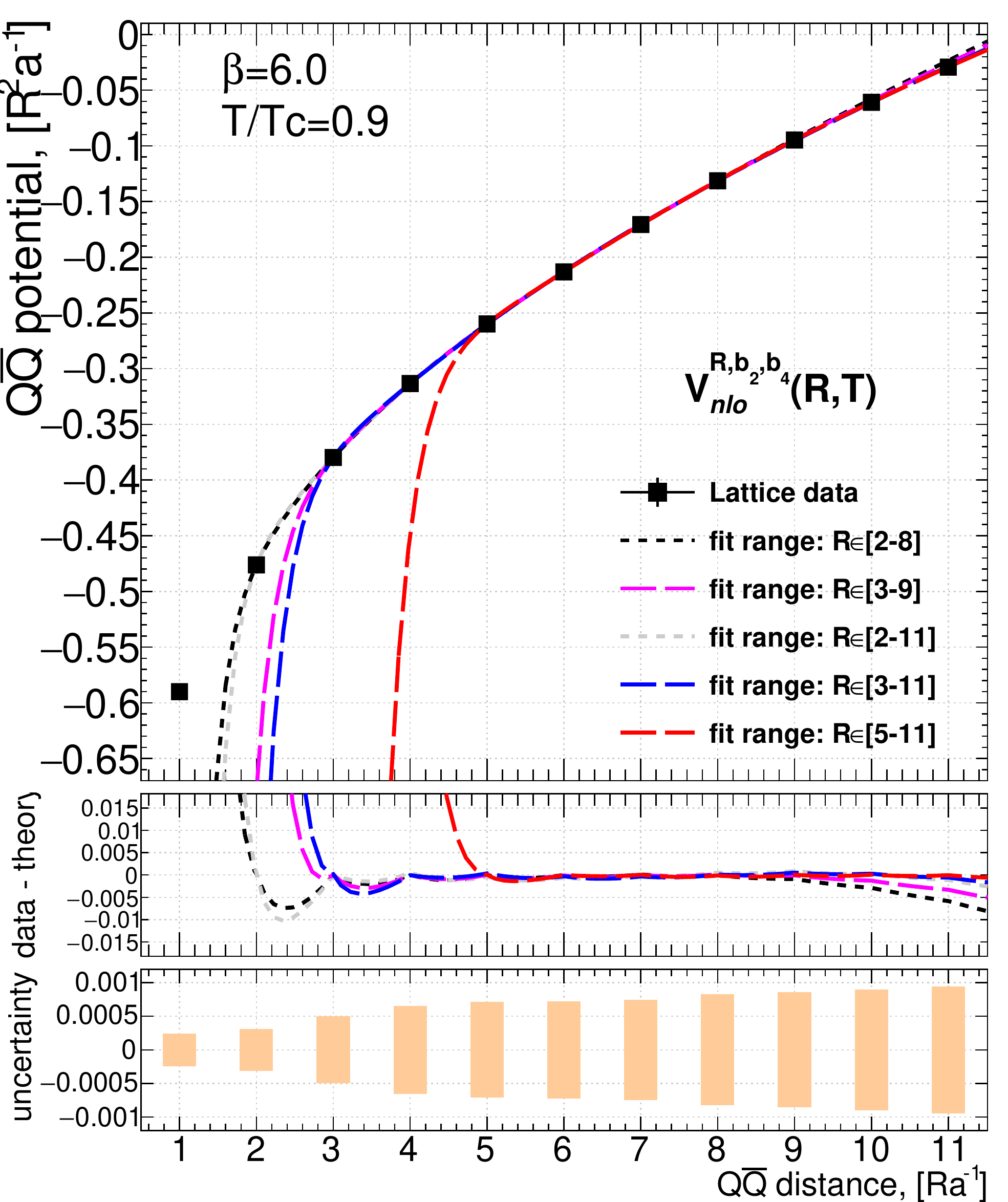}
\caption{The quark-antiquark $Q \bar{Q}$ potential  $T/T_c=0.9$. The lines correspond to the fits to self-interacting NG potential with two boundary terms $V_{n\ell o}^{b_2,b_4}$ given by Eq.~\eqref{Pot_NG_LO_NLO_Boundb2b3},in addition the rigidity of the string has been included in the fitting ansatz $V_{n\ell o}^{R,b_2,b_4}$ Eq.~\eqref{Pot_NG_LO_NLO_Rigid_Boundb2b3}.}
\label{Fig7_Fits_Pot_T09_NG_NLO_Rigid_Boundb2b3}
\end{figure}
    
  Since the leading nonvanshing boundary terms appear at the order of four derivative term Eq.~\eqref{boundaryaction} in L\"uscher-Weisz action Eq.~\ref{LWaction}, It may be more convenient to discuss the corresponding effects ~\eqref{Pot_Boundb2} in conjunction with the NLO form of the potential of the NG action Eqs.~\ref{Pot_NG_NLO} with the renormalization of the string tension included.

 For the purpose of the discussion of the numerical data of the static meson potential, We define the following possible combinations of LO and NLO Nambu-Goto static potential with boundary terms, 

  
\begin{align}
V_{n\ell o}^{b_2}&=V_{n\ell o}+V^{b_2},\label{Pot_NG_LO_NLO_Boundb2}\\
V_{\ell o,n\ell o}^{b_4}&=V_{\ell o,n \ell o}+V^{b_4},\label{Pot_NG_LO_NLO_Boundb3}\\
V_{n \ell o}^{b_2,b_4}&=V_{n\ell o}+V^{b_2}+V^{b_4},\label{Pot_NG_LO_NLO_Boundb2b3}
\end{align}  
  where subscripted $V_{\ell o, n\ell o}$ denotes  either the LO NG static potential or the NLO Eq.~\eqref{NGpert}. The $Q\bar{Q}$ potential data are fitted to the static potential with a two possibly interesting combinations of the boundary terms $V_{n\ell o}^{b_2}$ and $V_{n \ell o}^{b_2,b_4}$ given by Eq.~\eqref{Pot_NG_LO_NLO_Boundb2} and  Eq.~\eqref{Pot_NG_LO_NLO_Boundb2b3}, respectively. The inspection of each boundary term allows for understanding of the relevance of each boundary parameters to the fit arrangement. The corresponding returned values of $\chi^{2}$ and fit parameters are enlisted in Table~\ref{T4_Pot_T09_NG_NLO_boundb2},~\ref{T5_Pot_T09_NG_NLO_boundb3} and ~\ref{T5_Pot_T09_NG_NLO_boundb2b4}, considering various fit intervals. 
  The fit to the static potential $V_{n\ell o}^{b_2}$ of the model Eq.~\eqref{Pot_NG_LO_NLO_Boundb2} returns values for the parameter $b_2$ which appear to vary dramatically with the considered range. The values of $\chi^{2}$ are still high when considering the entire fit-interval $R \in [0.5,1.1]$ fm. Even though, none trivial improvements in the values of $\chi^{2}$ are retrieved as shown in Table.~\ref{T4_Pot_T09_NG_NLO_boundb2}  compared to that obtained by merely considering the NG string potential  Eqs.~\eqref{Pot_NG_NLO} (Table.~\ref{T3_Pot_NG_LO_NLO}). The fits to the static potential with the boundary term $V_{n\ell o}^{b_2}$ produce acceptable $\chi^{2}_{\rm{dof}}$ value for shorter fit intervals commencing from $R \in [0.7,1.2]$ fm. 

  With the interchange of the term $V_{n\ell o}^{b_4}$ in place of $V_{n\ell o}^{b_2}$ in the string model Eq.~\eqref{Pot_NG_LO_NLO_Boundb2}, the fits are not surprisingly good. However, the interesting observation is that the fits at shorter distances commencing from $R=0.3$ fm upto $R=0.7$ fm, for both models $V_{n\ell o}^{b_2}$ and $V_{n \ell o}^{b_2,b_4}$, are remarkably good with a returned $\sigma_{0}a^{2}=0.042$ (Table~\ref{T5_Pot_T09_NG_NLO_boundb2b4}). 
  
  Inspection of Tables~\ref{T4_Pot_T09_NG_NLO_boundb2} and ~\ref{T5_Pot_T09_NG_NLO_boundb2b4} indicates that fits to the NLO form of (NG) string with boundary term $V_{n \ell o}^{b_{2}}$ produce very close value of $\sigma_{0}a^{2}$ as the pure NG string $V_{n \ell o}$. The same observation holds for the fits to $V_{n \ell o}^{b_4}$ and $V_{n \ell o}^{b_2,b_4}$. Acceptable value of $\chi^{2}$ are returned $\sigma_{0}a^{2}=0.0397$ over the fit interval $[0.7,1.2]$ fm.

  Despite of the reductions in the minima of $\chi^{2}_{dof}$, the string models with boundary terms have no significantly different behavior with respect to the string tension parameter. This is consistent with the modular transforms where the inverse of the cyliner's modular parameter does not produce terms linear in $R$. Therof, the boundary corrections to the static potential do not contribute to the renormalization of the string tension.

\subsubsection{Rigidity terms in Polyakov-Kleinert action}
  
  Establishing a precise string description of the $Q\bar{Q}$ potential data and a correct thermodynamic behavior for the string tension at high temperature have been a long withstanding issues in many investigated gauge models. The consideration of the boundary terms solely does not provide an optimal fits and other possible string properties may be questioned in this context. The possible rigidity/stiffness/self-repulsion of QCD flux tube; or the resistance to sharp transverse-bending should manifest by the onset of the excited fluctuations at high temperatures.

  In order to unambiguously appreciate the changes on the fits when the rigidity of the string is taken into account, we discuss the modified static potential of rigid string in conjunction with both the leading and the next-to-leading approximations to NG action $V_{\ell o}^{R}$ and $V_{n \ell o}^{R}$ Eqs.~\eqref{Pot_NG_LO_NLO_Rigid}, separately.
  
  More variants of string models can be attained by including other combinations of the rigid terms such as,
\begin{align}
V_{n\ell o}^{R}&=V_{n\ell o}+V^{R},\label{Pot_NG_LO_NLO_Rigid}\\
V_{n\ell o}^{R,b_2}&=V_{n\ell o}+V^{R}+V^{b_2},\label{Pot_NG_LO_NLO_Rigid_Boundb2}\\
V_{n\ell o}^{R,b_4}&=V_{n\ell o}+V^{R}+V^{b_4},\label{Pot_NG_LO_NLO_Rigid_Boundb3}\\
V_{n \ell o}^{R,b_2,b_4}&=V_{n\ell o}+V^{R}+V^{b_2}+V^{b_4},\label{Pot_NG_LO_NLO_Rigid_Boundb2b3}
\end{align}

 the above compilations are particular choices of terms from the most general formalism L\"uscher-Weisz action.
  
  We proceed in the fit analysis of the $Q\bar{Q}$ static potential data without fixing the value of the string tension. The rigidity factor, $\alpha_{r}$ which weighs the extrinsic curvature tensor, and the ultraviolet cutoff $\mu$ are taken as a free fit parameters as well. Table.~\ref{T7_Pot_T09_NG_NLO_Rigid} summarizes values of $\chi^{2}$ obtained from fits to $V_{n \ell o}^{R}$ Eq.~\eqref{Pot_NG_LO_NLO_Rigid}. We remark the following points:
    
   Figure~\ref{Chi2AlphaSigma} plots the returned $\chi^{2}$ values versus both the string tension and rigidity. The plot indicates the quality of the fit in the parameteric space $(\alpha_{r},\sigma_{0})$, the oscillatory nature of $\chi^{2}_{dof}$ when the rigidity is included and attainment of the global minima of $\chi^{2}$  at $\sigma_0=0.044$.

   Drawing comparison between the returned $\chi^{2}$ in Table.~\ref{T2_Pot_NG_LO_NLO} and Table~\ref{T7_Pot_T09_NG_NLO_Rigid} reveals significant improvement in the fit behavior with the rigidity term $V_{n\ell o}^{R}$ of the string model Eq.~\eqref{Pot_NG_LO_NLO_Rigid} over the pure NG string potential given by Eq.~\eqref{Pot_NG_LO} and \eqref{Pot_NG_NLO}.
    
   The residuals are reduced on the fit interval $R \in [0.7,1.1]$ fm, the returned $\chi^{2}_{dof}$ indicate good values for $R>0.7$ fm. Remarkably, the value of the returned string tension $\sigma a^{2}=0.0442(3)$ on the interval $R\in[0.7,1.1]$ fm and  $\sigma a^{2}=0.044(2)$ on the interval $R\in[0.8,1.1]$ fm  is shifted above the value obtained from considering fits to merely the static potential of the pure NG string. The fit to Eq.~\eqref{Pot_NG_LO_NLO_Rigid} results in a value of the string tension which, within the numerical uncertainities, is equivalent to that reproduced at zero temperature measurements~\cite{Koma:2017hcm}.

   The consideration of the two-parameter rigid string model $(\mu,\alpha_{r})$ with the next-to-leading order NG potential $V_{n\ell o}^{R}$ of Eq.~\eqref{Pot_NG_LO_NLO_Rigid} results in a smaller $\chi^{2}$ compared to the fit with the two parameter $(\mu,\alpha)$ string models $V_{n\ell o}^{b_2}$ of formula Eq.~\eqref{Pot_NG_LO_NLO_Boundb2}. Nevertheless, the fit with boundary action models (Table.~\ref{T5_Pot_T09_NG_NLO_boundb2b4})compares to the rigid string when considering larger parametric space,i.e., the three parameter $(\mu, b_2,b_4)$ model of $V_{n\ell o}^{b_2,b_4}$ Eqs.~\eqref{Pot_NG_LO_NLO_Boundb2b3}.
  
   The plot in Fig.~\ref{Fig6_Fits_Pot_T09_NG_NLO_Rigid_Boundb2}(a) is the fit of the static potential of the rigid string for the fit intervals over the given in Table~\ref{T7_Pot_T09_NG_NLO_Rigid}. A descending sequence of the values of rigidity parameter $\alpha_r$ are returned. The values of $\alpha$ decreases from $\alpha=1.0$ to $\alpha=0.16$ as one includes smaller distances into the fit range over intermediate distances $[0.8,11]$ to $[0.5,11]$ fm.

   Large uncertainties in the rigidity parameter $\alpha_r$ are returned from the fits with the decrease of minimal source separation $R_{m}$ of the fit range $R\in[R_{m},R_{M}]$. This is perhaps owing to higher-order terms in the perturbative expansion of the extrinsic curvature in the rigid string action.

   The renormalization of the string tension has been explicitly given by German~\cite{German:1989cz,German:1991tc} long ago. The misfortune that we are lacking an ansatz for the potential for the two-loop static quark-antiquark~\cite{German:1989vk,German:1989cz,German:1991tc} at finite temperature scales and dimension. We evaluate the thermodynamic properties of this string gas at two-loop orders and examine the effects on the relatively short distance physics in a separate report~\cite{Bakry2020}.   

   Despite of the outstanding match between the static potential curve with $\alpha_r=0.6$ and the data over the range $R \in [0.7,1.1]$, the plot in Fig.~\ref{Fig6_Fits_Pot_T09_NG_NLO_Rigid_Boundb2}(a) exhibits palpable deviations if short distances $R<0.5$ fm are not included in the fit interval.

   Nevertheless, the reduction in the residual from the fits to Eq.~\eqref{Pot_NG_LO_NLO_Rigid} and the subsequent retrieve of the correct normalization of the string tension grasp a clue that the rigid properties are nontrivial ingredient in a faithful representation of the physics of the QCD flux-tube. 

\begin{figure*}[!hptb]
\centering
\includegraphics[scale=0.75]{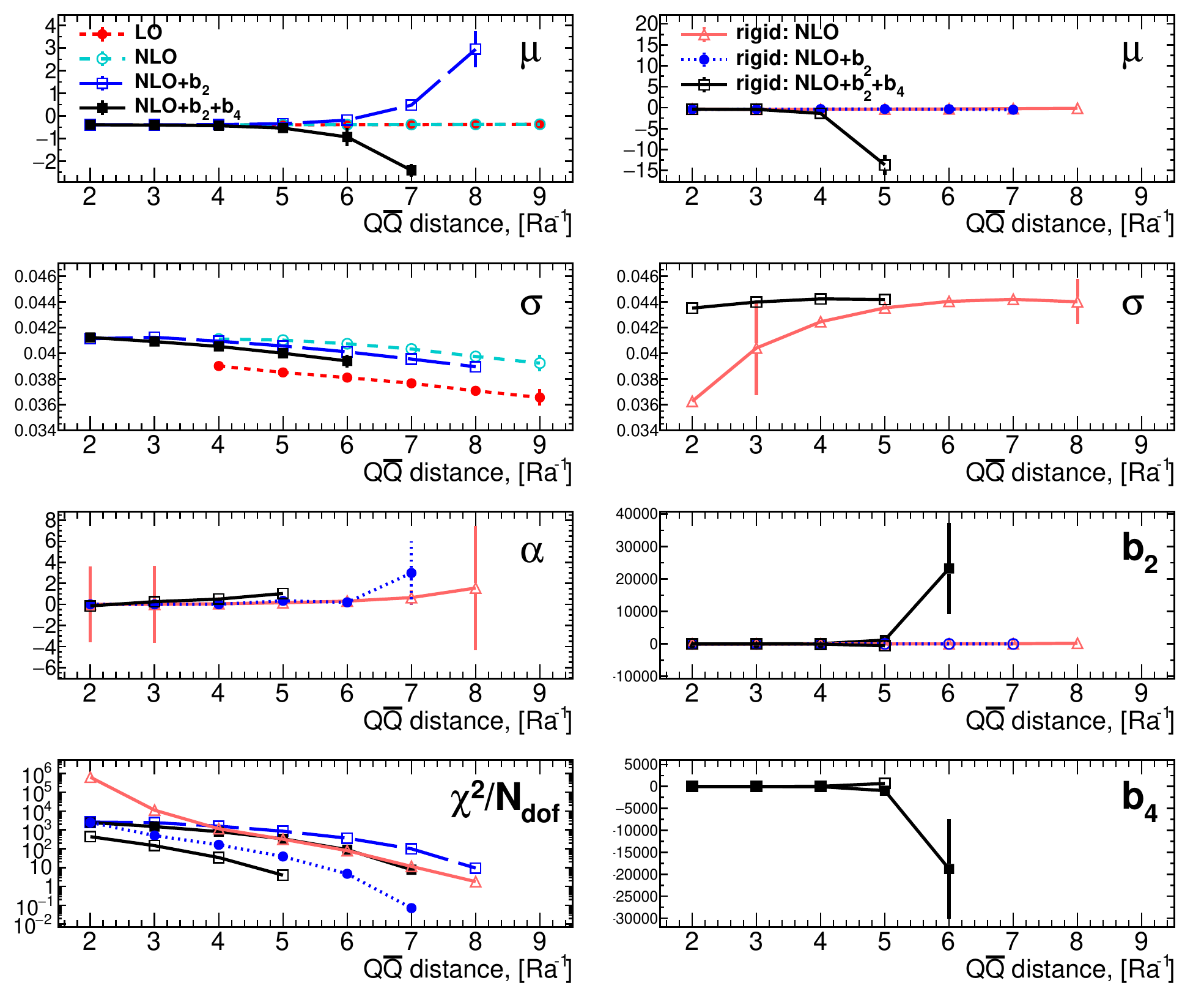}
\caption{Parameter chart of returned from the fit of each corresponding string model versus the fit rang lower bound $R_{m}$ of the interval $ R \in [R_{m}, R_{M}]$ whereas the upper $[R_{m}]$ bound is kept fixed at $R_{M}=1.1$ fm. Each point type correspond to the parameter value depicted in the corresponding pure NG string model Eq.~\eqref{Pot_NG_LO}, Eq.~\eqref{Pot_NG_NLO}, the NLO NG model with boundary terms Eq.~\eqref{Pot_NG_LO_NLO_Boundb2}, Eq.~\eqref{Pot_NG_LO_NLO_Boundb2b3}, and the rigid string models Eq.~\eqref{Pot_NG_LO_NLO_Rigid},  Eq.~\eqref{Pot_NG_LO_NLO_Rigid_Boundb2} and Eq.~\eqref{Pot_NG_LO_NLO_Rigid_Boundb2b3}. The last subfigure on the left shows the $\chi^2$ corresponding to each model.}  
  \label{ParameterChart}
\end{figure*}
\begin{figure*}[!hptb]
\centering
\subfigure[]{\includegraphics[scale=0.37]{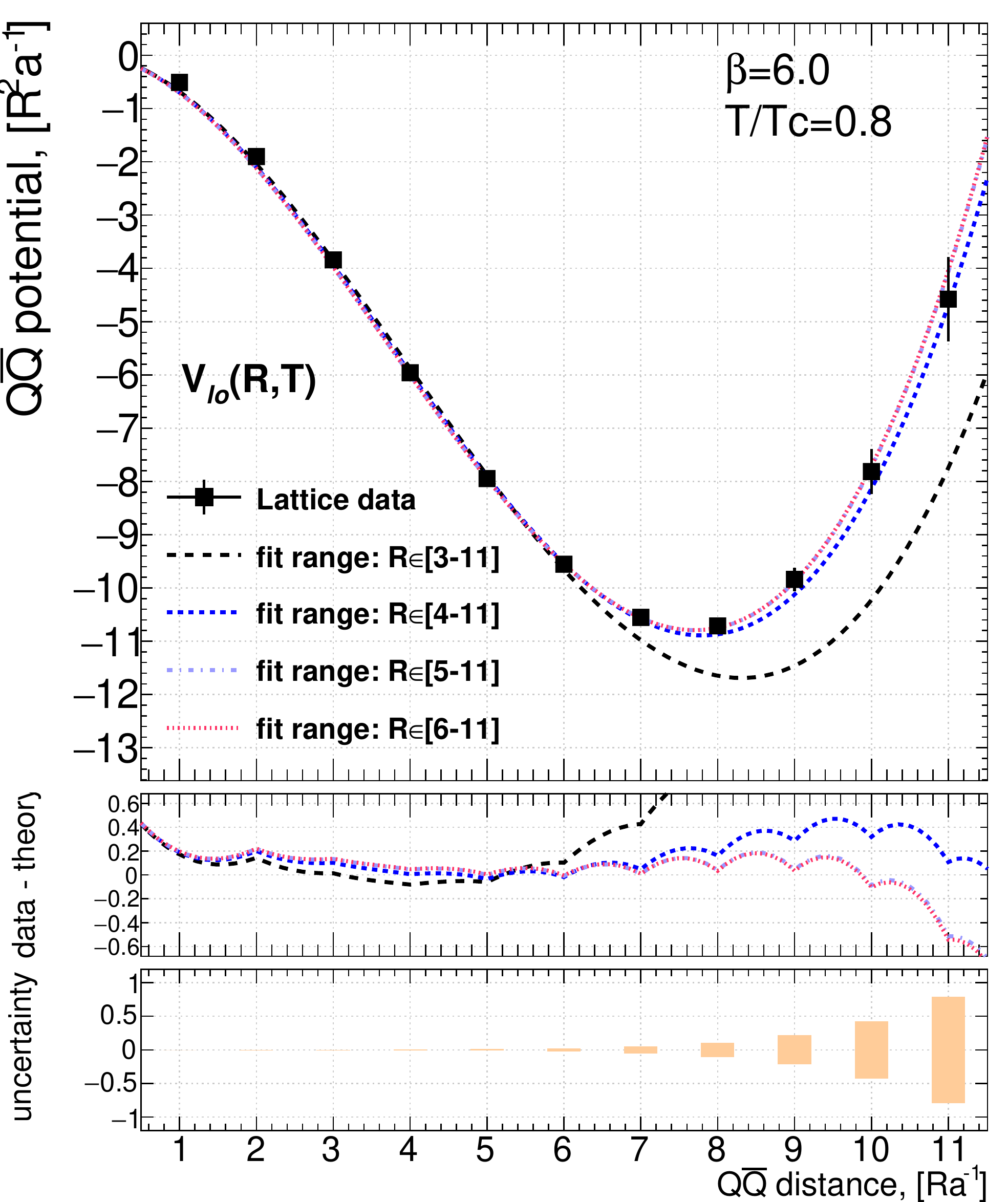}}
\subfigure[]{\includegraphics[scale=0.37]{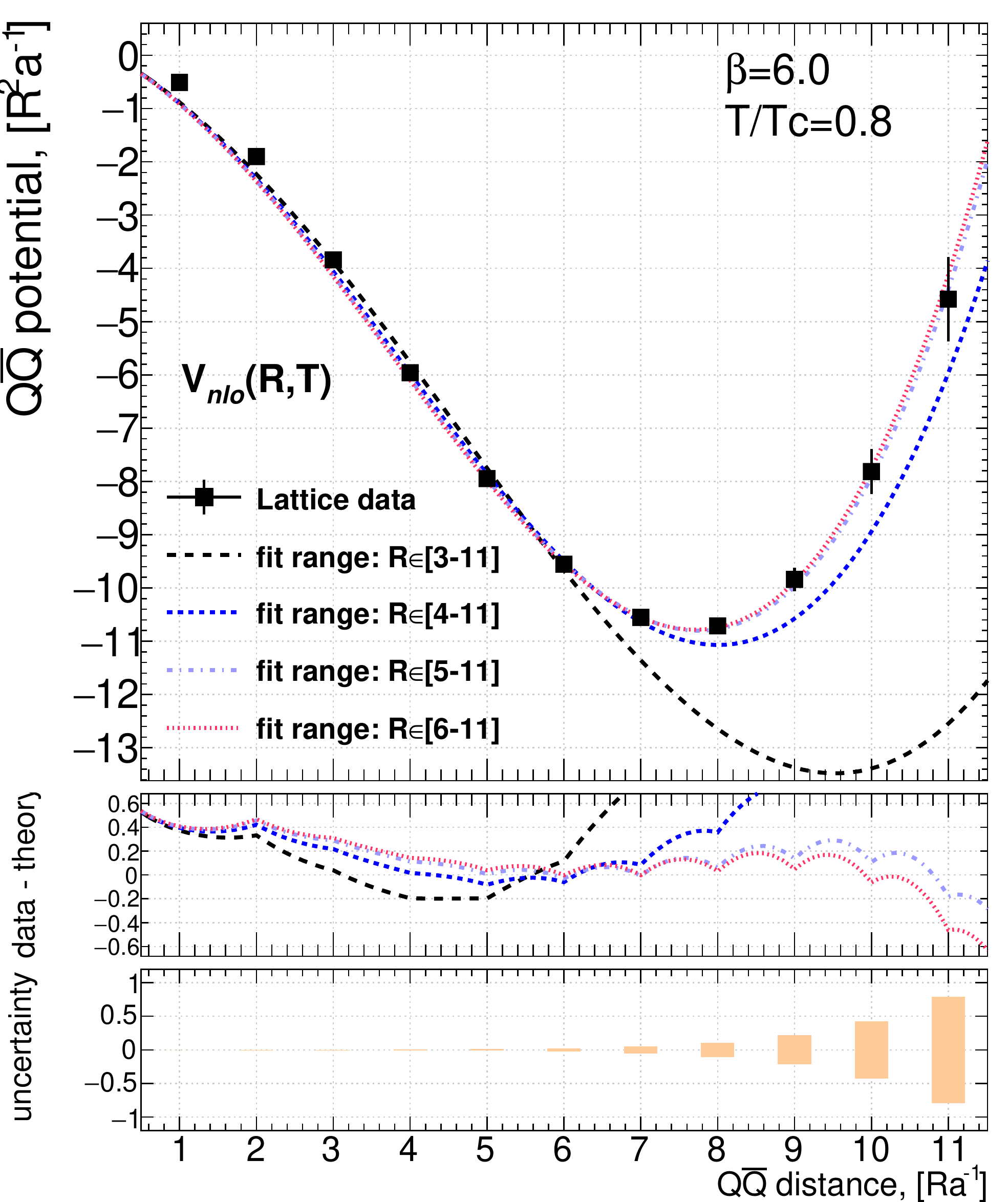}}
\caption{The quark-antiquark $Q \bar{Q}$ potential data at temperature $T/T_c=0.8$ the lines correspond to the fits to the corresponding string models. The potential data on the Y-axis have been scaled by the square of source separation distance $R^2$ and is given in Lattice units (a) The LO Nambu-Goto string model $V_{\ell o}$ given by Eq.~\eqref{Pot_NG_LO} for the depicted fit interval $R \in [R_{m},R_{M}]$ (b) The corresponding fits to the NLO Nambu-Goto string model $V_{n\ell o}$ Eq.~\eqref{Pot_NG_NLO}.}
\label{Fits_Pot_T08_NG_LO_NLO}
\end{figure*}

\begin{figure*}[!hptb]
\subfigure[]{\includegraphics[scale=0.37]{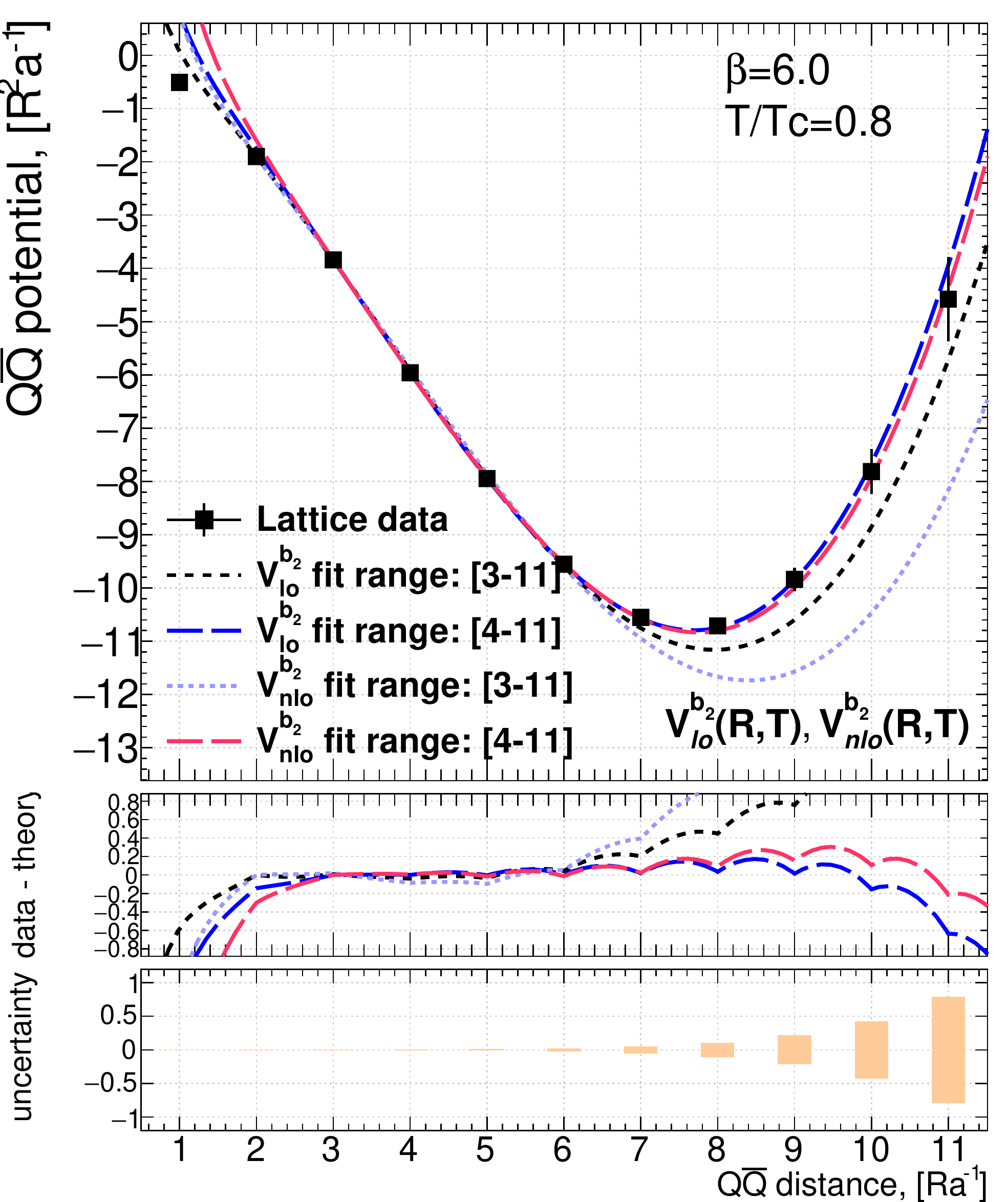}}
\subfigure[]{\includegraphics[scale=0.37]{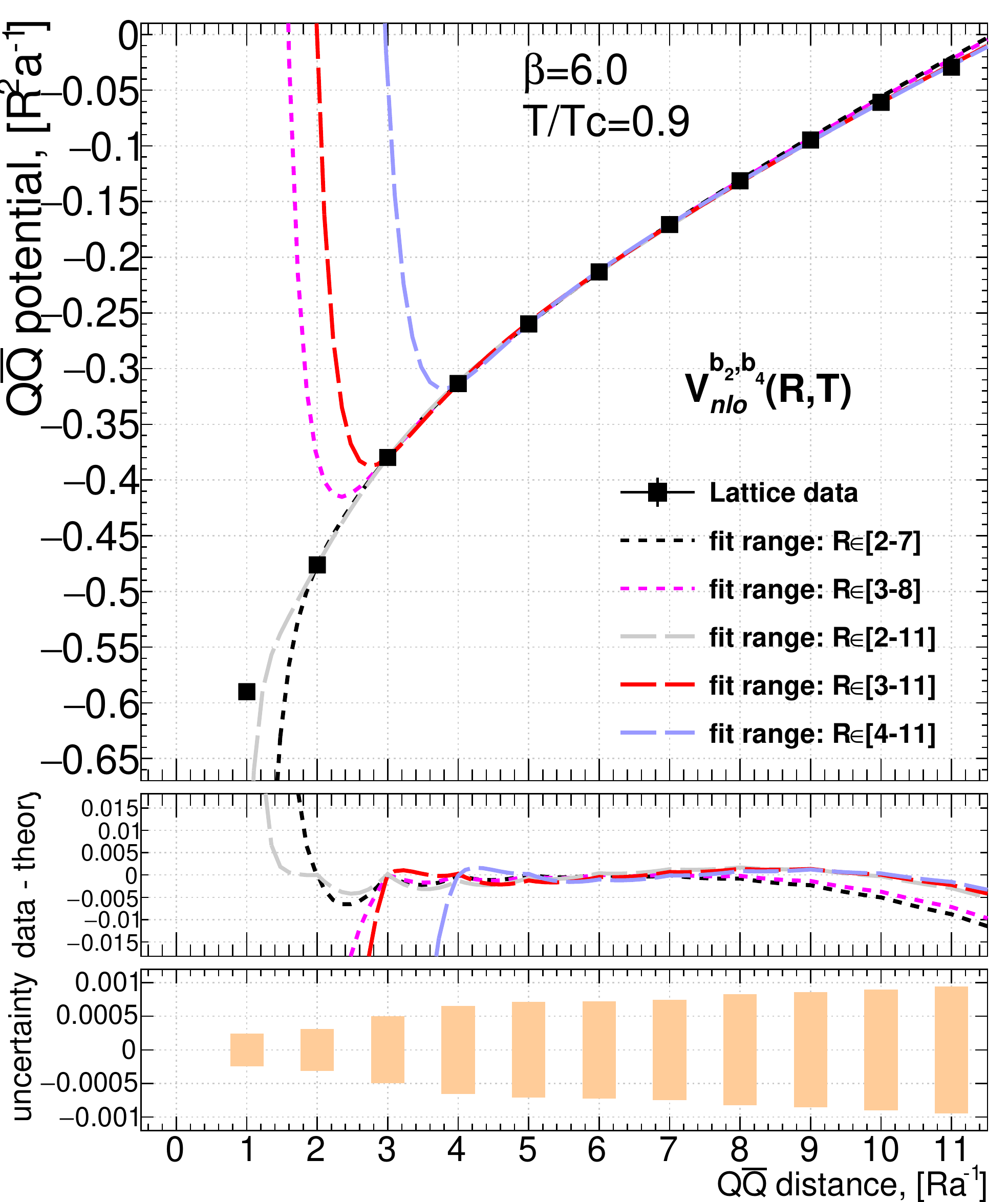}}
\caption{The quark-antiquark $Q \bar{Q}$ potential data at temperature $T/T_c=0.8$ the lines correspond to the fits to the corresponding string models. The potential data on the Y-axis have been scaled by the square of source separation distance $R^2$ and is given in Lattice units (a) Compares the fits from both LO and NLO Nambu-Goto string with boundary term $V^{b_2}$, the fitted models are given by $V^{b_2}_{\ell o}$ Eq.~\eqref{Pot_NG_LO_NLO_Boundb2} and Eq.~\eqref{Pot_NG_LO_NLO_Boundb2} (b) Similar to (a); however, the fits are for both boundary terms $V^{b_2,b_3}$. The fitted models are given by $V^{b_2,b_3}_{\ell o}$ Eq.~\eqref{Pot_NG_LO_NLO_Boundb2b3} and $V^{b_2,b_3}_{n\ell o}$ Eq.~\eqref{Pot_NG_LO_NLO_Boundb2b3} }
\label{Fits_Pot_T08_NG_LO_NLO_Boundb2b3}
\end{figure*}

    As discussed above, the boundary corrections to the static potential are mitigating the short distance mismatch with the data; whereas, the rigidity term evidently returns acceptable values of $\chi^{2}$ and string tension over long distances. An optimization of both models is expected, thereof, to provide a prospect for an extended distance scale of validity.

   The models given by Eq.~\eqref{Pot_NG_LO_NLO_Rigid_Boundb2}, Eq.~\eqref{Pot_NG_LO_NLO_Rigid_Boundb3} and Eq.~\eqref{Pot_NG_LO_NLO_Rigid_Boundb2b3} define selected compilations grasping both aspects of the rigidity and boundary effects. Tables.~\ref{T8_Pot_T09_NG_NLO_Rigid_Boundb2b3}, \ref{T5_Pot_T09_NG_NLO_boundb3} and \ref{T7_Pot_T09_NG_NLO_Rigid_Boundb2} enlist the parameters and $\chi^{2}$ returned from the fits of the numerical data of the static potential to these models, respectively. 	

   Inspection of the above mentioned tables reveals that the two models given by the interchanging the two boundary terms $V^{b_2}$ and $V^{b_4}$ in  Eq.~\eqref{Pot_NG_LO_NLO_Rigid_Boundb2}, Eq.~\eqref{Pot_NG_LO_NLO_Rigid_Boundb3} are yielding almost comparable values of $\chi^2$ for each given fit interval. 

   However, the model encompassing both of the boundary terms Eq.~\eqref{Pot_NG_LO_NLO_Rigid_Boundb2} provide the best fit for the targeted distance which is the intermediate distance scale [5,11], $\chi^{2}=3.96$. Interestingly, the fit over short distances interval in the second panel of Table.~\ref{T8_Pot_T09_NG_NLO_Rigid_Boundb2b3} is as well indicating a good $\chi^{2}$ for most fit intervals. The plot of the static potential in Figs.~\ref{Fig6_Fits_Pot_T09_NG_NLO_Rigid_Boundb2}(a), Fig.~\ref{Fig6_Fits_Pot_T09_NG_NLO_Rigid_Boundb2}(b) and Fig.~\ref{Fig7_Fits_Pot_T09_NG_NLO_Rigid_Boundb2b3} respectively illustrates the subsequent diminish of the errors $|Data-Model|$ as the boundary terms are included to rigid model.
   
   The panel of Fig.~\ref{ParameterChart} congregates subfigures each show the variation in the parametric subspace versus the below bound $R_{m}$ of the fit interval $[R_{m},R_{M}]$ whereas the last point has kept fixed at $R_{M}=1.1$ fm. 

   The returned zero-temperature string tension $\sigma_{0}a^{2}$ unless otherwise the rigidity is considered is decreasing with the increase of lower bound of the fit $R_{M}$ to achieve acceptable reduction of $\chi^{2}$. The remarkable feature is that the contrary of this behavior is observed when we string models with rigidity are considered.

  The rigidity parameter $\alpha_r$ appears to assume stable values $\alpha_r \le 1.0$ in the range $R_{m} \in [0.4,0.6]$ fm. At $R=0.3$ fm, the large uncertainities are suppressed owing to the additional of boundary terms of couplings $b_2$ and $b_4$. 
  
  The optimal $\chi^{2}$ is attained at $R_{m}=0.5$ fm for the rigid string model with two boundary terms of Eq.~\eqref{Pot_NG_LO_NLO_Rigid_Boundb2b3}. The increase in $\chi^{2}$ is almost dramatic for all other models $R \le 0.5$ fm. However, good $\chi^{2}$ indicating the long distance behavior of effective bosonic strings are obtained from distances $R_{m}=0.9$ and decreases according to a certain selection from the extended parametric space $(\alpha_r,b_2,b_4)$ to $R_{m}=0.5$ fm. 


\subsection{Temperature scale near the Plateau $T/T_c=0.8$}
  The analysis of the pure gluonic configuration near the end of QCD plateau is very interesting since a small change in the temperature could produce essential different effects on the properties of the confining force. It ought to be instructive to extend the above reported analysis to the lower temperature scale $T/T_c=0.8$ where the thermal fluctuations are expected to be milder.
  
  In the following we question the string self-interaction and the rigidity together with the boundary terms at each selected source separation intervals. Our target is to illuminate which terms in the interaction potential which persist to provide a good match with the numerical data regardless of the temperature scale in addition to the fit ansatz which nolonger have palpable effects on the fit behavior with the decrease of the temperature.

\begin{table}[!hpt]
	\begin{center}
		\begin{ruledtabular}
			\begin{tabular}{ccccccccccc}
				\multirow{2}{*}{} & &\multirow{2}{*}{$V_{Q\bar{Q}}(R,T)$ $fm^{-1}$} &\multirow{2}{*}{$e(R)$}\\
				&$n=R/a$ &&&\\
				\hline                                             
				\multirow{12}{*}{\begin{turn}{90}$~T/T_{c}=0.8~~~$ \end{turn}}		
				&1 & -5.09326&	0.000697444 \\                 
				\multicolumn{1}{c}{} &2&-4.75149&	0.00095430 \\
				\multicolumn{1}{c}{} &3&-4.26808&	0.00151333  \\
                                \multicolumn{1}{c}{} &4&-3.72477&	0.0025312   \\
                                \multicolumn{1}{c}{} &5&-3.17865&	0.00419748  \\
                                \multicolumn{1}{c}{} &6&-2.65406&	0.0067545   \\
                                \multicolumn{1}{c}{} &7&-2.1536 &       0.0106855   \\     
                                \multicolumn{1}{c}{} &8&-1.67372&	0.0169562   \\        
                                \multicolumn{1}{c}{} &9&-1.21469&	0.0269818   \\
                                \multicolumn{1}{c}{} &10&-0.781636&	0.0424626   \\
                                \multicolumn{1}{c}{} &11&-0.378375&	0.0654182   \\
                                \multicolumn{1}{c}{} &12&0.0&	        0.0986779\\
			\end{tabular}		
		\end{ruledtabular}
	\end{center}

	\caption{
	  The quark-antiquark potential  color source separation distances $R$ and temperature $T/T_{c}=0.8$, the lattice parameters $\beta=6.0$, $N_t=10$ time slices and spatial volume $N_s=36^3$. The avaraging is taken for two Polyakov lines separated by distance $R$ Eq.~\eqref{PolyakovCorrelators}.}
\label{T9_Pot_Cor_T08}	      	
\end{table}   

%
   
\begin{table}[!hpt]
	\begin{center}
		\begin{ruledtabular}
			\begin{tabular}{cc|ccccccccc}
			  \multirow{2}{*}{$V_{NG}$} &Fit
				&\multicolumn{2}{c}{Fit Parameters, $T/T_{c}=0.8$} &\multicolumn{2}{c}{}\\
				&Interval
				&\multicolumn{1}{c}{$\sigma a^{2}$} &\multicolumn{1}{c}{$\mu$} 
			       &\multicolumn{1}{c}{$\chi^{2}$}\\
				\hline
                                \multirow{10}{*}{\begin{turn}{90}\hspace{1cm}$V_{{\ell o}}$ \end{turn}}
				&$[R_{m},R_{M}]$  &    &          &     \\
\multicolumn{1}{c}{} &[2,11]     &0.02520(8)& -0.4000(2)&21875.7    \\
\multicolumn{1}{c}{} &[3,11]     &0.0392(1) &-0.4586(5)&795.844\\
\multicolumn{1}{c}{} &[4,11]     &0.0435(2) &-0.482(1) &18.6047\\
\multicolumn{1}{c}{} &[5,11]     &0.045(3)  &-0.49(2)  &00\\
\multicolumn{1}{c}{} &[3,6]      &0.0373(2) &-0.4523(5)&399.716	\\
\multicolumn{1}{c}{} &[3,7]      &0.0381(1) &-0.4553(5)&566.884\\
\multicolumn{1}{c}{} &[3,10]     &0.0391(1) &-0.4583(4)&779.503 \\
\multicolumn{1}{c}{} &[4,10]     &0.0435(2) &-0.4816(9)&18.5844 \\
			\end{tabular}		
		\end{ruledtabular}
	\end{center}
          \caption{The $\chi^{2}$ values and the corresponding fit parameters; string tension and cutoff $(\sigma,\mu)$, returned from fits to the free string potential $V_{\ell o}$ given by Eq.~\eqref{Pot_NG_LO}.}
          \label{T10_Pot_T08_NG_LO}          
\end{table}
\begin{table}[!hpt]
	\begin{center}
		\begin{ruledtabular}
			\begin{tabular}{cc|ccccccccc}
			  \multirow{2}{*}{$V_{NG}$} &Fit
				&\multicolumn{2}{c}{Fit Parameters, $T/T_{c}=0.8$} &\multicolumn{2}{c}{}\\
				&Interval
				&\multicolumn{1}{c}{$\sigma a^{2}$} &\multicolumn{1}{c}{$\mu$} 
			       &\multicolumn{1}{c}{$\chi^{2}$}\\
				\hline
                                \multirow{10}{*}{\begin{turn}{90}\hspace{1cm}$V_{{n\ell o}}$ \end{turn}}
				&$[R_{m},R_{M}]$  &    &          &     \\
\multicolumn{1}{c}{} &[2,11]  &0.0240(2)&-0.3208(8)   &  340026  \\
\multicolumn{1}{c}{} &[3,11]  &0.0325(2)&-0.4084(6)   &  4609.75 \\
\multicolumn{1}{c}{} &[4,11]  &0.0422(2)&-0.463(1)    &  119.701\\
\multicolumn{1}{c}{} &[5,11]  &0.0449(3)&-0.481(2)    &  2.60999 \\
			\end{tabular}		
		\end{ruledtabular}
	\end{center}
\caption{The $\chi^{2}$ values and the corresponding fit parameters; string tension and cutoff $(\sigma,\mu)$, returned from fits to the self-interacting string potential at next to leading order $V_{n\ell o}$ given by Eq.~\eqref{Pot_NG_NLO}.}
\label{T10_Pot_T08_NG_NLO}  
\end{table}   
\begin{table*}[!hpt]
	\begin{center}
		\begin{ruledtabular}
			\begin{tabular}{cc|cccccccccc}
			  \multirow{2}{*}{\begin{turn}{90}\hspace{1cm}$ V_{{n\ell o}}^{b_2,b_4} $ \end{turn}  } &\tiny{Fit Interval}
				&\multicolumn{2}{c}{\tiny{Fit Parameters, $T/T_{c}=0.8$}} &\multicolumn{2}{c}{}\\
				&$R\in I$
				&\multicolumn{1}{c}{$\sigma_0 a^{2}$} &\multicolumn{1}{c}{$\mu$(LU)} 
			  &\multicolumn{1}{c}{$b_2$(LU)} &\multicolumn{1}{c}{$b_4$(LU)}&\multicolumn{1}{c}{$\chi^{2}_{\rm{dof}}$}\\
                          &&&&\\
				\hline
\multirow{10}{*}{\begin{turn}{90}\hspace{2cm}$V_{{\ell o}}^{b_{2}}$ \end{turn}}				
&(a)~~~~~~~~~~~~~~~~  &    &          &     \\
\multicolumn{1}{c}{} &[2,5]  &0.0400(2)  &-0.4689(9)     &-0.138(1) &0.0 & 50.9115  \\
\multicolumn{1}{c}{} &[2,11]  &0.0422(1) &-0.4767(6)	 &-0.148(1) &0.0&197.566   \\
\multicolumn{1}{c}{} &[3,11]  &0.0452(2) &-0.489(1)      &-0.607(9)	 &0.0 & 5.91231\\\hline
\multirow{10}{*}{\begin{turn}{90}\hspace{2cm}$V_{{\ell o}}^{b_{4}}$ \end{turn}}				
 &(b)~~~~~~~~~~~~~~~~  &    &          &     \\
\multicolumn{1}{c}{} &[2,5]  &0.0376637, 0.000209043&-0.456079, 0.000749928&0.0&-0.0481111,0.000478584 &107.329\\
\multicolumn{1}{c}{} &[3,7]  &0.0444179, 0.000296234&-0.486774, 0.00141769&0.0&-0.213006, 0.00896163&1.93\\
\multicolumn{1}{c}{} &[2,11] &0.040609, 0.000129929&-0.466179, 0.000499672&0.0&-0.0536944,0.000366839&451.37\\
\multicolumn{1}{c}{} &[3,11] &0.0445357, 0.000226426&-0.487308, 0.00111589&0.0&-0.215779, 0.00766315&2.96961\\\hline
\multirow{10}{*}{\begin{turn}{90}\hspace{2cm}$V_{{\ell o}}^{b2,b_{4}}$ \end{turn}}
 &(c)~~~~~~~~~~~~~~~~  &    &          &     \\
\multicolumn{1}{c}{} &[2,11]&0.0451733, 0.00025135&-0.496249, 0.00150301&-0.436979, 0.0205995&0.104738, 0.00747759&1.37428\\
\multicolumn{1}{c}{} &[3,11]&0.0450192, 0.000425963&-0.494044, 0.00514882&-0.327809, 0.24461&0.0244465, 0.17942&1.17367\\
			\end{tabular}		
		\end{ruledtabular}
	\end{center}
        \caption{The $\chi^{2}$ values and the corresponding fit parameters; string tension and cutoff $(\sigma,\mu)$ together with the boundary parameters $(b_2,b_4)$ returned from fits three possible combinations of the boundary terms at LO level (a) $V_{\ell o}^{b_2}$, (b) $V_{\ell o}^{b_4}$ and (c) $V_{\ell o}^{b_2,b_4}$ given by Eq.~\eqref{Pot_NG_LO_NLO_Boundb2},Eq.~\eqref{Pot_NG_LO_NLO_Boundb3} and Eq.~\eqref{Pot_NG_LO_NLO_Boundb2b3}, respectively.}
\label{T11_Pot_T08_NG_LO_Boundb2b3}  
\end{table*}

\begin{table*}[!hpt]
	\begin{center}
		\begin{ruledtabular}
			\begin{tabular}{cc|cccccccccc}
			  \multirow{2}{*}{\begin{turn}{90}\hspace{1cm}$ V_{{n\ell o}}^{b_2,b_4} $ \end{turn}  } &\tiny{Fit Interval}
				&\multicolumn{2}{c}{\tiny{Fit Parameters, $T/T_{c}=0.8$}} &\multicolumn{2}{c}{}\\
				&$R\in I$
				&\multicolumn{1}{c}{$\sigma_0 a^{2}$} &\multicolumn{1}{c}{$\mu$(LU)} 
			  &\multicolumn{1}{c}{$b_2$(LU)} &\multicolumn{1}{c}{$b_4$(LU)}&\multicolumn{1}{c}{$\chi^{2}_{\rm{dof}}$}\\
                         &&&& \\
				\hline
\multirow{10}{*}{\begin{turn}{90}\hspace{2cm}$V_{{n\ell o}}^{b_{2}}$ \end{turn}}				
&(a)~~~~~~~~~~~~~~~~  &    &          &     \\
\multicolumn{1}{c}{} &[2,11]  & 0.0393(2)  &-0.4521(7) &-0.3189(7)&0.0 &1005.75  \\
\multicolumn{1}{c}{} &[2,5]   &0.0332(3)   &-0.427(1) &-0.3077(8) &0.0 &259.723   \\
\multicolumn{1}{c}{} &[3,11]  &0.0452(2)   &-0.489(1) &-0.607(9)  &0.0 &5.91231  \\\hline
\multirow{10}{*}{\begin{turn}{90}\hspace{2cm}$V_{{n\ell o}}^{b_{4}}$ \end{turn}}
&(b)~~~~~~~~~~~~~~~~  &    &          &     \\
\multicolumn{1}{c}{} &[2,11]&0.0357542, 0.000149607&-0.427408, 0.000648593&0.0&-0.119262, 0.000259866&2447.1\\
\multicolumn{1}{c}{} &[2,5] &0.0262228, 0.000337028&-0.387406, 0.00151795 &0.0& -0.115406, 0.000226769&688.624\\
\multicolumn{1}{c}{} &[3,6] &0.0429852, 0.00039019 &-0.470346, 0.0019032  &0.0& -0.424624, 0.00898908&6.57496\\
\multicolumn{1}{c}{} &[3,7] &0.0436562, 0.000301773&-0.473489, 0.00150711 &0.0&-0.436586, 0.00787107&13.9662\\
\multicolumn{1}{c}{} &[3,11]&0.0442668, 0.000228102&-0.476414, 0.00117044 &0.0&-0.448646, 0.00685283&23.7994\\\hline
\multirow{10}{*}{\begin{turn}{90}\hspace{2cm}$V_{{n\ell o}}^{b_2,b_{4}}$ \end{turn}}
&(c)~~~~~~~~~~~~~~~~  &    &          &     \\
\multicolumn{1}{c}{} &           &             &        &    & \\
\multicolumn{1}{c}{} &[2,11]&0.0456057, 0.000249587&-0.495268, 0.0015125&-0.901874, 0.0186775&0.217754, 0.00695054&1.91735 \\
\multicolumn{1}{c}{} &[3,11]&0.0459421, 0.000418268&-0.50015, 0.00509954&-1.14027, 0.238561&0.394536, 0.176489&0.912238\\
			\end{tabular}		
		\end{ruledtabular}
	\end{center}
        \caption{The $\chi^{2}$ values and the corresponding fit parameters at NLO level in NG action; string tension and cutoff $(\sigma,\mu)$ together with the boundary parameters $(b_2,b_4)$ returned from the fits to three possible combinations of the boundary terms (a) $V_{n\ell o}^{b_2}$, (b) $V_{n\ell o}^{b_4}$ and (c) $V_{n\ell o}^{b_2,b_4}$ given by Eq.~\eqref{Pot_NG_LO_NLO_Boundb2}, Eq.~\eqref{Pot_NG_LO_NLO_Boundb3} and Eq.~\eqref{Pot_NG_LO_NLO_Boundb2b3}, respectively.}
\label{T12_Pot_T08_NG_LO_Boundb2b3}  
\end{table*}
\begin{table}[!hpt]
	\begin{center}
		\begin{ruledtabular}
			\begin{tabular}{cc|cccccccccc}
			  \multirow{2}{*}{\begin{turn}{90}\hspace{1cm}$  $ \end{turn}  } &\tiny{Fit Interval}
				&\multicolumn{2}{c}{\tiny{Fit Parameters, $T/T_{c}=0.8$}} &\multicolumn{2}{c}{}\\
				&$R\in I$
				&\multicolumn{1}{c}{$\sigma_0 a^{2}$} &\multicolumn{1}{c}{$\mu$(LU)} 
	       		  &\multicolumn{1}{c}{$\alpha_r$} &\multicolumn{1}{c}{$\chi^{2}$}\\
                          &&&&\\
				\hline
\multirow{10}{*}{\begin{turn}{90}\hspace{1cm}$V^{R}_{{\ell o}}$ \end{turn}}				
&\tiny{$[R_{m},R_{M}]$}  &    &          &    & \\
\multicolumn{1}{c}{}&(a)~~~~~~~~~  &    &          &    & \\
\multicolumn{1}{c}{} &[3,11]     &0.0392(1)    &-0.4586(4)  &0.000139251	   &795.844;2 \\
\multicolumn{1}{c}{} &[4,11]     &0.0436(2)    &-0.4816(9)  &0.000190263           &18.6047 \\
\multicolumn{1}{c}{} &[5,11]     &0.045	     &-0.5(2)       &                      &0.747119 \\\hline
\multirow{10}{*}{\begin{turn}{90}\hspace{2cm}$V_{{n\ell o}}^{R}$ \end{turn}}				
&(b)~~~~~~~~~  &    &          &     \\
\multicolumn{1}{c}{} &[3,11] &0.0326(1) & -0.4085(7) &  0.000334052    & 4609.75 \\
\multicolumn{1}{c}{} &[4,11] &0.0422(2) & -0.464(1)  &  0.000259934    & 23.94\\
\multicolumn{1}{c}{} &[5,11] &0.0449(1) & -0.481(2)  &  0.00349991     & 2.61\\
\multicolumn{1}{c}{} &[6,11] &0.0463(6) &-0.46(6)    & 0.138479  &0.312231 \\
			\end{tabular}		
		\end{ruledtabular}
	\end{center}
        \caption{The $\chi^{2}$ values and the corresponding fit parameters at both LO and NLO level in NG action; string tension and cutoff $(\sigma,\mu)$ together with the rigidity parameter $\alpha_r$ returned from the fits to (a) $V_{\ell o}^{R}$ and (b) $V_{n\ell o}^{R}$  given by Eq.~\eqref{Pot_NG_LO_NLO_Rigid}, respectively.}
\label{T12_Pot_T08_NG_LO_NLO_Rigid}  
\end{table}

\begin{figure*}[!hpt]
\centering
\subfigure[]{\includegraphics[scale=0.8]{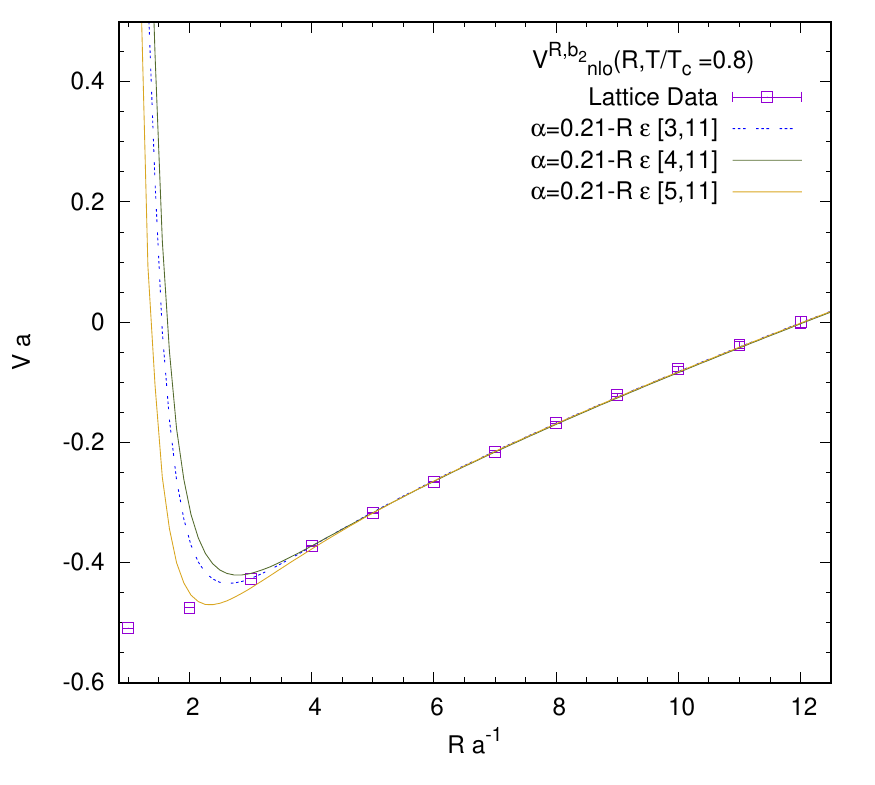}}
\subfigure[]{\includegraphics[scale=0.8]{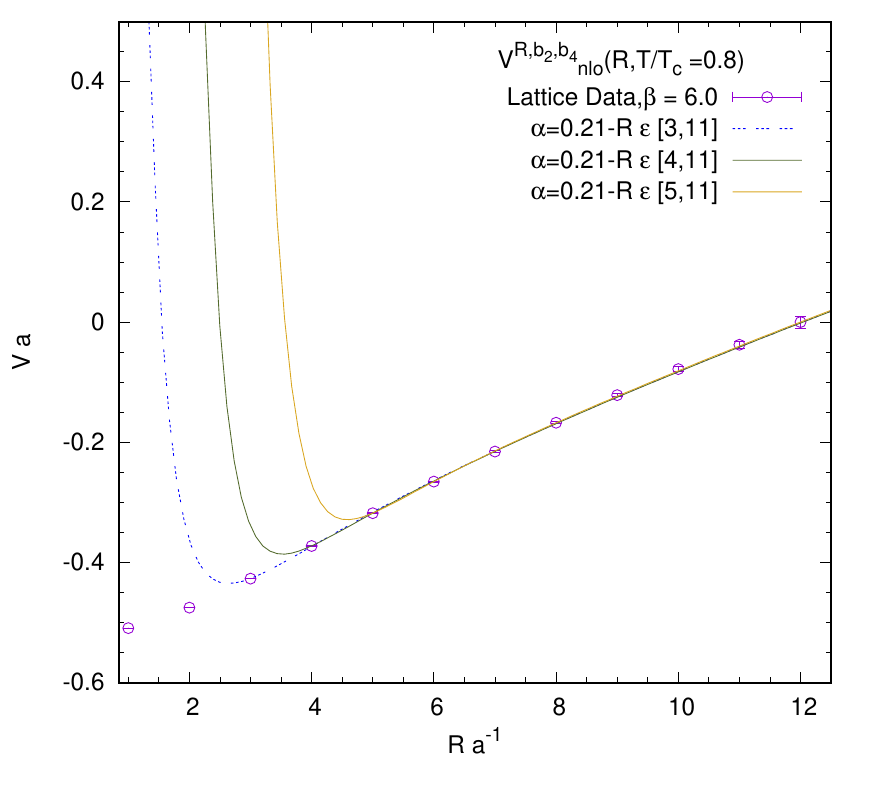}}
\caption{The quark-antiquark $Q \bar{Q}$ potential  $T/T_c=0.8$, (a)Compares the rigid string models with the boundary parameter $V_{\ell o}^{R,b_2}$ given by Eq.~\eqref{Pot_NG_LO_NLO_Rigid} and Eq.~\eqref{Pot_NG_LO_NLO_Rigid_Boundb2} for the depicted fit interval $R \in [R_{m},R_{M}]$ (b) Similar to (a); however, the corresponding fits are considered for two boundary parameters $(b_2,b_4)$ together with the leading order rigid string model $V_{\ell o}^{R,b_2,b_4}$ and the corresponding self-interacting model $V_{n\ell o}^{R,b_2,b_4}$ Eq.~\eqref{Pot_NG_LO_NLO_Rigid_Boundb2b3}.}
\label{Fits_Pot_T08_NG_LO_NLO_Boundb2b3_Rigid}
\end{figure*}

\begin{table*}[!hpt]
	\begin{center}
		\begin{ruledtabular}
			\begin{tabular}{cc|cccccccccc|cccccccccc|}
			  \multirow{2}{*}{\begin{turn}{90}\hspace{1cm}$ V_{{\ell o}}^{R,b_2} $ \end{turn}  } &\tiny{Fit Interval}
				&\multicolumn{2}{c}{\tiny{Fit Parameters, $T/T_{c}=0.8$}} &\multicolumn{2}{c}{}\\
				&$R\in I$
				&\multicolumn{1}{c}{$\sigma_0 a^{2}$} &\multicolumn{1}{c}{$\mu$(LU)} 
			  &\multicolumn{1}{c}{$\alpha$(LU)} &\multicolumn{1}{c}{$b2$}&\multicolumn{1}{c}{$b_3$}&\multicolumn{1}{c}{$\chi^2$}\\
                          \\
				\hline
\multirow{10}{*}{\begin{turn}{90}\hspace{2.5cm}\tiny{$V_{{\ell o}}^{R,b2}$} \end{turn}}				
&\tiny{$[R_{m},R_{M}]$}  &     &          &   &&&&  \\
\multicolumn{1}{c}{} &     &\\
\multicolumn{1}{c}{} &2-11 &0.0423778,	0.0055167&-0.476726,	0.0409872&-0.0175304,	0.56730& -0.159085,	0.620026&0.0&188.181;\\
\multicolumn{1}{c}{} &3-11    &0.0452433,	0.000546988&-0.489922,	0.0148488&-0.0415493,	0.0793197&-0.335264,	0.13425&0.0&1.12352;\\\hline
\multirow{10}{*}{\begin{turn}{90}\hspace{2.4cm}\tiny{$V_{{n\ell o}}^{R,b_2}$} \end{turn}}				
&2-7     &0.04(1)	&-0.44(9)&0.0165(1.0)&-0.326 (1.3)&0.0&707.179\\
\multicolumn{1}{c}{} &2-11    &0.039(6) &-0.45(4) &  0.0170071	(0.54)   &-0.33	(0.6) &0.0&979.948&\\
\multicolumn{1}{c}{} &3-11    &0.045(6) &-0.48(4)& 0.0192 (0.6)& -0.614308(0.6)&0.0&5.9\\
			\end{tabular}		
		\end{ruledtabular}
	\end{center}
	\caption{The $\chi^{2}$ values and the corresponding fit parameters returned from fits to the leading order (NLO) static potential with boundary terms Eq.~\eqref{Pot_NG_LO_NLO_Boundb2} and Eq.~\eqref{Pot_NG_LO_NLO_Boundb2b3}.}
\label{Tboundary}	
\end{table*}


\subsubsection{The pure Nambu-Goto action}
   A large value of $\chi^2$ is returned for fits of color sources separations commencing from $R=0.4$. For separations distance $R\leq 0.4$ fm the NG string description is showing increasingly significant deviations from the LGT data due to the short distance physics and the one dimensional idealization of NG string. In Refs.~\cite{Caselle:2012rp,Vyas:2010wg} the intrinsic thickness of the flux-tube has been discussed. 

   As Table~\ref{T10_Pot_T08_NG_LO} depicts, Excluding the point $R=0.4$ fm dramatically decreases the returned value of $\chi^{2}$ for both the leading order and the next to leading approximation Eqs.\eqref{Pot_NG_LO} and Eqs.~\eqref{Pot_NG_NLO}, respectively. The returned values of the string tension parameter quickly reaches stability even by the exclusion of further points at short distances $R=0.5$ fm and $R=0.6$ fm from the fit range. At this temperature, the string tension settles at a stable value of $\sigma_{0}\, a^{2}=0.0445$ measured in lattice units.

   The measured value of $\sigma_{0} a^{2}$ at $T/T_c=0.8$ making use of the fits of LO and NLO approximations are the same within the standard deviation of the measurements. We take this value of the string tension as a reference value for the zero-temperature string tension $\sigma_0=0.0445$ measured also in ~\cite{Koma:2017hcm}. The numerical data for the $Q\bar{Q}$ potential match both the free leading-order NG string and the NLO self-interacting form of Eq.\eqref{Pot_NG_LO}Eq.\eqref{Pot_NG_NLO}. Approximately the same difference in the value of the string tension is retrieved for fit domains involving short to large $Q\bar{Q}$ separation distances.
   
   Considering the fit of the same data of $Q\bar{Q}$ potential to the two-loop expression of the NG string Eq.~\eqref{Pot_NG_NLO}, the value of $\chi_{\rm{dof}}^2$ does not apprise mismatches for source separation distances commencing from $R>0.5$ fm. As shown in Table.~\ref{T10_Pot_T08_NG_NLO}, for different fit ranges with a fixed end point at $R=1.2$ fm, the fit return acceptable values of $\chi^2$ with subtle changes in the free fit parameter $\sigma_{0} a^{2}$. The absence of the mismatch between Eq.~\eqref{Pot_NG_NLO} and the numerical data at this temperature scale does not rule out the validity at this temperature scale.

   This points out to the minor role of the higher order modes at the end of the QCD plateau $T/T_{c}=0.8$. The pale out of the thermal effects together with a flat plateau region at this temperature is present as well in the string tension measurements~\cite{PhysRevD.85.077501} and the more recent Monte-Carlo measurements ~\cite{Koma:2017hcm} which reproduces the same value of $0.044$ of the zero temperature string tension.
   
   It is worth noting, on the otherhand, that the NLO terms alone esclates the fit on intervals $R .le. 0.5$ apprising an increase in the $\chi^{2}$ values by around 6 times larger. The enclosure of the fourth derivative term of the NLO term in the NG action appears, thereof, neither to alter the poor parameterisation behavior nor to indicate significant changes on the value of the string tension shown in Table.~\ref{T10_Pot_T08_NG_NLO}.

\subsubsection{Boundary terms in L\"uscher-Weisz action}    
   The values of $\chi^2$ returned from the fits to $V_{\ell o}^{b_2}$ and $V_{n \ell o}^{b_4}$ Eq.~\eqref{Pot_NG_LO_NLO_Boundb2} are enlisted in Table~\ref{T12_Pot_T08_NG_LO_Boundb2b3}. The consideration of the boundary terms $V_{\ell o}^{b_2}$ or $V_{n \ell o}^{b_4}$ persist to provide good $\chi^{2}$ values over short distance intervals $R \in [0.3,0.7]$ fm. The same observation holds at the other temperature $T/T_{c}=0.9$. These are well-known deviations from the free bosonic string over short distance even at zero temperature as well. This may suggest a role to the boundary terms for the deviation from the free non-interacting model over short distances $R<0.5$ fm.    

 However, the possible fit ansatz $V_{n \ell o}^{b_2}$, $V_{n \ell o}^{b_4}$ or $V_{n \ell o}^{b_2,b_4}$ at NLO in NG string and $V_{ \ell o}^{b_2}$, $V_{\ell o}^{b_4}$ or $V_{ \ell o}^{b_2,b_4}$ at leading order does not produce good $\chi^{2}$ for any fit interval involving $R=0.2$ fm. Moreover, the first three fit ansatz involving the string self-interaction seem to escalate the fit over this short distance and temperature. The minimal residuals of the fits over the interval $R \in [0.2,1.1]$ fm is provided by the ansatz $V_{\ell o}^{b_2,b_3}$ which produces $\chi^{2}_{dof}=139.7/(9-5)$.
 
 
  Figure~\ref{Fits_Pot_T08_NG_LO_NLO_Boundb2b3} illustrates the fitted potential curves and the recieved small residuals over short distances as $R=0.3$ fm when the boundary term included into both LO and NLO NG string potentials. The potential curves are scaled by the square of the separation distance $R^{2}$ to magnify the long distances fits and the corresponding residuals. The plot exhibits the large value of the residuals and deviations over long distances when the point at $R=0.2$ fm is included into the fit intervals. These deviations are much less compared to the deviations over the same fit interval considering the pure NG string  without boundary terms.

\subsubsection{Rigidity terms in Polyakov-Kleinert action}
  Unlike the relative reduction in the square of the residuals $\chi^{2}$ at $T/T_c=0.9$ over all distances, very small values of the rigidity parameter is returned at this temperature $T/T_c=0.8$. The inclusure of the rigidity terms as well as self-interactions at this temperature do not return any sigficant improvement in the fit behavior at short distances at this temperature scale.  

  Effects such as the string's rigidity and self-interactions seem to become noticeable at higher temperatures and energies, the question should be posed here is whether these terms with the returned values at higher temperature $T/T_c=0.9$ are consistent at $T/T_c=0.8$. 
  
  In Table we inlisted the returned values $\chi^{2}$ at the intermediate distances $0.5 \le R \le 1.1$ fm the self-interactions and rigidity do not escalates the fits. At the temperature $T/T_c=0.8$ and string tension value of $\sigma_{0} a^{2}=0.0445$,  Fig(12-a) shows the two solid curves which are the best fits over intervals $R \in [4,11]$ and $R \in [5,11]$ with good values $\chi^{2}$, respectively. The fit parameters are given
  
\begin{equation}
\left(
\begin{array}{ccc}
    & \text{Estimate} & \text{Standard Error}  \\
  R_{0}  & -0.441441 & 0.000680461  \\
 b_2  & -1.03459 & 0.03637  \\
  b_4  & 0.0 & 0.0  \\
\end{array}
\right),
\end{equation}

and 
\begin{equation}
\left(
  \begin{array}{ccc}
 \text{} & \text{Estimate} & \text{Standard Error} \\
 R_0 & -0.436602 & 0.0023778  \\
 b_2 & -0.579979 & 0.217104  \\
 b_4 & 0.0 & 0.0  \\
  \end{array}
  \right).
\end{equation}
Taking into account the additional degree of freedom endowed by $b_4$, the fits to Eq.~\eqref{Pot_NG_LO_NLO_Rigid_Boundb2b3} are returning $\chi^{2}=3.2$ over fit intervals $R \in [5,11]$ the parameter values
\begin{equation}
\left(
\begin{array}{ccc}
 \text{} & \text{Estimate} & \text{Standard Error}  \\
 R_0 & 0.819389 & 0.41822  \\
 b_2 & 103.955 & 34.7415  \\
 b_4 & -125.381 & 41.8451  \\
\end{array}
\right)
\end{equation}
   the rigidity parameter in both cases is $\alpha_{r}=0.21$ which is the same measured at $T/T_{c}=0.9$ (within the numerical uncertainties as shown in Table.~\ref{T12_Pot_T08_NG_LO_NLO_Rigid}).
  
\subsection{String tension}   
    A discussion concerning the order of the phase transition and the value of the temperature at the critical point would be out of the scope of the present discussion. However, a correct string tension dependancy on the temperature entails that $\sigma(T) a^{2}$ at higher temperature would fall into the same theoretical curve fixed by the plateau value of $\sigma_0 a^{2}$. This is equivalent to say that that all fits to the $Q\bar{Q}$ potential are returning the same value of $\sigma_{0}a^{2}$ measured at zero temperature.
   
\begin{figure}[!hptb]
\centering
\subfigure[]{\includegraphics[scale=0.42]{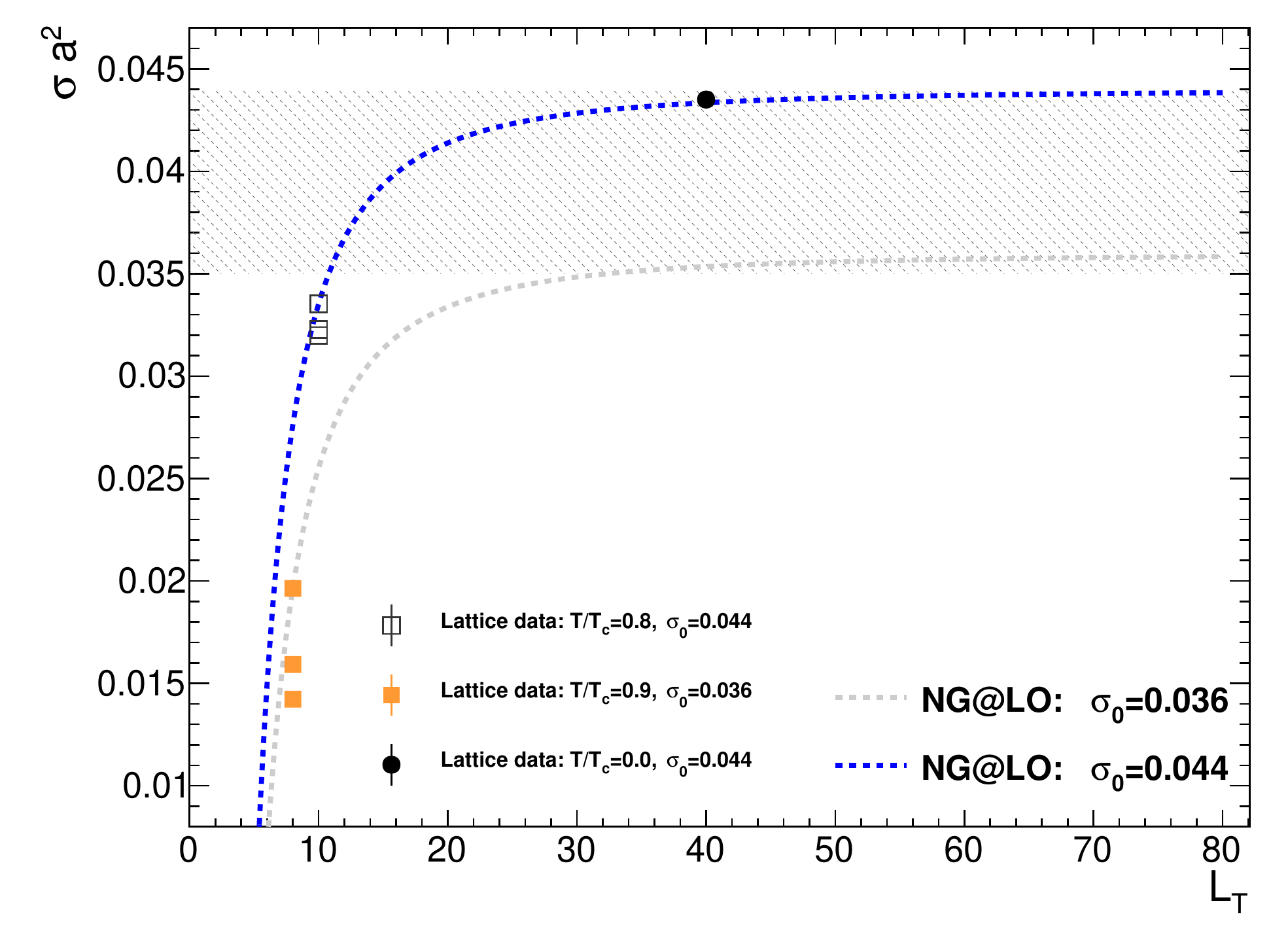}}
\subfigure[]{\includegraphics[scale=0.42]{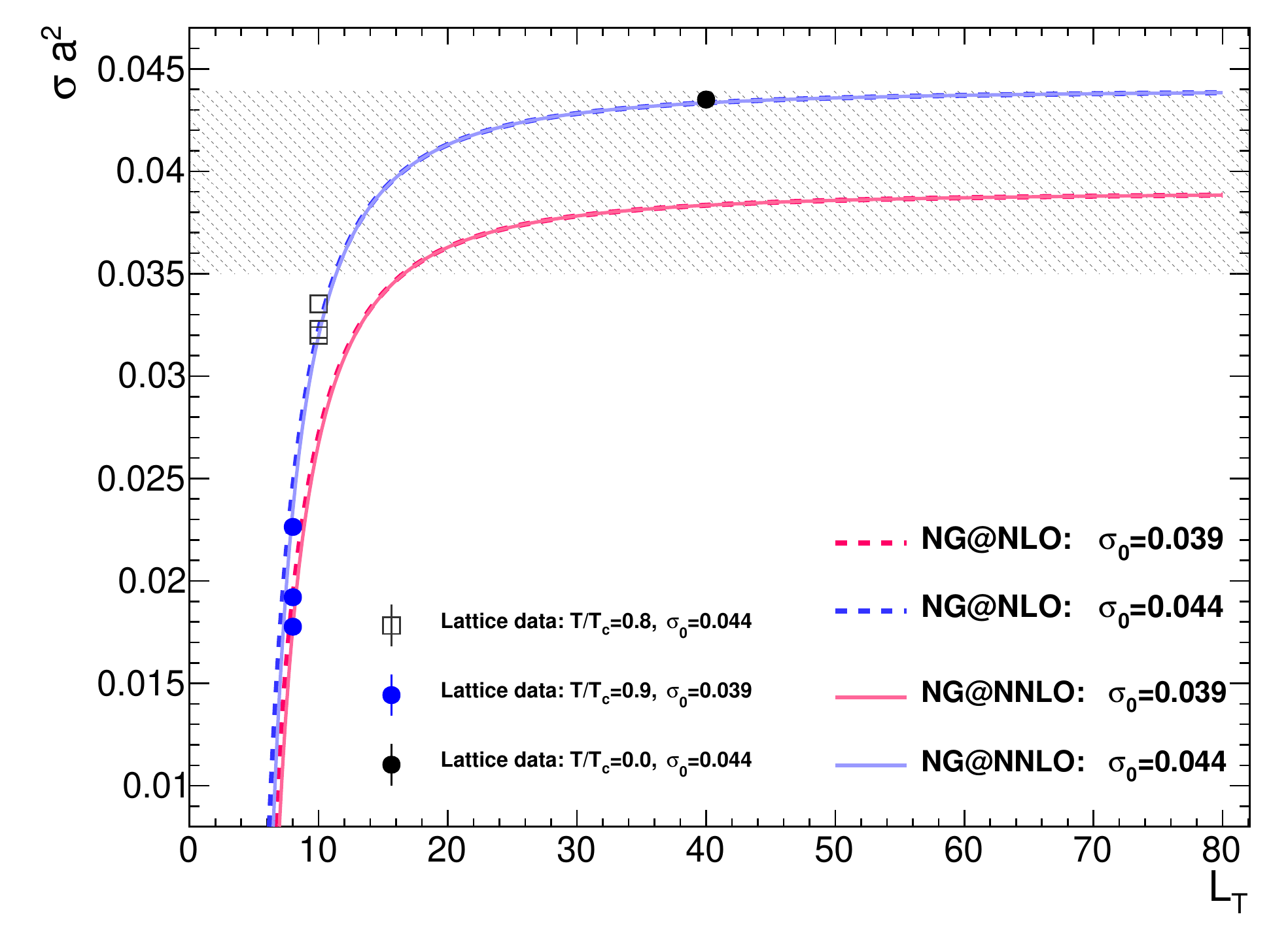}}
\caption{The temperature-dependence of the string tension for Nambu-Goto string action at LO, NLO and NNLO perturbative expansion. The dashed lines correspond to $\sigma_{0}a^{2}=0.044$ and solid line corresponds to  $\sigma_{0}a^{2}=0.039$.}
\label{TensionT}
\label{NG_String_Tension}
\end{figure}

\begin{figure}[!hptb]
\centering
\includegraphics[scale=0.85]{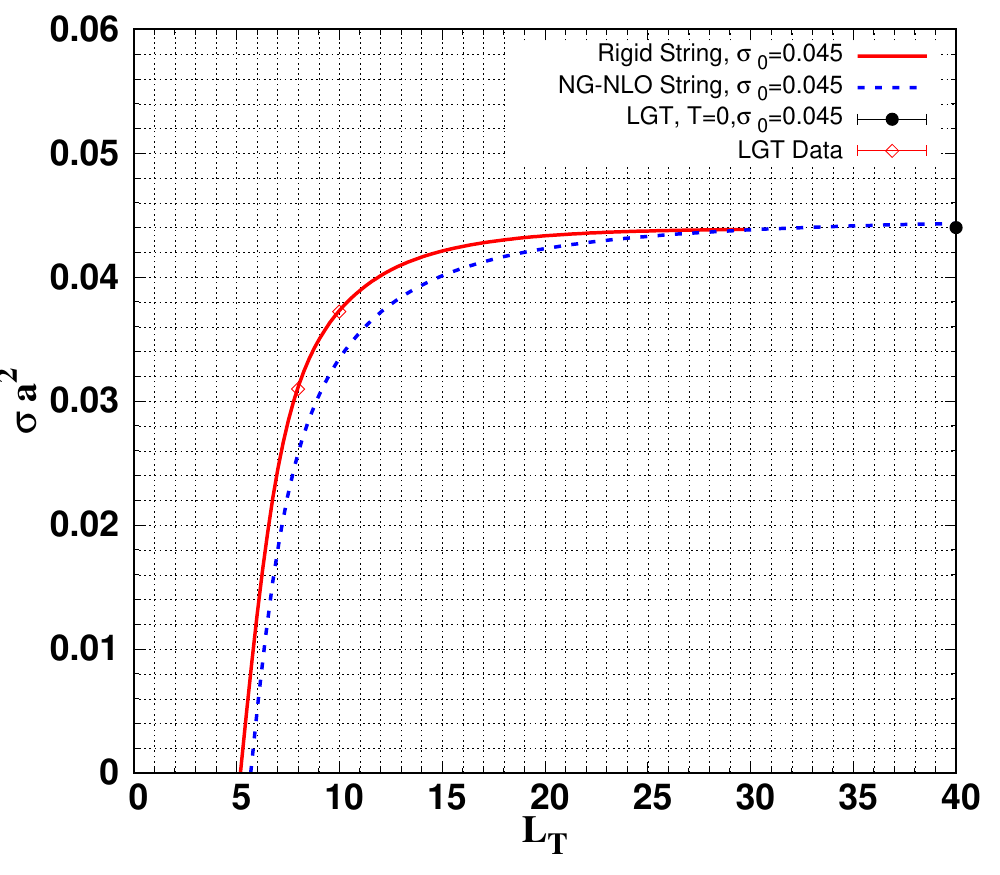}
\caption{The solid line shows the temperature-dependence of the string tension for rigid string at NLO in NG perturbative expansionfor two values of the parameter $\alpha=0.3$ and $\alpha=0.6$. The dashed lines correspond to $\sigma_{0}a^{2}=0.0445$ and solid line corresponds to rigid string model at one loop order.}
\label{NG_Rigid_String_Tension}
\end{figure}

    In Fig~\ref{NG_String_Tension} each theoretical curve is a plot corresponding to the respective order in the NG power expansion. Each of these curves is well defined by the plateau value of $\sigma_0 a^{2}$ and endows the string tension dependency on the temperature. Usually, this envolves a measurement extracted from the lattice data through the slop of the linearly rising potential at low-enough temperature.
    
   The perturbative string tension up to the fourth power in the temperature is layed out through Eq.\eqref{Tension_NG_LO} and Eq.\eqref{Tension_NG_NLO}. The exact temperature-dependent NG string tension is given by Eq.~\eqref{Tension_NG_Exact}. A probable role of more higher-order terms of the power expansion of NG string action may be discussed in the context of the string tension dependency on the temperature. 

   At the leading order NG string-tension the lattice data point at $T/T_c=0.8$ produces $\sigma(0.8T_c) a^{2}=0.038$ at the fixed parameter $\sigma_0 a^2=0.045$, which fits well into the LO of NG string Eq.~\eqref{Tension_NG_LO} curve as depicted in Fig.~\ref{NG_String_Tension}. However, at the temperature $T/T_c=0.9$ the string tension yields a value  $\sigma_{\ell o}(0.9T_c)a^{2}=0.0245952$ for the same fixed parameter $\sigma_0a^2=0.045$. This value deviates by $15 \%$  from that returned from the fit of LO potential of NG string $\sigma(0.9T_c)a^{2}=0.0192051$ for fits over the largest interval $R \in [0.9,1.1]$ fm.

   These deviations at $T/T_c=0.9$ reduce to $\sigma_{n \ell o}(T)a^{2}=0.0234638$ at NLO Eq.~\eqref{Tension_NG_NLO} compared to that retrieved from the fits over the same interval. The data point largly deviates from LO of NG string curve as shown in Fig.~\ref{NG_String_Tension}. Moreover, very small correction are received from the term proportional to the six power in the temperature. At the fourth and sixth order the string tension is $\sigma(T)a^{2}= $ and $\sigma(T)a^{2}=0.0234638$, respectively.
    

   The deviations of the theoretical lines from the plateau value at $\sigma_{0}a^{2}=0.045$ indicate that all orders in NG action do not provide the correct behavior of the temperature-dependent string tension in the present four-dimensional Yang-Mills model. In addition to that, the boundary corrections $(V^{b_2},V^{b_4})$ Eq.~\eqref{Pot_Boundb2} and Eq.~\eqref{Pot_Boundb4} do not contribute to the renormalization of the string tension, since the modular transform of the potential does not produce terms linearly proprtional to the string length $R$.

   Nevertheless, the extrinsic curvature terms in PK action do redefine the temperature-dependence of the string tension. The corrections to the NG string tension are uniquely determined by the value of the rigidity parameter.

   In Tables~\ref{T7_Pot_T09_NG_NLO_Rigid}, the enlisted parameters of the rigid string model at $T/T_c=0.9$ show returned average value of $\sigma_0 a^{2}=0.044$ for fits over both intervals $R \in [0.7,1.1]$ fm and $R \in [0.8,1.1]$ fm. Similarly, in Table~\ref{T7_Pot_T09_NG_NLO_Rigid_Boundb2} where the rigid string fit ansatz includes boundary correction $V^{b_2}$ Eq.~\eqref{Pot_NG_LO_NLO_Rigid_Boundb2} a stable value is returned over $R \in [0.6,1.1]$ fm and $R \in [0.7,1.1]$ fm.

   The string tension dependency on the temperature in rigid string model is given by the series expansion Eq.~\ref{StringTensionRigid}, with the asymptotic forms given by Eq.~\eqref{LT} and Eq.~\eqref{HT} at low and high temperatures, respectively.

    Figure.~\ref{NG_Rigid_String_Tension} compares the fourth power NG string tension at next to leading order given by Eq.\eqref{Tension_NG_NLO} and the corresponding rigid string-tension given by Eq.~\eqref{HT} versus the temperature for the returned $\alpha=0.3$.

   The rigid string is showing a more flat region at the end of the plateau region rather than the pure NG string. This observation has been reported in model independent calculations of Ref.~\cite{PhysRevD.85.077501}. For rigidity parameter $\alpha=0.3$ the string tension at $\sigma_0 a^{2}=0.0445$ at both temperature scale $T/T_c=0.8$ and $T/T_c=0.9$. It would be desirable to include more lattice data~\cite{PhysRevD.85.077501} at lower and higher temperature to well-establish the flatness of QCD transition curve, which we report in the future. 

\section{Summary and Concluding Remarks}   
   In this work, we compared the static quark-antiquark potential of effective bosonic string model of confinement beyond free Nambu-Goto approximation to the Mont-Carlo lattice data~\cite{talk}. The study mainly targets the color source separation $R=0.5$ fm to $R=1.2$ fm, where it is well-known that the free string poorly describe the string tension and quark-antiquark potential in the vicinity of the critical point.

  The Nambu-Goto (NG) action at two-loop order have been set into comparison with the corresponding  $SU(3)$ Yang-Mills lattice data in four dimensions. The theoretical predictions laid down by the effects of boundary terms in  L\"uscher-Weisz (LW) action and extrinsic curvature in Polyakov-Kleinert (PK) action have been also considered.

  Both the LO and the NLO approximations of Nambu-Goto string show a good fit behavior for the data corresponding to the $Q\bar{Q}$ potential near the end of the QCD plateau region, namely, at $T/T_{c}=0.8$. The fit returns almost the same parameterization behavior with negligible differences for the measured zero temperature string tension $\sigma_{0}a^{2}$. The returned value of this fit paramter is in agreement with the measurements at zero temperature~\cite{Koma:2017hcm}.

   We detect signatures of two boundary terms of the Lüscher-Weisz (LW) string action. The (LW) string with boundary action is yielding a static potential which is in a good agreement with the static potential lattice data as well, however, for color source separation as short as $R=0.3$ fm.
  
  However, at higher temperature near the deconfinement point $T/T_{c}=0.9$ the fits to the Nambu-Goto string model considering either LO or NLO approximation poorly describes the lattice data of the static $Q\bar{Q}$ potential data for the fit region span the distances under scrutiny $R \in [0.5,1.2]$ fm.

  Nevertheless, the fits show reduction of the residuals for the next-to-leading order approximation of the NG string on each corresponding fit interval. In both LO and NLO of NG string good $\chi^{2}_{\rm{dof}}$ is attained by the exclusion of the data points at short distances, i.e., $R \in[0.9,1.2]$ fm.

  The effective description based only on Nambu-Goto model does not accurately describe the $Q\bar{Q}$ potential data which occur as a deviation from the standard value of the string tension and the static potential data. The fit to the Casimir energy of the self-interacting string returns a value of the zero temperature string tension $\sigma_{0} a^{2}=0.041$ which deviates by $11\%$ of that measured at $T/T_{c}=0.8$ and zero temperature. This motivated discussing other effects such as the interaction with the boundaries and stiffness of the flux-tube.

  The inclusion of leading boundary term of L\"uscher-Weisz action in the approximation scheme reduces the residuals of the at all the considered source separations, however, deviations from the value of the zero temperature string tension $\sigma_{0}a^{2}$ do not diminish.

  The fit of the static potential considering boundary terms of LW action and contributions from the extrinsic curvature of PK action show a significant improvement compared to that considering merely the ordinary Nambu-Goto string for the intermediate and asymptotic color source separation distances $R \in [0.5,1.2]$. The fits reproduce an acceptable value of $\chi^{2}$ and a zero temperature string tension $\sigma_{0}a^{2}$ measured at $T/T_c=0.8$ or at $T=0$ ~\cite{Koma:2017hcm}, thus, indicating a correct temperature dependence of the string tension.
   
  The following enlists intervals over which the optimal value of $\chi^{2}_{\rm{dof}}$ is returned from the fit of the corresponding string model:
  \begin{itemize}
    
  \item The model  encompassing one boundary term $V^{b_2}_{n \ell o}$ given by Eq.~\eqref{Pot_NG_LO_NLO_Boundb2} return best fit on the interval $R\in[0.8,1.1]$ fm. The rigid string model $V^{R}_{n \ell o}$ of Eq.~\eqref{Pot_NG_LO_NLO_Rigid} retrieves best fit over interval $R\in[0.7,1.1]$ fm.

  \item  Considering both rigidity and boundary term $b_2$ the model $V^{R,b_2}_{n \ell o}$ of Eq.~\eqref{Pot_NG_LO_NLO_Rigid_Boundb2} extendsbest fit to the interval $R \in [0.6,1.1]$ fm. Eventually, the next boundary-correction $b_4$ of the model $V^{R,b_2,b_4}_{n \ell o}$ of Eq.~\eqref{Pot_NG_LO_NLO_Rigid_Boundb2b3} reproduces best fit over the interval $R \in [0.5,1.1]$ fm.
    
 \end{itemize}


\begin{acknowledgments}
   We are thankful to S. Brandt, C. Bonati, M. Casselle, Ph. de Forcrand and T. Filk for their very useful comments. This work has been funded by the Chinese Academy of Sciences President's International Fellowship Initiative grants No.2015PM062 and No.2016PM043, the Recruitment Program of Foreign Experts, the Polish National Science Centre (NCN) grant 2016/23/B/ST2/00692, NSFC grants (Nos.~11035006,~11175215,~11175220) and the Hundred Talent Program of the Chinese Academy of Sciences (Y101020BR0).
\end{acknowledgments}

\bibliography{Biblio2}%

\end{document}